\documentclass[letterpaper,10pt]{article}
\usepackage{amsfonts}
\usepackage{amssymb, amsmath, amsthm,  graphicx}
\usepackage[margin=1in]{geometry}
\usepackage[bottom]{footmisc}
\usepackage{commath}
\usepackage[utf8]{inputenc}
\usepackage{natbib}
\usepackage{booktabs, multirow}
\usepackage{url}
\usepackage[normalem]{ulem}
\usepackage{lscape}
\useunder{\uline}{\ul}{}
\usepackage{accents}
\usepackage[flushleft]{threeparttable}
\usepackage{adjustbox,appendix}

\renewcommand{\baselinestretch}{1.5}
\DeclareMathOperator*{\argmax}{\arg\max}

\DeclareMathOperator*{\argmin}{\arg\min}

\newtheorem{thm}{Theorem}

\newtheorem{lemma}{Lemma}
\newtheorem{corollary}{Corollary}
\newtheorem{assum}{Assumption}

\newtheorem{assumL}{Assumption}

\def\E{\mathbb{E}}
\def\P{\mathbb{P}}
\def\R{\mathbb{R}}

\def\b{\boldsymbol}

\begin{document}

\title{ Sharpe Ratio  Analysis in High Dimensions:\\Residual-Based Nodewise Regression in Factor Models}
\author{\textsc{Mehmet Caner\thanks{
North Carolina State University, Nelson Hall, Department of Economics, NC 27695. Email: mcaner@ncsu.edu. }}
\and \textsc {Marcelo Medeiros%
\thanks{Department of Economics, Pontifical Catholic University of Rio de Janeiro - Brazil. Email: mcm@econ.puc-rio.br}}
\and \textsc{Gabriel F. R. Vasconcelos%
\thanks{Head of Quantitative Research, BOCOM BBM Bank. Av. Barão de Tefé, 34 - 20º e 21º floors - Rio de Janeiro - RJ, 20220-460. Email: gabrielvasconcelos@bocombbm.com.br.
We are very grateful to the co-editor, Torben Andersen, the Associate Editor and an anonymous referee for very insightful comments and suggestions which led to a much improved version of the manuscript. We thank Vanderbilt Economics Department seminar guests and the participants of the World Congress of the Econometric Society for useful comments. Finally, we are thankful for the comments by Harold Chiang, Maurizio Daniele, Anders Kock, Srini Krishnamurthy, and Michael Wolf. Medeiros acknowledges the partial financial \textnormal{sup}port from CNPq and CAPES.
}}}

\date{\today}

\maketitle


\begin{abstract}
We provide a new theory for nodewise regression when the residuals from a fitted factor model are used. We apply our results to the analysis of the consistency of Sharpe Ratio estimators when there are many assets in a portfolio. We allow for an increasing number of assets as well as time observations of the portfolio. Since the nodewise regression is not feasible due to the unknown nature of idiosyncratic errors, we provide a feasible-residual-based nodewise regression to estimate the precision matrix of errors which is consistent  even when number of assets, $p$, exceeds the time span of the portfolio, $n$. In another new development, we also show that the precision matrix of returns can be estimated consistently, even with an increasing number of factors and $p>n$.  We show that: (1) with $p>n$, the Sharpe Ratio estimators are consistent in global minimum-variance and mean-variance portfolios; and  (2) with $p>n$, the maximum Sharpe Ratio estimator is consistent when the portfolio weights sum to one; and (3) with $p<<n$, the maximum-out-of-sample Sharpe Ratio estimator is consistent.
\end{abstract}

\section{Introduction}

One of the key issues in finance, especially in empirical asset pricing, is the trade-off between the returns and the risk of a portfolio. One important way to quantify such trade-off is via the Sharpe Ratio. 

We contribute to this literature by studying the case when the number of assets, namely $p$, grows with the time span of the portfolio, $n$. To obtain the Sharpe Ratio, and also its maximum, we make use of the asset return's precision matrix. In order to get an estimate of the precision matrix for asset returns in a large portfolio, we propose that an approximate factor model governs the dynamics of excess returns. Hence, asset returns (excess returns over a risk-free asset) can be explained by an increasing but known number of factors with unknown idiosyncratic errors entering the linear relation in an additive way. One major difference with the previous literature is that, in our case, the precision matrix has to be sparse. Therefore, this is a hybrid method that combines factor models with high-dimensional econometrics.

The first step in getting the Sharpe Ratio and its maximum involves the estimation of the precision matrix of the idiosyncratic terms (errors). Estimating the such precision matrix is not an easy task, and the simple nodewise regression idea as in \cite{mein2006} is not feasible. Therefore, we provide a simple, feasible residual-based nodewise regression method to estimate the precision matrix of errors in a factor model setup even if $p>n$. This feasible residual-based nodewise regression is a new idea, and it is shown to be consistently estimating the precision matrix of the errors which is our first contribution. Next, we obtain consistent estimators to the precision matrix of asset returns, even if $p>n$, which is our second technical contribution. Although, we focus on factor models in asset pricing, our methodology can be applied to any situation where the interest is the precision matrix of the errors of a linear regression model. 

Next, by using the precision matrix estimator for returns we can link our technical analysis to the financial econometrics literature. We make three contributions towards Sharpe Ratio analysis. First, we consider the Sharpe Ratios in the global minimum-variance portfolio and Markowitz mean-variance portfolio. We develop consistent estimators even if $p>n$, and both dimensions diverge. Second, we consider the rate of convergence and consistency of the maximum Sharpe Ratio when the portfolio weights are normalized to one. Recently, \cite{maller2002}, and \cite{maller2016} analyze the limit with a fixed number of assets and extend that approach to a large number of assets, but a number less than the time span of the portfolio. Their papers make a key discovery: in the case of weight constraints (summing to one), the formula for the maximum Sharpe Ratio depends on a technical term, unlike the unconstrained maximum Sharpe Ratio case. Practitioners could obtain the minimum Sharpe Ratio instead of the maximum if they are using the unconstrained formula. Our paper extends their paper by analyzing two issues. First, the case if $p>n$, with both quantities growing to infinity, and second, by handling the uncertainty created by this technical term, which we can estimate and use to obtain a new constrained and consistent maximum Sharpe Ratio. The assumption of constant loadings in the factor model is clearly a constraint for portfolio analysis over longer horizons. However, the setup where $p>n$ provides the statistical tools for us to analyze portfolios in short horizons and small samples as high-dimensional asymptotics can be seen as a good approximation for situations when $n$ is small but $p$ is large compared to $n$. Third, only in the case of $p<<n$, we obtain the consistency of our nodewise-based maximum-out-of-sample Sharpe Ratio estimate, with both $p,n$ growing to infinity and $p/n \to 0$. We also provide an analysis of the Sharpe Ratio with only portfolio weights estimated in the formula. In that way, we can see the effect of estimated portfolio on getting the optimal Sharpe Ratio. Our analysis shows this is possible when $p<n$ only.

\subsection{The Sparsity of the Precision Matrix}

There are several reasons motivating the assumption of sparsity of the precision matrix of the errors from the factor model. In technical terms, this is a convenient and widely used asymptotic tool when we want to consider high dimensional problems when $p>n$. The sparsity assumption on the precision matrix of errors gives rise to a direct way of estimating the precision matrix for the returns via Sherman-Morrison-Woodbury formula. We solve two technical issues with this assumption. First, consistent estimation of the precision matrix of returns is possible, yielding consistent estimation of the Sharpe Ratio and it's maximum, even in constrained case. Also, as far as we know, in the case of $p>n$, we do not know any other consistent estimation results for global minimum variance and Markowitz portfolios, as well as the constrained maximum Sharpe Ratio in the literature.

The sparsity assumption on the precision matrix of the errors from a factor model can be also justified in situations of interest in the empirical finance literature. First, even though we do not assume normality of the errors here, in this particular case the conditional independence of two errors given all the other errors, is represented by a zero entry in the precision matrix of errors. This is explained in p.1436-1439 of \cite{mein2006}. So, in the case of normally distributed data, sparsity can be thought as a conditional independence restriction. When the errors follow an elliptical distribution, conditional uncorrelatedness of two errors amount to a zero cell in the precision matrix as discussed in Section 2.4 of \cite{flw2018}. The authors claim that sparse precision matrix may be more useful when we estimate a network of stocks, by taking out common factors from returns and analyzing the conditional independence among idiosyncratic components (errors). Finally, there are a number of recent papers in the literature showing that after removing common factors, the covariance matrix of the errors is ``almost'' block diagonal, yielding a sparse precision matrix; see, for example, \citet{fan2016factors} and \citet{dBmMrR2018}. When the covariance matrix is block-diagonal, the precision matrix can be computed by inverting the estimated covariance matrix, which in turn can be consistently estimated by several different methods. However, even in this case, there are potential benefits of estimating the precision matrix directly as shown in our simulations and empirical exercise; see also \citet{ieee}.

\subsection{A Brief Review of the Literature and Main Takeaways}

In terms of the literature on nodewise regression and related methods, the most relevant papers are as follows. \cite{mein2006} establish the nodewise regression approach and provide an optimality result when data are normally distributed. \cite{chang2019} extend the nodewise regression method to time-series data and build confidence intervals for the elements in the precision matrix. However, the goal of \cite{chang2019} only centers on the elements of the precision matrix, and there is no connection to factor models. Furthermore, their results are based on the precision matrix of observed data and not on the residuals of a first-stage estimator. Finally, the authors do not consider the case of maximum Sharpe Ratio, and it is not clear if their results are directly applicable  to financial applications. \cite{canerkock2018} establish uniform confidence intervals in the case of high-dimensional parameters in heteroskedastic setups using nodewise regression, but, as in the previous paper, there is no connection to factor models in empirical finance. \cite{caner2019} provide the variance, the risk, and the weight estimation of a portfolio via nodewise regression. They take the nodewise regression directly from \cite{mein2006} and apply it to returns. However, they assume that the precision matrix of returns is sparse. Hence, it is more restrictive and less realistic than the method we propose. We combine factor models with the sparsity of the precision matrix of errors. As a consequence, our method is much more connected to typical empirical asset pricing models. Furthermore, we do not impose any sparsity on the precision matrix of returns. \cite{caner2019} also has no proofs about the estimation of the Sharpe Ratio.


In terms of recent contributions to the literature on factor models and sparse regression, we highlight \cite{jFrMmM2021}. The authors consider the combination of factor models and sparse regression in a very general setting. More specifically, they analyze a panel data model with a factor structure and idiosyncratic terms that are sparsely related. They also provide an inference procedure designed to test hypotheses on the entries of the covariance matrix of the residuals of pre-estimated models, including principal component regressions. Our paper differs from theirs in several directions. First, \cite{jFrMmM2021} considers only the covariance matrix and not the precision matrix. Second, their approach is not based on nodewise regressions. Finally, Sharpe Ratio estimation and portfolio allocation are not considered. A seminal paper is by \cite{gagl2016}, where they analyze time-varying risk premia in large portfolios with factor models. They develop a structural model, and can tie that to factor models, and after that, they can estimate time-varying risk-premia. One of their main assumptions is that the maximum eigenvalue of covariance matrix of errors in the factor structure can diverge. Also, they assume sparsity of covariance matrix of errors and observed factors in the factor model. We also use diverging eigenvalue assumption in Assumption \ref{as7}(i) in our paper, as well as an increasing number of factors here, but with the assumption of sparsity on the precision matrix of errors. \cite{gos2019} develop a diagnostic test for omitted factors in factor models. They rely on residuals rather than errors for their tests. As clear in their analysis, working with residuals pose major difficulties. We also face the similar difficulty in our paper. Then, \cite{gos2020} analyze large conditional factor models. They analyze conditional risk premia even when the number of assets dominate the time span of the portfolio.

In a recent paper, \cite{flw2018} use sparse precision matrix estimation with hidden factors. Their approach uses a Dantzig based constrained estimator for precision matrix. The main differences are that the type of estimator depends on magnitude of coefficients in the precision matrix, with larger coefficients, and that the rate of estimation slows down considerably as seen in their equation (2.12)-result 2. Also, they assume bounded-finite $l_{\infty}$ matrix norm, which is restrictive. We allow diverging matrix $l_{\infty}$ norm. Also they do not apply their results to Sharpe Ratio analysis in high dimensions as we do.

Recently, important contributions have been obtained in this area by using shrinkage and factor models. \cite{lw2017} propose a nonlinear shrinkage estimator in which small eigenvalues of the sample covariance matrix are increased and large eigenvalues are decreased by a shrinkage formula. Their main contribution is the optimal shrinkage function, which they find by minimizing a loss function. The maximum out-of-sample Sharpe Ratio is an inverse function of this loss. Their results cover the independent and identically distributed case and when $p/n \to (0,1) \cup (1,+\infty)$. For the analysis of mean-variance efficiency, \cite{ao2019} make a novel contribution in which they take a constrained optimization, maximize returns subject to the risk of the portfolio, and show that it is equivalent to an unconstrained objective function, where they minimize a scaled return of the portfolio error by choosing optimal weights. To obtain these weights, they use lasso regression and assume a sparse number of nonzero weights of the portfolio, and they analyze $p/n \to (0,1)$. They show that their method maximizes the expected return of the portfolio and satisfies the risk constraint. Their paper is an important result on its own. One key paper in the literature is by \cite{fan2011} which assumes an approximate factor model, but, on the other hand, the authors assume conditional sparsity-diagonality of the covariance matrix of errors. \cite{fan2011} show for the first time how to build a precision matrix of returns in a large portfolio via factor models. Therefore, it is a key paper in the high-dimensional econometrics literature.

Regarding other papers, Ledoit and Wolf (2003,2004)\nocite{lw2003,lw2004} propose a linear shrinkage estimator of the covariance matrix and apply it to portfolio optimization. \cite{lw2017} shows that nonlinear shrinkage performs better in out-of-sample forecasts. \cite{lai2011}, and \cite{garlappi2007} approach the same problem from a Bayesian perspective by aiming to maximize a utility function tied to portfolio optimization. Another avenue of the literature improves the performance of the portfolios by introducing constraints on the weights. This type of literature is in the case of the global minimum-variance portfolio. Examples of works investigating this problem include \cite{jag2003} and \cite{fanli2012}. We also see a combination of different portfolios proposed by \cite{kan2007}, and \cite{tu2011}. Very recently, \cite{dlz2020} extended factor models to assumptions that are more consistent with principal components analysis. They provide consistent estimation of the risk of the portfolio under the sparsity of covariance of errors with a fixed number of factors. \cite{bgs2018}, \cite{bddgl2009}, \cite{cr1983}, \cite{dgnu2009}, \cite{fls2015} analyze the mutual fund industry, sparsely constructed Markowitz portfolio, arbitrage and factor models in large portfolios, sparsely constructed mean-variance portfolios, and risks of large portfolios, respectively.

\subsection{Organization of the Paper}

This paper is organized as follows. Section \ref{fnwe} considers our assumptions and feasible precision matrix estimation for errors. Section \ref{fnwr} provides the feasible precision matrix estimate for asset returns.
Section \ref{sratio} analyzes consistency of the Sharpe Ratio in a portfolio with large number of assets in three different scenarios.
Section \ref{sims} provides simulations that compare several methods. Section \ref{emp} presents an out-of-sample forecasting exercise. The main proofs are in the Supplement A, common proofs used for Theorems 3-8 are in Supplement B, Supplement C contains proofs related to section 4.4,  and the  Supplement  D has a proof of mean-variance efficiency of a large portfolio  in case of out-of-sample context, and some extra simulation results.

\subsection{Notation}

Let $\| \b\nu \|_{l_1}, \| \b\nu \|_{l_2}, \| \b\nu \|_{\infty}$ be the $l_1, l_2, l_{\infty},$ norms of a generic vector $\b\nu$.  Let
 $\| \b v \|_n^2:= n^{-1} \sum_{t=1}^n v_t^2$ which is the prediction norm for an $n \times 1$ vector $\b v$. Let $\textnormal{Eigmin}(\b A)$ represents the minimum eigenvalue of a matrix $\b A$, and $\textnormal{Eigmax}(\b A)$ represent the maximum eigenvalue of the matrix $\b A$. For a generic matrix
$\b A$, let $\| \b A \|_{l_1}, \| \b A \|_{l_{\infty}}, \| \b A \|_{l_2}$,  be the $l_1$ induced matrix norm (i.e. maximum absolute column sum norm), $l_{\infty}$ induced matrix norm (i.e. maximum absolute row sum norm), spectral matrix norm, respectively.  $\| \b A \|_{\infty}$ is maximum absolute value of element of a matrix, and also  a norm (but not a matrix norm). Matrix norms have the additional desirable feature of submultiplicativity property. For further information on matrix norms, see p.341 of \cite{hj2013}.

\section{Factor Model and Feasible Nodewise Regression}\label{fnwe}

We start with the following model for the $j$th asset return (excess asset return) at time $t$, $y_{j,t}$, for $j=1,\cdots, p$, and time periods $t=1,\cdots,n$, such that
\begin{equation}
y_{j,t} = \b b_j' \b f_t + u_{j,t}.\label{rv1}
\end{equation}
where $\b b_j$ is a $K \times 1$ vector of factor loadings, $\b f_t$ is the $K \times 1$ vector of common factors to all assets' returns, and $u_{j,t}$ is the scalar error (idiosyncratic) term for asset return $j$ at time $t$. All the factors are assumed to be observed. This model is used by \cite{fan2011}. From this point on, when asset return is mentioned, it should be understood as excess asset return.

For the $j$th asset return we can rewrite (\ref{rv1}) in the vector form, for $j=1,\cdots, p$:
\begin{equation}
\b y_j = \b X' \b b_j + \b u_j,\label{rv2}
\end{equation}
where $\b X = (\b f_1, \cdots, \b f_n)$ is a $K \times n$ matrix, and $\b y_j = (y_{j,1}, \cdots, y_{j,n})'$ is a $n \times 1$ vector of returns of the $j$th asset.  We can also express the same relation in a matrix form as follows:
\begin{equation}
\b Y = \b B \b X + \b U,\label{rv3}
\end{equation}
where $\b Y$ is a $p\times n$ matrix, $\b B$ is a $p \times K$ matrix, and $\b U$ is a $p \times n$ matrix.~\footnote{
We can also  write the returns  for each period in time, $t=1,\cdots, n$
\[ 
\b y_t = \b B \b f_t + \b u_t,
\]
where $\b y_t =(y_{1,t}, \cdots, y_{j,t}, \cdots, y_{p,t})'$ is a $p \times 1$ vector.}
Define the covariance matrix of the $p \times 1$ vector of errors $\b u_t:=(u_{1,t}, \cdots, u_{j,t},\cdots, u_{p,t})'$ as $\b\Sigma_n:= \E\left[\b u_t \b u_t'\right]$.

We take $\{(\b f_t,\b u_t)\}_{t=1}^n$ to be a strictly stationary, ergodic, and strong mixing sequence of random variables.
Also, let ${\cal F}_{-\infty}^0, {\cal F}_{n}^{\infty}$ be the $\b\Sigma$- algebras generated by $\{(\b f_t, \b u_t)\}$, for $-\infty < t \le 0$, and $n \le t <\infty$, respectively. Denote the strong mixing coefficient as $\alpha(n):= \textnormal{sup}_{ {\cal A} \in {\cal F}_{-\infty}^0, {\cal B} \in {\cal F}_{n}^{\infty}}
| \P ({\cal A}) P ({\cal B}) - \P ( {\cal A} \cap {\cal B} )|.$ 

In Assumption \ref{as7} below, we assume that maximum eigenvalue of $ \b\Sigma_n$ can grow with sample size, this is due to $\b\Sigma_n$ being a $p \times p$ matrix where $p$ may grow with $n$. We will assume sparsity for the precision matrix of errors $\b\Omega:= \b\Sigma_n^{-1}$, but we do not subscript $\b\Omega$ with $n$ to avoid cumbersome notation. Each row of $\b\Omega$ will be denoted as a $1\times p$ vector $\b\Omega_j'$. We represent the indices of nonzero cells in $\b\Omega_j'$ as $S_j$, for $l=1,\cdots, p$,
\[
S_j:= \{ j: \Omega_{j,l} \neq 0 \},
\]
where $\Omega_{j,l}$ represents the $l$th element in the $j$th row of $\b\Omega$. Let $S_j^c$ represents the index set of all zero elements in the $j$th row of $\b\Omega$. Define the cardinality of the non-zero cells in the $j$th row of the precision matrix as $s_j:= |S_j|$, which can be nondecreasing in $n$, but we do not subscript that with $n$.
Denote the maximum number of nonzero elements across all rows $j=1,\cdots,p$ of the precision matrix $\b\Omega$ as $\bar{s}:= \max_{1 \le j \le p} s_j$, which is nondecreasing in $n$.

This last definition plays a key role in analysis of the rate of convergence of estimation errors. Note that, just to be clear, when $n \to \infty$, we allow $p \to \infty$, $K \to \infty$, and $\bar{s} \to \infty$. As in the literature, we do not subscript them by $n$. Also, we allow for $p>n$, when $n \to \infty, p \to \infty$, and $p/n \to \infty$ in our analysis in Theorems 1-7, which can be considered ultra-high dimensional portfolio analysis. 
For future references, we denote all of the asset returns except the $j$th one as
\begin{equation}
\b Y_{-j} = \b B_{-j} \b X + \b U_{-j},\label{-yj}
\end{equation}
where $\b Y_{-j}$, of dimension $(p-1) \times n$, is the $\b Y$ matrix without the $j$th row, $\b B_{-j}$ is the $(p-1)\times K$ matrix which is $\b B$ without the $j$th row, and $\b U_{-j}$ is the $(p-1) \times n$ matrix given by $\b U$ matrix without the $j$th row.

It has been well established in the literature that in case of known $U_j$, $\gamma_j$, which is essential input in nodewise regression, can be recovered with the following lasso problem, with a sequence $\lambda_n >0$,  for all $j=1,\cdots, p$,  
\begin{equation}
 \tilde{\b\gamma}_j= \argmin_{\b\gamma_j \in \R^{p-1}} \left[ \| \b u_j - \b U_{-j}' \b\gamma_j \|_n^2 + 2 \lambda_n \|\b \gamma_j \|_1\right].\label{rv8}
 \end{equation}
 The  main issue  with (\ref{rv8}) is, unlike nodewise regression in \cite{canerkock2018}, it is infeasible due to error terms regressed on each other. We now show how to turn this to feasible regression and still consistently estimate $\b\gamma_j$.

To get estimates for $\b b_j$ and $\b B_{-j}'$, \cite{fan2011} use the Ordinary Least Squares (OLS) and show that \footnote{See p.3347 of \cite{fan2011}.}
 \begin{equation}
 \widehat{\b b}_j - \b b_j = (\b X \b X')^{-1} \b X \b u_j.\label{rv11}
 \end{equation}
By equation (\ref{rv2}) we can define the OLS residual as 
\begin{eqnarray}
\widehat{\b u}_j &=& \b y_j- \b X' \widehat{\b b}_j 
 =  \b u_j - \b X' (\b X\b X')^{-1} \b X \b u_j \nonumber \\
& = & \b M_X \b u_j,\label{rv12}
\end{eqnarray}
where $\b X$ is a $K \times n$ matrix and 
\begin{equation}
\b M_X := \b I_n - \b X'(\b X \b X')^{-1} \b X.\label{rv12a}
\end{equation}
Then, by OLS, with $\widehat{\b B}_{-j}'$ and $\b B_{-j}'$ being $K \times (p-1)$ matrices such that $\widehat{\b B}_{-j}' - \b B_{-j}' = (\b X \b X')^{-1} \b X \b U_{-j}'$.

Define the residuals by transposing (\ref{-yj}) such that
\begin{eqnarray}
\widehat{\b U}_{-j}' &= &\b Y_{-j}' - \b X' \widehat{\b B}_{-j}' 
 =  \b U_{-j}' - \b X' (\widehat{\b B}_{-j}' - \b B_{-j}') \nonumber \\
& = & \b U_{-j }' - \b X' (\b X \b X')^{-1} \b X \b U_{-j}' \nonumber \\
& = & \b M_X \b U_{-j}'.\label{rv13}
\end{eqnarray}
Note that $\widehat{\b U}_{-j}'$ is a $n \times (p-1)$ matrix $\b M_X$ is a $n \times n$ matrix, and $\b U_{-j}'$ is a $n \times (p-1)$ matrix.
Next, 
 use (\ref{rv12}) and (\ref{rv13}):
\begin{equation}
\widehat{\b u}_j = \widehat{\b U}_{-j}'\b\gamma_j + \b\eta_{xj},\label{rv14}
\end{equation}
where
\begin{equation}
\b\eta_{xj}:= \b M_X \b \eta_j,\label{rv14a}
\end{equation}
is a $n \times 1$ vector, with $\b \eta_j:= \b u_j - \b U_{-j}' \b \gamma_j$. Of course, the key difficulties are how the new $\b\eta_{xj}$ and the usage of the residuals affect the consistent estimation of $\b\gamma_j$. We define a feasible nodewise estimator
\begin{equation}
\widehat{\b\gamma}_j= \argmin_{\b\gamma_j \in \R^{p-1}} \left [\| \widehat{\b u}_j - \widehat{\b U}_{-j}'\b\gamma_j \|_n^2 + 2 \lambda_n \|\b\gamma_j \|_1 \right].\label{rv15}
\end{equation}
Then, to define $\widehat{\b\Omega}_j'$, which is the $j$th row of the precision matrix estimator, we need
\begin{equation}
\widehat{\tau}_j^2:= \widehat{\b u}_j'(\widehat{\b u}_j - \widehat{\b U}_{-j}'\widehat{\b\gamma}_j)/n.\label{rv16}
\end{equation}
Now, to form the $j$th row of $\widehat{\b\Omega}$, set the $j$th element in the $j$th row as
\begin{equation}
 \widehat{\Omega}_{j,j}:=1/\widehat{\tau}_j^2.\label{ojj}
 \end{equation}

\begin{equation}
 \widehat{\b\Omega}_{j,-j}':= -\frac{1}{\widehat{\tau}_j^2}\widehat{\b\gamma}_j'.\label{o-jj}
 \end{equation}
We want to show that for each $j=1,\cdots, p$, $\widehat{\b\Omega}_j'$ is consistent. We can write $\widehat{\b\Omega}_j'= \widehat{\b C}_j'/\widehat{\tau}_j^2$ with
$\widehat{\b C}_j'$ being an $1 \times p$ matrix of ones in $j$th cell and $-\widehat{\b\gamma}_j'$ in the other cells.

\subsection{Assumptions and a Key Result}

In this part, we provide the assumptions that will be needed for consistency for the $j$th row of the precision matrix estimator. Let $u_{j,t}$ be the $j$ the element of the $p\times 1$ vector $\b u_t$. Similarly, $\b u_{-j,t}$ is the $(p-1)\times 1$ vector of errors in $t$th time period, except the $j$th term in $\b u_t$. Define $\eta_{j,t}:= u_{j,t}- \b u_{-j,t}'\b\gamma_j$.

\begin{assum}\label{as1}
(i). $\{\b u_t \}_{t=1}^n, \{ \b f_t \}_{t=1}^n$ are sequences of (strictly) stationary and ergodic random variables. Furthermore, $\{\b u_t \}_{t=1}^n, \{\b f_t \}_{t=1}^n$ are independent. $\b u_t$ is a ($p \times 1$) zero mean random vector with covariance matrix $\b\Sigma_n$ ($p \times p$). $\textnormal{Eigmin} (\b\Sigma_n) \ge c > 0$, with $c$ a positive constant, and $\max_{1 \le j \le p} \E\left[u_{j,t}^2\right] \le C < \infty$. 
 (ii). For the strong mixing variables $\b f_t, \b u_t$:
$\alpha (t) \le \exp (-C t^{r_0})$, for a positive constant $r_0>0$.
\end{assum}

\begin{assum}\label{as2}
There exists positive constants $r_1$, $r_2$, $r_3 >0$ and another set of positive constants $B_1$, $b_2$, $b_3$, $s_1$, $s_2$, $s_3 >0$, and for $t=1,\cdots, n$, and
$j=1,\cdots, p$, with $k=1,\cdots, K$

(i).  
\[ 
\P\left[|u_{j,t} | > s_1\right] \le \exp[-(s_1/B_1)^{r_1}].
\]

(ii). \[
\P\left[| \eta_{j,t} | > s_2 \right]\le\exp [-(s_2/b_2)^{r_2}].
\]

(iii). \[ 
\P\left[|f_{k,t} | > s_3\right]\le \exp [-(s_3/b_3)^{r_3}].
\]

(iv). There exists $0< \gamma_1 < 1$ such that  $\gamma_1^{-1} = 3 r_1^{-1} + r_0^{-1}$, and we also assume
$3 r_2^{-1} + r_0^{-1} > 1$, and $3 r_3^{-1} + r_0^{-1} >1$.
\end{assum}

Define $\gamma_2^{-1}:= 1.5 r_1^{-1} + 1.5 r_2^{-1} + r_0^{-1}$, and $\gamma_3^{-1}:= 1.5 r_1^{-1} + 1.5 r_3^{-1} + r_0^{-1}$, let
$\gamma_{\min}:=\min (\gamma_1, \gamma_2, \gamma_3)$.

\begin{assum}\label{as3}
(i). $[\ln (p)]^{(2/\gamma_{min}) -1 } = o(n)$, and (ii). $K^2 = o(n)$, (iii). $K = o(p)$.
\end{assum}

\begin{assum}\label{as4}
(i). $\textnormal{Eigmin}[\textnormal{cov}(\b f_t)] \ge c > 0$, with $\textnormal{cov}(\b f_t)$ being the covariance matrix of the factors $\b f_t$, $t=1,\cdots, n$.
(ii). $ \max_{1 \le k \le K } \E\left[f_{kt}^2\right] \le C < \infty$, $\min_{1 \le k \le k} \E\left[f_{kt}^2\right] \ge c > 0$.
(iii). $\max_{1 \le j \le p}\E\left[\eta_{j,t}^2\right] \le C < \infty$.
\end{assum}

\begin{assum}\label{as5} $\bar{s}$, $K$, $p$, and $n$ are such that
(i). $ K^2 \bar{s}^{3/2} \frac{\ln(p)}{n} \to 0.$
(ii). $ \bar{s} \sqrt{\frac{\ln(p)}{n}} \to 0.$
\end{assum}

Note that Assumptions \ref{as1}-\ref{as3} are standard assumptions and are used in \cite{fan2011} as well. Also, we get $0 < \gamma_2 < 1, 0 < \gamma_3 < 1$ given Assumption \ref{as2}(iv). Furthermore, by Assumption 3, $\sqrt{\frac{\ln(p)}{n}} = o(1)$. Note that, Stationary GARCH models with finite second moments and continuous error distributions, as well as causal ARMA processes with continuous error distributions, and a certain class of stationary Markov chains satisfy our Assumptions \ref{as1}-\ref{as2} and are  discussed in p.61 of \cite{chang2019}. \cite{chang2019} also uses similar assumptions. 

Assumption \ref{as4}(i)-(ii) is also used in \cite{fan2011}, and the nodewise error assumption \ref{as4}(iii) is used in \cite{canerkock2018}. Assumption \ref{as5} shows the interaction of sparsity of the precision matrix with factors. They both contribute negatively to biases that our analysis will show below. 

Before the next theorem, we define $\lambda_n$ formally. Let $C>0$ be a generic positive constant, then 
\begin{equation}
\lambda_n:= C \left[ K^2 \bar{s}^{1/2} \frac{ln p}{n} + \sqrt{\frac{ln p}{n}}\right]=o(1),\label{lar}
\end{equation}
where we specify tuning parameter in Lemma A.5 in Supplement, and the asymptotic negligibility is by Assumption 5. Note that in tuning parameter $\lambda_n$, the first term involving $K^2$ is due to nodewise regression via factor models.  In \cite{caner2019}, without factor models, they have the second term only $\sqrt{\frac{lnp}{n}}$. We now provide one of the main Theorems in the paper.  Theorem provides consistent estimates for the rows of the precision matrix of errors.

\begin{thm}\label{rthm1}
Under Assumptions \ref{as1}-\ref{as5}
\[ \| \widehat{\b\Omega} - \b\Omega \|_{l_{\infty}}:= \max_{1 \le j \le p} \| \widehat{\b\Omega}_j' - \b\Omega_j' \|_1=
\max_{1 \le j \le p} \| \widehat{\b\Omega}_j - \b\Omega_j \|_1 = O_p\left( \bar{s} \lambda_n\right) = o_p (1).\]
\end{thm}



\textbf{Remarks:} 
\begin{enumerate}
\item
Note that $\hat{\b\Omega}_j, \b\Omega_j$ are not columns of $\hat{\b\Omega}, \b\Omega$ respectively. $\hat{\b\Omega}_j, \b\Omega_j$ are column representation of row vectors $\hat{\b\Omega}_j', \b\Omega_j'$, respectively. 
\item
As long as Assumption \ref{as5} is maintained, the rate of approximation error in Theorem \ref{rthm1} matches the
case where $B=0$ in the factor model (i.e., there is no factor structure). This is the case considered in \cite{caner2019}. The number of  factors $K$ increases the approximation error through $\lambda_n$.
\end{enumerate}

\section{Precision Matrix Estimate for The Returns}\label{fnwr}

Assuming orthogonality between factors and the idiosyncratic errors, the $(p\times p)$ covariance matrix of the asset returns is defined as:
\begin{equation}
\b\Sigma_y = \b B\textnormal{cov}(\b f_t)\b B' + \b\Sigma_n.\label{sigmay}
\end{equation}
We start with the precision matrix formula for the asset returns, based on factor model that we used. Using Sherman-Morrison-Woodbury formula, as in  p.13 of \cite{hj2013}, $\b\Gamma:=\b\Sigma_y^{-1}$ is defined as:
\begin{equation}
\b\Gamma:= \b\Omega - \b\Omega \b B\left[ \{\textnormal{cov} (\b f_t)\}^{-1} + \b B' \b\Omega \b B\right]^{-1} \b B' \b\Omega,\label{smw1}
\end{equation}
and the precision matrix estimator for the returns is
\begin{equation}
\widehat{\b\Gamma}:= \widehat{\b\Omega} - \widehat{\b\Omega} \widehat{\b B}\left[\{\widehat{\textnormal{cov}(\b f_t)}\}^{-1} + \widehat{\b B}' \widehat{\b\Omega}_{sym} \widehat{\b B}\right]^{-1} \widehat{\b B}' \widehat{\b\Omega},\label{smw2}
\end{equation}
where $\widehat{\b\Omega}_{sym}:= \frac{\widehat{\b\Omega} + \widehat{\b\Omega}'}{2}$ is the symmetrized version of our feasible nodewise regression estimator for the precision matrix for errors. 
$\widehat{\textnormal{cov}(\b f_t)}= n^{-1} \b X \b X' - n^{-2} \b X \b 1_n \b 1_n' \b X'$ is the estimator for the covariance matrix of returns, and it is given in
p.3327 of \cite{fan2011} with $\b 1_n$ representing a $(n \times 1)$ vector of ones. Also, $\widehat{\b B}= (\b Y \b X')(\b X \b X')^{-1}$ is the least-squares estimator for the factor model in (\ref{rv3}). In addition, $\widehat{\b B}$ is a $(p \times K)$ matrix, and $\widehat{\textnormal{cov}(\b f_t)}$ is a $K \times K$ matrix. Note that we use a symmetric version of our precision matrix estimator for errors in the term in square brackets in equation (\ref{smw2}). There is a technical reason behind that. The proofs depend on the symmetry of the matrix in the square brackets in (\ref{smw2}), but the other parts in the proof do not need symmetry of the precision matrix estimator. Hence, we use both symmetrized, $\widehat{\b\Omega}_{sym}$ and standard (non-symmetric version) of the precision matrix estimator, $\widehat{\b\Omega}$.
We want to rewrite the precision matrix and it's estimator so that it's convenient to analyze them technically. In this respect, define
\[ 
\b L:= \b B\left[ \{\textnormal{cov}(\b f_t)\}^{-1} + \b B' \b\Omega \b B\right]^{-1} \b B',
\]
and
\[ 
\widehat{\b L}:= \widehat{\b B}\left[ \{\widehat{\textnormal{cov} (\b f_t)}\}^{-1} + \widehat{\b B}' \widehat{\b\Omega}_{sym} \widehat{\b B}\right]^{-1} \widehat{\b B}'.
\]
As a consequence,
\begin{equation}
\b\Gamma= \b\Omega - \b\Omega \b L \b\Omega, \quad \widehat{\b\Gamma}= \widehat{\b\Omega} - \widehat{\b\Omega} \widehat{\b L} \widehat{\b\Omega}.\label{sw3}
\end{equation}

We need to find $ \max_{1 \le j \le p} \| \widehat{\b\Gamma}_j - \b\Gamma_j \|_1$ where $\b\Gamma_j'$ and $\widehat{\b\Gamma}_j'$ are the $1\times p$ dimensional rows of the precision matrix of the returns and its estimator, respectively. $\b\Gamma_j$ and $\widehat{\b \Gamma}_j$ are simply transposes of these rows which are $p \times 1$. In this respect, using (\ref{sw3}) we have that
\begin{equation}
\max_{1 \le j \le p} \| \widehat{\b\Gamma}_j - \b\Gamma_j \|_1=
\max_{1 \le j \le p } \| \widehat{\b\Gamma}_j' - \b\Gamma_j' \|_1= \max_{1 \le j \le p} \|( \widehat{\b\Omega}_j' - \b\Omega_j')-(\widehat{\b\Omega}_j' \widehat{\b L} \widehat{\b\Omega} - \b\Omega_j' \b L \b\Omega)\|_1.\label{sw4}
\end{equation}

Our aim is to simplify and get rates of convergence for the right side term in (\ref{sw4}).
To get consistency and rate of convergence results for the precision matrix for returns, rather than the errors as in Theorem \ref{rthm1} above, we need the following assumption on factor loadings.

\begin{assum}\label{as6} The factor loadings are such that:

(i). $ \max_{1 \le j \le p} \max_{1 \le k \le K} |b_{jk}| \le C <\infty$.

(ii).  $\| p^{-1} \b B' \b B - \b \Delta \|_{l_2} = o(1)$ for some $K \times K$ symmetric positive definite matrix $\b\Delta$ such that $\textnormal{Eigmin} (\Delta)$ is bounded away from zero.
\end{assum}

Also, a strengthened assumption on sparsity compared to Assumption \ref{as5} is provided. 

\begin{assum}\label{as7}
Assume that

(i). $\textnormal{Eigmax}(\b\Sigma_n) \le C r_n$, with $C>0$ a positive constant, and $r_n \to \infty$ as $n \to \infty$, and
 $r_n/p \to 0$, and $r_n$ is a positive sequence.

(ii). 
 \[ 
 \bar{s} l_n \to 0,
 \]
where 
\begin{equation}
 l_n:= r_n^2  K^{5/2} \max\left(\bar{s} \lambda_n, \bar{s}^{1/2} K^{1/2} \sqrt{\frac{\max[\ln(p), \ln(n)]}{n}}\right) .\label{dn}
 \end{equation}
\end{assum}

Specifically, the rate $l_n$ is the rate of estimation error for $\|\widehat{\b L} - \b L \|_{l_{\infty}}$ as in Lemma A.13 in Supplement A. Note that Assumption \ref{as6} is used in \cite{fan2011}. Assumption \ref{as7}(i) is used in \cite{gagl2016}. Assumption \ref{as7}(i) allows for the maximal eigenvalue of $\b\Sigma_n$ to grow with $n$. In the special case of a diagonal $\b\Sigma_n$, due to Assumption \ref{as1}(i), the maximum  eigenvalue of a diagonal $\b\Sigma_n$ matrix is finite. However, a diagonal matrix of variance of errors case is empirically less relevant and less realistic. We expect the errors to be correlated across assets. For an example of where the maximum eigenvalue of $\b\Sigma_n$ may diverge, we show that this may be the case for block diagonal matrix structure for $\b\Sigma_n$ in (\ref{suffc}). Note that \cite{s1992} criticizes standard Arbitrage Pricing Theory since eigenvalue of the residual covariances must be bounded even when the number of assets diverge. Our Assumption \ref{as7}(i) moves away from maximum bounded eigenvalue assumption. Our residual covariances approximate error covariances very well and this can be seen in (\ref{pla63}) and (\ref{pla64}) in Supplement A.

Assumption \ref{as7}(ii) is a sparsity assumption which tradeoffs between maximal eigenvalue and the sparsity of the precision matrix. This assumption is needed to analyze the precision matrix for the asset returns. To give an example, ignoring constants, we can have $\bar{s}= \ln(n), K =\ln(n), p=2n$, and $r_n= n^{1/5}$, $\lambda_n = O[\max(\ln(n)^{7/2}/n, \sqrt{\ln(n)/n}]$. Then, Assumption \ref{as7}(ii) is satisfied 
\[ 
n^{2/5} (ln n)^{7/2} max((lnn)^{9/2}/n, (ln n)^{3/2}/n^{1/2}) \to 0.
\]

Next, we define sample mean of the asset returns and the population mean of asset returns.
Let $\widehat{\b\mu}:=\frac{1}{n}\sum_{t=1}^n\b y_t$, where $\b y_t$ is a $p \times 1$ vector of asset returns. Let $\b\mu:=\E[ y_t]$. Next theorem provides one of our main results, which is the consistent estimation of the precision matrix for asset returns. Since the precision matrix of asset returns is in the formula of the Sharpe Ratio, as will be shown in Section 4, this theorem is crucial for subsequent analysis.

\begin{thm}\label{rthm2}
(i). Under Assumptions \ref{as1}-\ref{as4}, and \ref{as6}-\ref{as7}
\[ \max_{1 \le j \le p} \| \widehat{\b\Gamma}_j - \b\Gamma_j \|_1 = O_p \left(\bar{s}l_n\right)= o_p (1).\]

(ii). Under Assumptions \ref{as1}-\ref{as6}
\[ \| \widehat{\b\mu} - \b\mu \|_{\infty} = O_p \left[\max\left(K \sqrt{\frac{\ln(n)}{n}}, \sqrt{\frac{\ln(p)}{n}}\right)\right] = o_p (1).\]
\end{thm}

\textbf{Remarks:} 

\begin{enumerate}
\item
This theorem merges two key concepts: factor models and nodewise regression in high dimensional models.
Theorem \ref{rthm2} clearly shows that there is a tradeoff between the maximal eigenvalue of the errors, the number of factors, and the sparsity of the precision matrix. 
Increasing the number of factors in our model badly affect the rate of estimation of the precision matrix of the returns.

\item
Although we focus on factor models in empirical asset pricing, the vector $\b f_t$ can be seen as any set of random variables satisfying Assumptions 1-5. 
\end{enumerate}

\subsection{Two examples relating precision matrix restrictions to covariance matrix}

We now illustrate how specific structures of the covariance matrix are compatible with the sparsity assumption for the precision matrix. We provide two examples for errors, one block-diagonal covariance matrix for errors, and the other one is the Toeplitz form for the covariance matrix of errors. Then, we provide how they affect the precision matrix and Assumption \ref{as7}(i).

\subsubsection{Block Diagonal Covariance Matrix for Errors} Suppose that there are $m=1,\cdots,M$ blocks in a $p \times p$ covariance matrix of the errors.
\[ 
\b\Sigma_n:= 
\begin{bmatrix}
\b\Sigma_{n,p_1} & \cdots & \b{0} & \cdots & \b{0}\\
\vdots & \ddots &  \cdots & \cdots & \vdots\\
\b{0} & \cdots & \b\Sigma_{n,p_m} & \cdots & \b{0} \\
\vdots &\cdots&\cdots&\ddots&\vdots\\
\b{0} & \cdots & \b{0}& \cdots & \b\Sigma_{n,p_M}
\end{bmatrix}.
\]

Each block $\b\Sigma_{n,p_m}$ is of dimension $p_m \times p_m$  and $\sum_{m=1}^M p_m = p$. Clearly, the inverse is sparse as well:
\[ 
\b\Omega:=\b\Sigma_n^{-1}=
\begin{bmatrix}
\b\Sigma_{n,p_1}^{-1} & \cdots & \b{0} & \cdots & \b{0}\\
\vdots & \ddots &  \cdots & \cdots & \vdots\\
\b{0} & \cdots & \b\Sigma_{n,p_m}^{-1} & \cdots & \b{0} \\
\vdots &\cdots&\cdots&\ddots&\vdots\\
\b{0} & \cdots & \b{0}& \cdots & \b\Sigma_{n,p_M}^{-1}
\end{bmatrix}.
\]	

The sparsity assumption -- Assumption \ref{as1} -- for $\b\Omega$ can be translated into $\b\Sigma_n$ as $\max_{1 \le m \le M} \max_{1 \le j \le p_m} s_{p_m} = \bar{s}$, where this is the maximum number of nonzero cells in a given row of a block, across all blocks. For Assumption \ref{as7} we need the following inequality from Corollary 6.1.5 of \cite{hj2013}, by seeing that spectral radius of a matrix is larger than or equal to absolute value of any  eigenvalue for any square matrix $\b A$. Therefore,
\begin{equation}
\textnormal{Eigmax} (\b A) \le \min\left(\| \b A \|_{l_1}, \| \b A \|_{l_{\infty}}\right).\label{evin}
\end{equation}		
For the same inequality also see Theorem 5.6.9a of \cite{hj2013}. Relating to Assumption \ref{as7}(i)
\[ \textnormal{Eigmax} (\b\Sigma_n) \le \max_{1 \le j_1 \le p} \sum_{j_2=1}^p | \Sigma_{n,j_1, j_2}|=
\max_{1 \le j_1 \le p} \sum_{j_2=1}^p\left|\E [u_{j_1,t} u_{j_2,t}]\right|,\]
where $\Sigma_{n,j_1, j_2}$ is the $j_1, j_2$ element of covariance matrix of errors.
By (\ref{evin}), this last inequality becomes
\[ 
\textnormal{Eigmax} (\b\Sigma_n)\le  \max_{1 \le m \le M} \max_{1 \le j_1 \le p_m} \sum_{j_2=1}^{p_m}\left|\E[u_{j_1,t} u_{j_2,t}]\right|.
\]
It is easy to see that using Assumption \ref{as1}(i), and under sufficient conditions for Assumption \ref{as7}(i), with $p_m \to \infty$ as $n \to \infty$
\begin{equation}
\max_{1 \le m \le M} \max_{1 \le j_1 \le p_{m}}  \max_{1 \le j_2 \le p_m}\left|\E[u_{ j_1,t} u_{j_2,t}]\right| \le C < \infty, \,
\max_{1 \le m \le M} \frac{p_m}{p} \to 0, \, r_n:= \max_{1 \le m \le M} \,p_m,\label{suffc}
\end{equation}
we get $\textnormal{Eigmax} (\b\Sigma_n) \le C r_n, r_n/p \to 0$. This allows the size of the blocks to be increasing with $p$, but the ratio of the maximum block size to total number of parameters should be small.

\subsubsection{Toeplitz Analysis}
In this case, the correlation among errors are $\E[u_{j,t} u_{i,t}]=\rho^{|i-j|}$, with $|\rho| < 1$. Then.
\[ 
\b\Sigma_n=
\begin{bmatrix}
1 & \rho & \rho^2 & \cdots & \rho^{p-1} \\
\rho & 1 & \cdots & \cdots & \cdots \\
\rho^2 & \cdots & 1 & \cdots & \cdots \\
\vdots & \cdots & \cdots & \ddots & \vdots\\
\rho^{p-1} & \cdots & \rho^2 & \rho & 1\\
\end{bmatrix}.
\]
We have the tri-diagonal inverse, with all other cells being zero except the main and two adjacent diagonals.
\[ 
\b\Sigma_n^{-1}= \frac{1}{1-\rho^2}
\begin{bmatrix}
1 & -\rho & 0 & 0 & \cdots & 0\\
-\rho & 1+\rho^2 & -\rho & 0 & \cdots & 0 \\
0 & -\rho & 1+\rho^2 & -\rho & \cdots & 0 \\
\vdots & \cdots& \ddots& \ddots & \ddots & \vdots\\
0 &\cdots & 0 & -\rho & 1+\rho^2 & -\rho\\ 
0  &     \cdots  &         0& & -\rho & 1
\end{bmatrix}
\]
Clearly $\bar{s}=3$, and the covariance matrix for errors is not sparse. For Assumption \ref{as7}(i), using (\ref{evin})
\[ 
\textnormal{Eigmax} (\b\Sigma_n) \le \| \b\Sigma_n \|_{l_{\infty}} = \max_{ 1 \le j_1 \le p} \sum_{j_2=1}^p\left| \rho ^{|j_2-j_1|}\right|.
\]
Clearly Assumption \ref{as7}(i) is satisfied since the sum  on the right side converges to a constant.

\subsection{Algorithm For Asset Return Based Precision Matrix Estimation}

Here we provide a practical algorithm to get the precision matrix estimator for asset returns, $\widehat{\b\Gamma}$, and it will depend on the residual-based nodewise regression estimator $\widehat{\b\Omega}$, and its symmetric version $\widehat{\b\Omega}_{sym}$.
\begin{enumerate}
\item
Use equation (\ref{rv12}) to set up the residual from a least squares based regression via known factors with $\b y_j$ as the $j$th asset returns ($n \times 1$)
\[ 
\widehat{\b u}_j = \b y_j - \b X' \widehat{\b b}_j,
\]
with $\widehat{\b b}_j= (\b X\b X')^{-1}\b X \b y_j$, and $\b X=(\b f_1, \cdots, \b f_t, \cdots, \b f_n): K \times n$ matrix with $\b f_t: K \times 1$ known factor vector.
\item
Form the transpose matrix of residuals for all asset returns except $j$th one, $\widehat{\b U}_{-j}'$, which is a $n \times p-1$ matrix as in (\ref{rv13})
\[ 
\widehat{\b U}_{-j}' = \b Y_{-j}' - \b X' \widehat{\b B}_{-j}',
\]
where $\widehat{\b B}_{-j}'= (\b X \b X')^{-1} \b X \b Y_{-j}'$, $(K \times p-1$ matrix),  which is the transpose of factor loading estimates, $\b Y_{-j}': (n \times p-1$) is the  transpose matrix of asset returns except the $j$th asset.
\item
Run (\ref{rv15}), nodewise regression of $\widehat{\b u}_{j}$ on $\widehat{\b U}_{-j}'$ via lasso, and get $\lambda_n$ from Cross-validation or Generalized Information Criterion as in Section 5.1.
\item
Use equation (\ref{rv16}) to get $\widehat{\tau}_j^2$.
\item
Now form $\widehat{\b\Omega}_j'$ which is a row in the precision matrix estimate for the errors with $1/\widehat{\tau}_j^2$ as $j$th element of that $j$th row, and put all other elements of the $j$th row, as $-\widehat{\b \Gamma}_j'/\widehat{\tau}_j^2$. 
\item
Run steps 1-5 for all $j=1,\cdots, p$. Stack all rows $j=1,\cdots, p$ to form $p \times p$ matrix: $\widehat{\b\Omega}$. Form symmetric version by
$\widehat{\b\Omega}_{sym}:= \frac{\widehat{\b\Omega} + \widehat{\b\Omega}'}{2}$.
\item
Form 
\[
\widehat{\b B}= (\b Y \b X')(\b X \b X')^{-1},
\]
which is a $p \times K$ matrix of OLS estimates, where $\b Y: p\times n$ matrix of all asset returns, where $j=1,\cdots, p$ represent all column-assets, and rows $t=1,\cdots, n$ time periods. Also form the covariance matrix estimate for factors
\[
\widehat{\textnormal{cov}(\b f_t)} = n^{-1} \b X \b X' - n^{-2} \b X \b 1_n \b 1_n' \b X',
\]
where $\b 1_n$ is $n \times 1$ column vector of ones.
\item
Now form the precision matrix estimate for all asset returns by (\ref{smw2}) and steps 6-7:
\[ 
\widehat{\b\Gamma}= \widehat{\b\Omega} - \widehat{\b\Omega} \widehat{\b B}\{ [\widehat{\textnormal{cov}(\b{f}_t)}]^{-1} + \widehat{\b B}'\widehat{\b\Omega}_{sym} \widehat{\b B}\}^{-1} \widehat{\b B}'\widehat{\b\Omega}.
\]
We use $\widehat{\b\Omega}_{sym}$ in the inverse in square brackets, so that we can use  specific inequalities for the inverse in our proof. $\widehat{\b\Omega} $ is the nodewise regression estimator, and $\widehat{\b\Omega}_{sym}$ is the symmetrized version.
\end{enumerate}

\section{Sharpe Ratio Analysis with Large Number of Assets}\label{sratio}

In this section, we apply the results, mainly the estimation of precision matrix of returns, to the analysis of the Sharpe Ratio with large number of assets. Specifically, we allow $p \to \infty$, when $n \to \infty$. There will be four themes in each subsection below. But all of these themes relate to the analysis of consistency of the Sharpe Ratio in portfolios with a large number of assets. All our theoretical analysis is without transaction costs, however in simulations and also in empirical exercise we consider the presence of transaction costs.

The first subsection analyzes the Sharpe Ratio of Global Minimum Variance (GMV) portfolio, and Markowitz Mean-Variance (MMV) portfolio. In the GMV portfolio, we choose the weights to minimize the variance of the portfolio and restricted to sum one. Short-sales are allowed. The Sharpe Ratio is then constructed by dividing the mean portfolio returns by its standard deviation. In MMV portfolio, weights are chosen exactly as GMV but we also impose a target for the portfolio mean return. 

The second subsection considers choosing the weights of the portfolio in such a way to maximize the Sharpe Ratio, subject to weights of the portfolio adding up to one. Short sales are allowed. The main difference between GMV in Section \ref{gmv2}, and the Constrained Maximum Sharpe Ratio in Section \ref{model}, is that weights are chosen to minimize the variance in GMV portfolio and then the Sharpe Ratio is computed and, in case of the Constrained Maximum Sharpe Ratio, weights are chosen to maximize the Sharpe Ratio directly. Both methods use the same constraint that the weights of the portfolio should add up to one. In case of the MMV portfolio in Section \ref{mwtzmv} weights are chosen first to minimize the portfolio variance under the conditions described earlier and then, the Sharpe Ratio is computed. The constraint of weights adding up to one is helpful in visualizing assets in percentage terms.

In the third subsection, we analyze the maximum out-of-sample Sharpe Ratio. Here, we do not have a constraint that all weights of the portfolio should add up to one as in Sections \ref{gmv2}, \ref{mwtzmv}, and \ref{model}. The analysis is out-sample unlike the GMV, MMV, and Constrained Maximum Sharpe Ratio portfolios. Weights are chosen to maximize the portfolio returns subject to a constraint of a given variance. But the maximum out-of-sample Sharpe Ratio use estimated weights, with population out-sample mean return vector and the out-sample covariance matrix of returns in the formula. Since the maximum eigenvalue of out-sample covariance matrix of returns is growing, this affects the estimation error rate. Specifically, Sections \ref{gmv2}, \ref{mwtzmv}, and \ref{model} allow $p>n$ and we still get consistency, when $n \to \infty, p \to \infty$. With the maximum-out-of-sample Sharpe Ratio we get consistency only when $p<n$ and $n \to \infty, p \to \infty$. 

In the fourth subsection, we consider the effect of estimated portfolio weights on obtaining the optimal Sharpe Ratio in large samples. Specifically, we estimate the weights and substitute this into the Sharpe Ratio formula, with keeping $\b\mu, \b\Sigma_y$ intact, and then try to show that this estimate is consistent. We show that it is possible only in the case of $p<n$, and this includes diverging number of assets and time span.

Before we state the theorems, we need the following sparsity assumption. Assumption \ref{as8}(i) below replaces Assumption \ref{as7}(ii). In Assumption \ref{as8}(ii), the first term shows square of the maximum Sharpe Ratio is lower bounded, (scaled by $p$),  to be positive. Scaling by $p$ is needed since the numerator is summed over $p$ terms. In a similar way, the second term in Assumption \ref{as8}(ii) imposes that the variance of the GMV portfolio (scaled) to be finite. The variance of the GMV portfolio is $\left[\frac{\b 1_{p}'\b\Gamma \b 1_p}{p}\right]^{-1}$. Let $c>0$ be a positive constant.

\begin{assum}\label{as8} Assume that 
(i). \[ K^{3} \bar{s} l_n = o(1),\]

(ii). \[ \frac{\b\mu'\b\Gamma \b\mu}{p} \ge c > 0, \quad \frac{\b 1_{p}'\b\Gamma \b 1_p}{p} \ge c > 0.\]
\end{assum}

\subsection{Commonly Used Portfolios with a Large Number of Assets}\label{gmv1}
Here, we provide consistent estimates of the Sharpe Ratio of the GMV and MMV portfolios when $p>n$.

\subsubsection{Global Minimum-Variance (GMV) Portfolio}\label{gmv2}
In this part, we analyze the Sharpe Ratio that we can infer from the GMV portfolio. This is the portfolio in which weights are chosen to minimize the variance of the portfolio subject to the weights summing to one. Specifically,
\begin{equation}
\b w_{nw} = \argmin_{\b w \in \R^p} \b w' \b\Sigma_y \b w, \quad \mbox{\textnormal{subject  to}} \quad \b w' \b 1_p =1.\label{35a}
\end{equation}
The solution to the above problem is well known and is given by
\[ 
\b w_{nw} = \frac{\b\Sigma_y^{-1} \b 1_p}{\b 1_p' \b\Sigma_y^{-1} \b 1_p}.
\]
Next, substitute these weights into the Sharpe Ratio formula, normalized by the number of assets
\begin{equation}
SR= \frac{\b w'_{nw} \b \mu}{\sqrt{\b w_{nw}'  \b\Sigma_y \b w_{nw}}} = \sqrt{p}\left(\frac{\b 1_p' \b\Sigma_y^{-1} \b\mu}{p}\right)\left(\frac{\b 1_p' \b\Sigma_y^{-1} 1_p}{p}\right)^{-1/2}.\label{5.1}
\end{equation}
We estimate (\ref{5.1}) by nodewise regression, noting that $\b\Gamma:=\b\Sigma_y^{-1}$,
\begin{equation}
\widehat{SR}_{nw} = \sqrt{p}\left(\frac{\b 1_p' \widehat{\b\Gamma} \widehat{\b \mu}}{p}\right)\left(\frac{\b 1_p' \widehat{\b\Gamma} \b 1_p}{p}\right)^{-1/2}.\label{5.2}
\end{equation}

The following theorem is also valid when  $p>n$ and establishes both consistency and rate of convergence in the case of the Sharpe Ratio in the global minimum-variance portfolio.

\begin{thm}\label{gmv}
Under Assumptions \ref{as1}--\ref{as4}, \ref{as6}, \ref{as7}(i), and  \ref{as8} with $\left|\b1_p'\b\Gamma\b\mu\right|/p \ge C >  0$,
\[ 
\left| \frac{\widehat{SR}_{nw}^2}{SR^2} - 1
\right| = O_p\left(K^{3/2}\bar{s}l_n\right)=o_p (1).
\]
\end{thm}

\textbf{Remarks:} 

\begin{enumerate}
\item 
We see that a large $p$ only affects the error by a logarithmic factor as in the definition of $l_n$ in (\ref{dn}). The estimation error increases with the non-sparsity of the precision matrix.
\item
In the case of non-sparse precision matrix, we can only get consistency when $p<<n$. To show this, in case of non-sparse precision matrix $\bar{s}=p$, where all the rows of precision matrix consists of non-zero cells. Then, using (\ref{lar})(\ref{dn}) and Assumption 3, after simplifying expressions, we must have that
\[ 
K^{3/2} \bar{s} l_n = K^{3/2} p l_n = r_n^2 K^4 p \max[p^{3/2} K^2 \ln(p)/n + p \sqrt{\ln(p)/n}, p^{1/2} K^{1/2} \sqrt{\ln/n}] \to 0,
\]
to get consistency. 
\item
Condition $|\b 1_p'\b\Gamma\b\mu |/p \ge  C > 0$ is discussed in detail in Remark 3 of Theorem 7.
 \end{enumerate}
 
\subsubsection{Markowitz Mean-Variance (MMV) Portfolio}\label{mwtzmv}

\cite{mar52} portfolio selection is defined as finding the smallest variance given a desired expected return $\rho_1$. The decision problem is
\[ 
\b w_{mv} = \argmin_{\b w \in \R^p} (\b w' \b\Sigma_y \b w) \quad \mbox{\textnormal{such that}} \quad \b w'\b 1_p=1, \quad \textnormal{and}\quad \b w'\b \mu=\rho_1.
\]
The formula for optimal weight is
\begin{equation}
\b w_{mv} =  \left[\frac{D - \rho_1 F}{AD - F^2} \right] (\b\Sigma_y^{-1} \b 1_p/p) + \left[\frac{\rho_1 A - F}{AD - F^2} \right] (\b\Sigma_y^{-1} \b\mu/p),\label{38}
\end{equation}
where we use $A,F,D$ formulas $A:= \b 1_p'\b\Gamma \b 1_p/p, F := \b 1_p' \b\Gamma \b\mu/p, D:= \b\mu' \b\Gamma \b\mu/p$, with $\b\Gamma:= \b\Sigma_y^{-1}$. We define the estimators of these terms as $\widehat{A}:=\b 1_p' \widehat{\b \Gamma}\b 1_p/p, \widehat{F}:= \b 1_p' \widehat{\b \Gamma}\widehat{\b \mu}/p, \widehat{D}:=\widehat{\b \mu}' \widehat{\b \Gamma} \widehat{\b \mu}/p$.
The optimal variance of the portfolio in this scenario is normalized by the number of assets
\begin{equation}
V=\frac{1}{p} \left[\frac{A \rho_1^2 - 2 F \rho_1 + D }{AD-F^2} \right].\label{38aa}
\end{equation}
The estimate of that variance is
\[ \widehat{V} = \frac{1}{p} \left[ \frac{\widehat{A} \rho_1^2  - 2 \widehat{F} \rho_1 + \widehat{D}}{\widehat{A} \widehat{D} - \widehat{F}^2} \right].
\]
By our constraint, we obtain
\begin{equation}
\b w_{mv}'\b\mu = \rho_1.\label{38a}
\end{equation}
Using the variance $V$ above
\begin{equation}
SR_{mv} = \rho_1 \sqrt{p \left( \frac{AD - F^2}{A \rho_1^2 - 2 F \rho_1 + D}
\right)}.\label{30a}
\end{equation}
The estimate of the Sharpe Ratio under the MMV portfolio is
\[ 
\widehat{SR}_{mv} = \rho_1 \sqrt{ p \left( \frac{\widehat{A} \widehat{D} - \widehat{F}^2}{\widehat{A} \rho_1^2 - 2  \widehat{F} \rho_1 + \widehat{D} } \right) }.
\]

We provide the maximum Sharpe Ratio (squared) consistency in this framework when the number of assets is larger than the sample size. This is a novel result in the literature.

\begin{thm}\label{mmv}
Under Assumptions \ref{as1}-\ref{as4}, \ref{as6},\ref{as7}(i), and \ref{as8} with condition $\left|\b 1_p'\b\Gamma\b\mu /p\right| \ge C >  0$ and $AD - F^2 \ge C_1 > 0$, $A \rho_1^2 - 2 F \rho_1 + D \ge C_1 > 0$, with $\rho_1$ uniformly bounded away from zero and infinity, we have that
\[ 
\left|\frac{\widehat{SR}_{mv}^2}{SR_{mv}^2} - 1
\right| = O_p\left( K^3 \bar{s} l_n\right)=o_p(1).\]
\end{thm}

\textbf{Remarks:}
\begin{enumerate}
\item
Condition $A D - F^2 \ge C_1 > 0$ shows that the variance is bounded away from infinity, and $A \rho_1^2 - 2 F \rho_1 - D \ge C_1 > 0$ restricts the variance to be positive and bounded away from zero.
\item
We provide the rate of convergence of the estimators, which increases with $p$ in a logarithmic way as in $l_n$ definition in (\ref{dn}), and the non-sparsity of the precision matrix linearly affects affects the error.
\item
To get consistency when  there is non-sparse precision matrix, the same analysis in Remark 2 of Theorem \ref{gmv} applies, with $\bar{s}=p$, we need 
$p<n$.
\item
Number of factors slows the rate of convergence of estimation error to zero here. This is due to the fact that we have an extra constraint that is affected by number of factors compared with GMV Portfolio.
\end{enumerate}

\subsection{Maximum Sharpe Ratio: Portfolio Weights Normalized to One}\label{model}

In this section, we define the maximum Sharpe Ratio when the portfolio weights are normalized to one. This, in turn will depend on a critical term that will determine the formula below.
The maximum Sharpe Ratio is defined as follows, with $\b w$ as the $p \times 1$ vector of portfolio weights:
\[ 
\max_{\b w} \frac{\b w'\b\mu}{\sqrt{\b w' \b\Sigma_y \b w}},\quad  \textnormal{subject to} \quad \b 1_p' \b w =1,
\]
where $\b 1_p$ is a vector of ones. This maximum Sharpe Ratio is constrained to have portfolio weights that sum to one. \cite{maller2016} shows that depending on a scalar, it has two solutions. When $\b 1_p' \b\Sigma_y^{-1} \b\mu > 0$, with $\b\Gamma:= \b\Sigma_y^{-1}$, we have the square of the maximum Sharpe Ratio:
\begin{equation}
MSR^2= \b\mu' \b\Sigma_y^{-1} \b\mu.\label{1}
\end{equation}

\noindent When $\b 1_p' \b \Sigma_y^{-1} \b \mu >0$, \cite{maller2002} show 
\[ {\b w}_{c,1}:= \frac{\b \Sigma_y^{-1} \b \mu}{\b 1_p' \b \Sigma_y^{-1} \b \mu}.\]

\noindent On the other hand, when $\b 1_p' \b\Sigma_y^{-1}\b\mu \le  0$, we have
\begin{equation}
MSR_c^2 = \b\mu' \b\Sigma_y^{-1}\b\mu - (\b 1_p' \b\Sigma_y^{-1} \b\mu)^2/(\b 1_p' \b\Sigma_y^{-1}\b 1_p).\label{e1}
\end{equation}
This is equation (6.1) of \cite{maller2016}. Equation (\ref{1}) is used in the literature, and this is the formula when the weights do not necessarily sum to one given a return constraint as in \cite{ao2019}.  In case of $\b 1_p' \b \Sigma_y^{-1} \b \mu \le 0$, in equations (2.7)-(2.10) of \cite{maller2002}, there is an approximation to optimal portfolio weights. To be specific, with a positive   $\delta >0$, optimal portfolio weights, which is $(p \times 1)$ vector: 
\[ \b w_{c,2}:= (\delta \b u_{max}', 1 - \delta \b 1_{p-1}' \b u_{\max})',\]
where 
\[ \b u_{max}:= \frac{(\b A_{p-1}' \b A_{p-1})^{-1} \b A_{p-1}' \b z_{\max}}{\sqrt{\b z_{max}' \b A_{p-1} (\b A_{p-1}' \b A_{p-1})^{-2} \b A_{p-1}' \b z_{\max}}}\]
is a $(p-1) \times 1$ matrix with $\b A_{p-1}:= (I_{p-1},-1_{p-1})': p \times p-1$ matrix, with $1_{p-1}$ a $(p-1)$ column vector of ones, and 
\[ z_{\max}:= \b \Sigma_y^{-1} \left(  \b I_p - \frac{\b 1_p \b 1_p' \b \Sigma_y^{-1}}{\b 1_p' \b \Sigma_y^{-1} \b 1_p} 
\right) \frac{\b \mu}{MSR_c}\]
is of dimension $p \times 1$.

When $\delta \to \infty$, the weights can provide the maximum Sharpe Ratio: $MSR_c$, as discussed in p.504 of \cite{maller2002}.

These equations can be estimated by their sample counterparts, but in the case of $p>n$, $\widehat{\b\Sigma}_n$ is not invertible, so we need to use new tools from high-dimensional statistics.  We use the nodewise regression precision matrix estimate of \cite{mein2006}. This estimate is denoted by $\widehat{\b\Omega}$. $\widehat{\b \Omega}$ is incorporated into the precision matrix of returns $\hat{\b \Gamma}$. 

We will also introduce the maximum Sharpe Ratio, which addresses the uncertainty regarding whether we should analyze $MSR$ or $MSR_c$. This is
\[ 
(MSR^*)^2 = MSR^2 1_{ \{\b 1_p' \b\Sigma_y^{-1}\b\mu > 0 \}} + MSR_c^2  1_{ \{\b1_p' \b\Sigma_y^{-1}\b\mu \le 0 \}}.
\]
Note also that with $\b 1_p' \b \Sigma_y^{-1} \b \mu =0$, $MSR=MSR_c$. 
The estimators for $MSR, MSR_c, MSR^*$ will be introduced in the next subsection.


\subsubsection{Consistency and Rate of Convergence of Constrained Maximum Sharpe Ratio Estimators}\label{cmsr}

First, when $\b 1_p' \b\Sigma_y^{-1} \b\mu > 0$, we have the square of the maximum Sharpe Ratio as in (\ref{1}).  Namely, the estimate of the square of the maximum Sharpe Ratio is:
\begin{equation}
\widehat{MSR}^2= \widehat{\b\mu}' \widehat{\b\Gamma} \widehat{\b\mu}.\label{est1}
\end{equation}

\begin{thm}\label{nwsr}
Under Assumptions \ref{as1}-\ref{as4}, \ref{as6},\ref{as7}(i), \ref{as8}  with $\b 1_p'\b\Gamma\b\mu  > 0$,
\[ 
\left| \frac{\widehat{MSR}^2}{MSR^2} -1 \right| =O_p\left( K^2 \bar{s} l_n\right)= o_p(1).
\]
\end{thm}

\textbf{Remarks:}
\begin{enumerate}
\item
We allow $p>n$ and $p$ can grow exponentially in $n$. We also allow for time-series data and establish a rate of convergence. The number of assets, on the other hand, can also increase the error on a logarithmic scale, as can be seen in (\ref{dn}). So assumption on sparsity of the precision matrix helps us derive this  result.
\item
When there is no sparsity of the precision matrix, i.e. $\bar{s}=p$, we can still get consistency but for $p<<n$. To see this, consider the error rate in Theorem \ref{nwsr} above, with $l_n$ definition in (\ref{dn})
\[K^2 \bar{s} l_n =  K^2 p l_n  \to 0.\]
This implies that to get consistency we need $p<n$.
\end{enumerate}

If $\b 1_p' \b\Sigma_y^{-1}\b\mu \le 0 $, the Sharpe Ratio is minimized, as shown on p.503 of \cite{maller2002}. The new maximum Sharpe Ratio in the case when $\b 1_p' \b\Sigma_y^{-1}\b \mu \le 0$ is in Theorem 2.1 of \cite{maller2002}.
The square of the maximum Sharpe Ratio when $\b 1_p' \b\Sigma_y^{-1}\b\mu \le 0$ is given in (33).

An estimator in this case is
\begin{equation}
\widehat{MSR}_c^2= \widehat{\b\mu}' \widehat{\b\Gamma} \widehat{\b\mu} - (\b 1_p' \widehat{\b\Gamma} \widehat{\b\mu})^2/(\b 1_p' \widehat{\b\Gamma}\b 1_p).\label{e2}
\end{equation}

The optimal portfolio allocation for such a case is given in (2.10) of \cite{maller2002}, and shown in $\b w_{c,2}$ here in Section 4.2. The limit for such estimators when the number of assets is fixed ($p$ fixed) is given in Theorems 3.1b-c of \cite{maller2016}. 

\begin{thm}\label{tme1}
If $\b 1_p' \b\Gamma\b\mu \le 0$, and under Assumptions \ref{as1}-\ref{as4},\ref{as6},\ref{as7}(i), \ref{as8}  with $AD - F^2 \ge C_1 > 0$, where $C_1$ is a positive constant,
\[ 
\left|
\frac{\widehat{MSR}_c^2}{MSR_c^2} - 1
\right| = O_p\left(K^2 \bar{s} l_n\right)= o_p (1).
\]
\end{thm}

\textbf{Remarks:} 
\begin{enumerate}
\item 
In Theorem \ref{tme1}, we allow $p>n$, and time-series data are allowed, unlike the iid or normal return cases in the literature when dealing with large $p,n$. 
\item
Case of non-sparse precision matrix proceeds in the same way as Remark 2 of Theorem \ref{nwsr}. To have consistency, we need $p<n$, with non-sparse case $\bar{s}=p$.
\end{enumerate}

We provide an estimate that takes into account uncertainties about the term $\b 1_p'\b\Sigma_y^{-1}\b\mu$. Note that the term can be consistently estimated, as shown in Lemma \ref{tl3} in  Supplement B. A practical estimate for a maximum Sharpe Ratio that will be consistent is:
\[ 
\widehat{MSR}^* = \widehat{MSR} 1_{ \{\b 1_p'\widehat{\b\Gamma} \widehat{\b\mu} > 0 \} } + \widehat{MSR}_c 1_{ \{\b1_p' \widehat{\b\Gamma} \widehat{\b\mu} < 0 \} },
\]
where we excluded the case of $\b 1_p' \widehat{\b\Gamma} \widehat{\b\mu} =0$ in the estimator. That specific scenario is very restrictive in terms of returns and variance.
Note that under a mild assumption, when $\b 1_p'\b\Gamma\b\mu > 0$, we have $\b 1_p'\widehat{\b\Gamma}\widehat{\b\mu} >0$, and when $\b 1_p'\b\Gamma\b\mu < 0$, we have $\b 1_p' \widehat{\b\Gamma} \widehat{\b\mu}<0$ with probability approaching one in the proof of Theorem \ref{tme2}. Note that $\b\Gamma:= \b\Sigma_y^{-1}$.

\begin{thm}\label{tme2}
Under Assumptions \ref{as1}-\ref{as4},\ref{as6},\ref{as7}(i), \ref{as8},  with $AD - F^2 \ge C_1 > 0$, where $C_1$ is a positive constant, and assuming
$|\b 1_p' \b\Gamma \b\mu |/p \ge C >  2 \epsilon > 0$, with a sufficiently small positive $\epsilon > 0$, and $C$ being a positive constant,
\[ \left|
\frac{(\widehat{MSR}^*)^2}{(MSR^*)^2} - 1
\right| = O_p ( K^2 \bar{s} l_n)=  o_p (1).\]
\end{thm}

\textbf{Remarks:} 
\begin{enumerate}
\item 
In the case of $p>n$, we only consider consistency since standard central limit theorems (apart from those in rectangles or sparse convex sets) do not apply, and ideas such as multiplier bootstrap and empirical bootstrap with self-normalized moderate deviation results do not extend to this specific Sharpe Ratio formulation.
\item
The case of non-sparse precision matrix with $\bar{s}=p$ proceeds in the same way as in Remark 2 after Theorem \ref{nwsr}.
\item
Condition $|\b 1_p'\b\Gamma\b\mu |/p \ge  C> 2 \epsilon > 0$ shows that apart from a small region around 0, we include all cases. This is similar to the $\beta-\min$ condition in high-dimensional statistics used to achieve model selection. Note 
\[ \left|\b 1_p'\b\Gamma\b\mu /p \right| = \left|\sum_{j=1}^p \sum_{k=1}^p \Gamma_{j,k} \mu_k/p\right|,\]
which is a sum measure of roughly theoretical mean divided by standard deviations. It is difficult to see how this double sum in $p$ will be a small number, unless
the terms in the sum cancel out one another. Therefore, we exclude that type of case with our assumption. Additionally, $\epsilon$ is not arbitrary, from the
proof this is the upper bound on the $| \widehat{F} - F|$ in Lemma \ref{tl3} in Supplement B, and it is of order
\[
\epsilon= O(K\bar{s}l_n)=o(1),
\]
where the asymptotically small term follows Assumption \ref{as8}.
\end{enumerate}

\subsection{Maximum Out-of-Sample Sharpe Ratio}\label{mos}

This section analyzes the maximum out of Sharpe Ratio that is considered in \cite{ao2019}. To obtain that formula, we need the optimal calculation of the weights of the portfolio. The optimization of the portfolio weights is formulated as
\begin{equation}
\b w_{mos} = \argmax_{\b w\in\R^p} \b w'\b \mu \quad \textnormal{subject to} \quad \b w'\b\Sigma_y \b w \le \sigma^2, \label{www}
\end{equation}
where we maximize the return subject to a specified positive and finite risk constraint, $\sigma^2>0$. Equation (A.2) of \cite{ao2019}
defines the estimated maximum out-of-sample ratio when $p<n$, with the inverse of the sample covariance matrix, $\widehat{\b\Sigma}_y^{-1} = [\frac{1}{n} \sum_{t=1}^n\b y_t \b y_t']^{-1}$ used as an estimator for the precision matrix estimate:
\[
\widehat{SR}_{moscov} := \frac{\b\mu' \widehat{\b\Sigma}_y^{-1} \widehat{\b\mu}}{\sqrt{\widehat{\b\mu}' \widehat{\b\Sigma}_y^{-1} \b\Sigma_y \widehat{\b\Sigma}_y^{-1} \widehat{\b\mu}}}.
\]
The theoretical version is written as, by definition of  $\b\Gamma:= \b\Sigma_y^{-1}$,
\[ 
SR^* := \sqrt{\b\mu'\b\Gamma\b\mu}.
\]
Then, equation (1.1) of \cite{ao2019} shows that when $p/n \to r_1 \in (0,1)$, the above plug-in maximum out-of-sample ratio cannot consistently estimate the theoretical version.  
The optimal weights of a portfolio are given in (2.3)
of \cite{ao2019} in an out-of-sample context given a risk level. This comes from maximizing the expected portfolio return subject to its variance being constrained by the square of the risk, where this is shown in (\ref{www}).
Since $\b\Gamma:= \b\Sigma_y^{-1}$, the formula for weights is
\[ 
\b w_{mos} = \frac{\sigma \b\Gamma \b\mu}{\sqrt{\b\mu' \b\Gamma\b\mu}}.
\]
The estimates that we will use
\[ \widehat{\b w}_{mos} = \frac{\sigma \widehat{\b \Gamma} \widehat{\b \mu}}{\sqrt{\widehat{\b \mu}' \widehat{\b \Gamma} \widehat{\b \mu}}}.\]

Our maximum out-of-sample Sharpe Ratio estimate using the nodewise estimate $\widehat{\b\Gamma}$ is:
\[ 
\widehat{SR}_{mos} := \frac{\widehat{\b w}_{mos}' \mu}{\sqrt{\widehat{\b w}_{mos}' \Sigma_y \widehat{\b w}_{mos}}}=
\frac{\b\mu' \widehat{\b\Gamma} \widehat{\b\mu}}{\sqrt{ \widehat{\b\mu}' \widehat{\b\Gamma}' \b\Sigma_y \widehat{\b\Gamma} \widehat{\b\mu}}}.
\]

Below we provide a sparsity assumption for the case of maximum out of sample Sharpe Ratio.

\begin{assum}\label{as9}
\[ 
p \bar{s} l_n = o(1).
\]
\end{assum}

\begin{thm}\label{msros}
Under Assumptions \ref{as1}-\ref{as4},\ref{as6}, \ref{as7}(i), \ref{as8}, \ref{as9}
\[ \left| \left[\frac{\widehat{SR}_{mos}}{SR^*}\right]^2 - 1\right| = O_p ( K^2 \bar{s} l_n)=o_p(1).\]
\end{thm}

\textbf{Remarks:}
\begin{enumerate}
\item
Note that p.4353 of \cite{lw2017} shows that the maximum out-of-sample Sharpe Ratio is equivalent to minimizing a certain loss function of the portfolio. The limit of the
loss function is derived under an optimal shrinkage function in Theorem 1. After that, they provide a shrinkage function even in the cases of $p/n \to r_1 \in (0,1) \cup (1,+\infty)$. Their proofs allow for iid data, which is restrictive since it does not allow for correlation in returns across time.
\item
We cannot have $p>n$ in this theorem, due to Assumption \ref{as9}, this shows the difficulty of maximum out of sample estimation. Mainly, $\textnormal{Eigmax} (\b\Sigma_y) = O(p)$ caused this problem in the proofs and provide the need for Assumption \ref{as9}.
\item
$p \bar{s} l_n = o(1)$ can be also obtained in non-sparse precision matrix, although the conditions will be more restrictive. To see this, we now have $\bar{s}=p$ in non-sparse case. So 
\[ 
p \bar{s} l_n = p^2 r_n^2 K^{5/2} \max[p \lambda_n, p^{1/2} K^{1/2} \sqrt{\ln(n)/n}] \to 0,
\]
by (\ref{lar})(\ref{dn}). This implies that we need $p<n$.
\item
The case of large non-negative weights can be handled with our analysis. This is the case of growing exposure, where the weights are depending on growing sparsity of the precision matrix, hence taking large values. For this, in Supplement D, our proof of Theorem D.1-analyzing mean of the portfolio- provides insight into this issue. Our Assumption 1 allows $\bar{s}$ to be nondecreasing in $n$.
\end{enumerate}

\subsection{Portfolio Estimation Based Sharpe Ratio Analysis}

In this section for the scenarios we considered in Sections 4.1-4.2, we form the estimate of the portfolio weights  and substitute that into the Sharpe Ratio. To understand the effects of only portfolio estimation for consistent estimation of Sharpe Ratio, we keep $\b\mu, \b\Sigma_y$ as constants in Sharpe Ratio estimates. We start with GMV  portfolio. The estimated portfolio weights are 
\[ 
\hat{\b w}_{nw}:= \frac{\hat{\b\Gamma}  \b1_p}{\b1_p' \hat{\b\Gamma}  \b1_p}.
\]

The Sharpe Ratio estimate of this portfolio is:
\[ \widehat{SR}_{nw,p}:= \frac{\hat{\b w}_{nw}' \b\mu}{\sqrt{\hat{\b w}_{nw}' \b\Sigma_y \hat{\b w}_{nw}}} = \frac{p^{1/2} \left(\b 1_p' \hat{\b \Gamma}' \b \mu/p\right)}{\sqrt{\b 1_p' \hat{\b \Gamma}' \b \Sigma_y
\hat{\b \Gamma} \b 1_p/p}}.\]
The optimized-target population Sharpe Ratio is given in (\ref{5.1}).

\begin{corollary}\label{cor1}
Under Assumptions \ref{as1}-\ref{as4},\ref{as6}, \ref{as7}(i), \ref{as8}, \ref{as9} with $|\b 1_p' \b \Gamma \b \mu|/p \ge C > 0$
\[ \left| \left[\frac{\widehat{SR}_{nw,p}}{SR}\right]^2 - 1\right| = O_p ( K \bar{s} l_n)=o_p(1).\]
\end{corollary}

Now we consider the Sharpe Ratio based on Markowitz portfolio. The estimated portfolio weights are
\[
\hat{\b w}_{mv}= \frac{\hat{D} - \rho_1 \hat{F}}{\hat{A} \hat{D} - \hat{F}^2} \left( \hat{\b \Gamma} \b 1_p/p
\right) + \frac{\rho_1 \hat{A} - \hat{F}}{\hat{A} \hat{D} - \hat{F}^2} \left( \hat{\b \Gamma} \hat{\b \mu}/p
\right).
\]
These are estimates by plugging in terms in equation (\ref{38}). Denote the Sharpe Ratio based on portfolio weight estimates
\[ \widehat{SR}_{mv,p} = \frac{\hat{\b w}_{mv}' \mu }{\sqrt{\hat{\b w}_{mv}' \b \Sigma_y \hat{\b w}_{mv}}}.\]
The optimal Sharpe Ratio is in (\ref{30a}) in this case.

\begin{corollary}\label{cor2}
Under Assumptions \ref{as1}-\ref{as4},\ref{as6}, \ref{as7}(i), \ref{as8}, \ref{as9} with $A \rho_1^2 - 2 F \rho_1 + D \ge C_1>0,
AD -F^2 \ge C_1 > 0$, and $|\b 1_p'\b  \Gamma \b \mu | \ge C >0$, with $\rho_1$ bounded away from zero and infinity,
\[ \left| \left[\frac{\widehat{SR}_{mv,p}}{SR_{mv}}\right]^2 - 1\right| = O_p ( K^{5/2} \bar{s} l_n)=o_p(1).\]
\end{corollary}

In case of constrained maximum Sharpe Ratio in section 4.2, when $\b 1_p' \b \Sigma_y^{-1} \b \mu >0$, we can establish the portfolio weight estimates
\[ \hat{\b w}_{c,1}= \frac{\hat{\b \Gamma} \hat{\b \mu}}{\b 1_p' \hat{\b \Gamma} \hat{\b \mu}}.\]

Constrained maximum Sharpe Ratio estimate when $\b 1_p' \b \Sigma_y^{-1} \b \mu >0$ is:

\[ \widehat{MSR}_{p}:= \frac{ \hat{\b w}_{c,1}' \b \mu}{\sqrt{\hat{\b w}_{c,1}' \b \Sigma_y \hat{\b w}_{c,1}}}
= \frac{\hat{\b \mu}' \hat{\b \Gamma}' \b \mu}{\sqrt{\hat{\b \mu}' \hat{\b \Gamma}' \b \Sigma_y \hat{\b \Gamma} \hat{\b \mu}}}.\]
The optimal Sharpe Ratio in this case is in (\ref{1}).

\begin{corollary}\label{cor3}
Under Assumptions \ref{as1}-\ref{as4},\ref{as6}, \ref{as7}(i), \ref{as8}, \ref{as9} with  $\b 1_p' \b \Sigma_y^{-1} \b \mu  \ge C >0$
\[ \left| \left[\frac{\widehat{MSR}_{p}}{MSR}\right]^2 - 1\right| = O_p ( K^{2} \bar{s} l_n)=o_p(1).\]
\end{corollary} 

The constrained maximum Sharpe Ratio weights when $\b 1_p' \b \Sigma_y^{-1} \b \mu \le 0$ are more complicated as seen in $\b w_c$ in Section 4.2. The estimate is:
\[ \hat{\b w}_{c,2}:= (\delta \hat{\b u}_{max}, 1 - \b 1_{p-1}' \hat{\b u}_{\max})', \]
with 
\[ \hat{\b u}_{\max}:= \frac{(\b A_{p-1}' \b A_{p-1})^{-1} (\b A_{p-1}'  \hat{\b z}_{max})}{\sqrt{\hat{\b z}_{max}' \b A_{p-1} (\b A_{p-1}' \b A_{p-1})^{-2} \b A_{p-1}' \hat{\b z}_{max}}}.\]

\[ \hat{\b z}_{max}:= \hat{\b \Gamma} \left( I_p - \frac{\b 1_p \b 1_p' \hat{\b \Gamma}}{\b 1_p' \hat{\b \Gamma} \b 1_p}
\right) \frac{\hat{\b \mu}}{\widehat{MSR}_c}.\]

Note that maximum Sharpe Ratio in this second constrained case is:
\[ \widehat{MSR}_{c,p}= \frac{\hat{\b w}_{c,2}' \b \mu}{\sqrt{\hat{\b w}_{c,2}' \b \Sigma_y \hat{\b w}_{c,2}}}.\]

Using $\hat{\b w}_{c,2}$ poses several challenges. 
Taking $\delta \to \infty$ to reach the optimal Sharpe Ratio is key but the rate may play a role and also the weights depend on $\hat{\b u}_{\max}$ term which depends on $\hat{\b z}_{max}$ that depends on precision matrix estimate $\hat{\b \Gamma}$, mean estimate $\hat{\mu}$,  and estimate $\widehat{MSR}_c$ from section 4.2. So, given Theorems 2 and 6, we think that consistency is plausible. However, given the lengthy material in this paper, this is beyond the scope of our theoretical analysis. Hence, similar corollaries for Theorems 6-7 cannot be handled in this paper.

An important fact that applies to all Corollaries here is that we can only have $p<n$ case, as discussed in Remark 3 of Theorem 8.

\section{Simulations}\label{sims}

\subsection{Models and Implementation Details}

In this section, we compare the nodewise regression with several models in a simulation exercise. The two aims of the exercise are to determine whether our method achieves consistency and how our method performs compared to others in the estimation of the constrained maximum Sharpe Ratio, the out-of-sample maximum Sharpe Ratio, and the Sharpe Ratio in global minimum-variance and Markowitz mean-variance portfolios.

The other methods that are used widely in the literature and benefit from high-dimensional techniques are the principal orthogonal complement thresholding (POET) from \cite{fan2013}, the nonlinear shrinkage (NL-LW) and the single factor nonlinear shrinkage (SF-NL-LW) from \cite{lw2017}, and the maximum Sharpe Ratio estimated and sparse regression (MAXSER) from \cite{ao2019}. All models except for the MAXSER are plug-in estimators, where the first step is to estimate the precision/covariance matrix, and the second step is to plug-in the estimate in the desired equation.

The POET uses principal components to estimate the covariance matrix and allows some eigenvalues of $\b\Sigma_n$ to be spiked and grow at a rate $O(p)$, which allows common and idiosyncratic components to be identified via principal components analysis and can consistently estimate the space spanned by the eigenvectors of $\b\Sigma_n$. However, \citet{fan2013} point out that the absolute convergence rate of the model is not satisfactory for estimating $\b\Sigma_n$, and consistency can only be achieved in terms of the relative error matrix.

Nonlinear shrinkage is a method that individually determines the amount of shrinkage of each eigenvalue in the covariance matrix for  a particular loss function. The main aim is to increase the value of the lowest eigenvalues and decrease the largest eigenvalues to stabilize the high-dimensional covariance matrix. This nonlinear method is a very novel and excellent idea.
\citet{lw2017} propose a function that captures the objective of an investor using portfolio selection. As a result, they have an optimal estimator of the covariance matrix for portfolio selection for many assets. The SF-NL-LW method extracts a single factor structure from the data before estimating the covariance matrix, which is simply an equal-weighted portfolio with all assets.

Finally, the MAXSER starts with estimating the adjusted squared maximum Sharpe Ratio  used in a penalized regression to obtain the portfolio weights. Of all the discussed models, the MAXSER is the only one that does not estimate the precision matrix in a plug-in estimator of the maximum Sharpe Ratio.

Regarding implementation, the POET and both models from \cite{lw2017} are available in the R packages POET \cite{POETR} and nlshrink \cite{nlshrinkR}. The SF-NL-LW needs some minor adjustments following the procedures described in \cite{lw2017}. For the MAXSER, we follow the steps for the non-factor case in \cite{ao2019}, and we use the package lars (\cite{larsR}) for the penalized regression estimation. We estimate the nodewise regression following the steps in Section 3.2 using the glmnet package \cite{glmnet2010} for penalized regressions. We used two alternatives to select the regularization parameter $\lambda$, a $10$-fold cross validation (CV), and the generalized information criterion (GIC) from \cite{zhang2010regularization}.

The GIC procedure starts by fitting $\widehat{\b\gamma}_j$ in  (\ref{rv15}) for a range of $\lambda_j$ that goes from the intercept-only model to the largest feasible model. This is automatically done by the glmnet package. Then, for the GIC procedure, we calculate the information criterion for a given $\lambda_j$ among the ranges of all possible tuning parameters
\begin{equation}
GIC_j (\lambda_j) = \frac{SSR (\lambda_j)}{n} + q(\lambda_j) \log(p-1) \frac{\ln[\ln(n)]}{n},
\end{equation}
where $SSR(\lambda_j)$ is the sum squared error for a given $\lambda_j$, $q (\lambda_j)$ is the number of variables, given $\lambda_j, $ in the model that is nonzero, and $p$ is the number of assets. The last step is to select the model with the smallest GIC. Once this is done for all assets $j = 1,\dots,p$, we can proceed to obtain $\widehat{\b\Gamma}_{GIC}$.

For the CV procedure, we split the sample into $k$ subsamples and fit the model for a range of $\lambda_j$ as in the GIC procedure. However, we will fit models in the subsamples. We always estimate the models in $k-1$ subsamples, leaving one subsample as a test sample, where we compute the mean squared error (MSE). After repeating the procedure using all $k$ subsamples as a test, we finally compute the average MSE across all subsamples and select the $\lambda_j$ for each asset $j$ that yields the smallest average MSE. We can then use the estimated $\widehat{\b\gamma}_j$ to obtain $\widehat{\b\Gamma}_{CV}$.

\subsection{Data Generation Process and Results}

The DGP is based on a simplified version of the factor DGP in \citet{ao2019}, for $j=1,\cdots, p$:
\begin{equation}
    \b y_j = \alpha_j + \sum_{k = 1}^K \beta_{j,k} f_k + \b e_j,
\end{equation}
where $\b y_j$ and $\b f_k$ are the monthly asset returns of asset $j$, factor returns of factor $k$ respectively, $\beta_{j,k}$ are the individual stock sensitivities to the factors, and $\alpha_j + e_j$ represent the idiosyncratic component of each stock. 
We start with two specifications that correspond to two tables. Table 1  corresponds to 1 factor: excess return of the market portfolio, hence $K=1$, and Table 2 corresponds to 3 factors from  the Fama \& French three factors, $K=3$. \footnote{The factors are book-to-market, market capitalization, and the excess return of the market portfolio.}  Let  $\b\mu_{f}$ and $\b\Sigma_f$ be the factors' sample mean and covariance matrix. The $\beta$, and $\alpha$ and covariance matrix of residuals: $\widetilde{\b\Sigma}_n$ are estimated using a simple least-squares regression using returns from the S\&P500 stocks that were part of the index in the entire period from 2008 to 2017. In each simulation, we randomly select $p$ stocks from the pool with replacement because our simulations require more than the total number of available stocks. We then used the selected stocks to generate individual returns with covariance matrix of errors: $\widehat{\b\Sigma}_n = \widetilde{\b\Sigma}_n \odot Toeplitz ({\rho})$, where $Toeplitz ({\rho})$ is the $p \times p$ matrix of the form, for (i,j)th element
\[ 
Toeplitz ({\rho})_{i,j}:= \rho^{|i-j|},
\]
with $\rho=0.25, 0.5, 0.75$. $\b A \odot \b B $ represents element by element multiplication (Hadamard product) of two square matrices $\b A, \b B$ of the same dimensions.

Tables 1-2 show the results. The values in each cell show the average absolute estimation error for estimating the square of the Sharpe Ratio. Each eight-column block in the table shows the results for a different sample size. In each of these blocks, the first four columns are for $p = n/2$, and the last four columns are for $p = 3n/2$. MSR, MSR-OOS, GMV-SR, and MKW-SR are the constrained maximum Sharpe Ratio, the out-of-sample maximum Sharpe Ratio, the Sharpe Ratio from the global minimum-variance portfolio, and the Sharpe Ratio from the Markowitz portfolio with target returns set to 1\%, respectively. Therefore, there are four categories to evaluate the different estimates. The MAXSER risk constraint was set to 0.04 following \cite{ao2019}. We ran 100 iterations in each simulation setup.
All bold-face entries in tables show category champions.

Both Tables show that our method achieves consistency, as shown in Theorems. Analyzing $K=3$, Table 2, with $\rho=0.50$ OOS-MSR (the Out Of Sample-Maximum Sharpe Ratio), and Generalized Information Criterion tuning parameter selection, the estimation error at $p=n/2$, with $n=100$ is 1.244, and this error declines to 0.585 at $p=n/2, n=200$, and then declines to 0.321 at $p=n/2, n =400$. So with jointly increasing $n,p$ we show that the error declines, as predicted by our theorems. The main reason is that errors grow with $\sqrt{\ln(p)}$, but decline with $n^{1/2}$ rate. So the number of assets in a large portfolio only affects the error logarithmically. To give another example from Table 2, with $\rho=0.50$, GMV-SR (Global Minimum Variance-Sharpe Ratio) and Cross Validation tuning parameter selection with our method, the estimation error is 0.352 with $p= 3n/2, n=100$, then this error declines to 0.213 with $p=3n/2, n=200$, and further declines to 0.143 with $p= 3n/2, n=400$.

Next, we consider which method achieves the smallest estimation error. Table 1 favors SF-NL-LW (Single Factor Non-Linear Shrinkage of Ledoit-Wolf) since it has a single factor built into this subset of their technique. We get better results in Table 2 ($K=3$) for our methods. We have 4 categories: MSR, OOS-MSR, GMV-SR, MKW-SR corresponding to our Theorems 3-9.
There are nine possibilities in each category (given we are either at $p=n/2$ or $p=3n/2$), representing three choices of sample sizes paired with 3 choices of different Toeplitz structures. 

We analyze each category. We start with Table 1. With $p=3n/2$ in OOS-MSR our NW-GIC method has the smallest errors 8 out of 9 categories. When $p=n/2$, MAXSER method dominates all others since it is specifically factor model designed to handle OOS-MSR with $p<n$.  In GMV-SR, with $p=n/2$, in 3 out of 9 cases, our NW-GIC dominates. In the other categories in Table 1, non-linear shrinkage method of Ledoit-Wolf (2017) does the best, but our methods come a very close second.

In Table 2, with $K=3$, our methods perform better than in Table 1. In the category of GMV-SR, with $p=3n/2$, out of 9 possible configurations, our methods have the smallest error in 7 cases. 
Our methods dominate in the same category, with $p=0.5n$, 5 out of 9 possibilities.  In the case of the category of MKW-SR (Markowitz-Sharpe Ratio), our theorems predict that our methods may suffer from a number of factors. We see that non-linear shrinkage methods are the best, and our methods are the second best in this category. In the constrained maximum Sharpe Ratio, (MSR)  non-linear shrinkage methods perform the best.

\begin{landscape}
\begin{table}[]\label{tab1}
\caption{Simulation Results -- Single Factor Toeplitz DGP with Real Factors}
\label{tab:sim-toe-sfact-real}
\begin{adjustbox}{max width=1.45\textwidth}
\begin{threeparttable}
\begin{tabular}{lccccccccccccccccccccccccccccc}
\hline
         &                                 &                                 &                                 &                                 &  &                                 &                                 &                                 &                                 &  & \multicolumn{9}{c}{{\ul \textbf{Toeplitz $\rho = 0.25$}}}                                                                                                                                                                                                                        &  &                                 &                                 &                                 &                                 &  &                                 &                                 &                                 &                                 \\
         & \multicolumn{9}{c}{n = 100}                                                                                                                                                                                                                                                      &  & \multicolumn{9}{c}{n = 200}                                                                                                                                                                                                                                                      &  & \multicolumn{9}{c}{n = 400}                                                                                                                                                                                                                                                      \\ \cline{2-10} \cline{12-20} \cline{22-30} 
         & \multicolumn{4}{c}{p = n/2}                                                                                                           &  & \multicolumn{4}{c}{p = 3n/2}                                                                                                          &  & \multicolumn{4}{c}{p = n/2}                                                                                                           &  & \multicolumn{4}{c}{p = 3n/2}                                                                                                          &  & \multicolumn{4}{c}{p = n/2}                                                                                                           &  & \multicolumn{4}{c}{p=3n/2}                                                                                                            \\ \cline{2-5} \cline{7-10} \cline{12-15} \cline{17-20} \cline{22-25} \cline{27-30} 
         & MSR                             & OOS-MSR                         & GMV-SR                          & MKW-SR                          &  & MSR                             & OOS-MSR                         & GMV-SR                          & MKW-SR                          &  & MSR                             & OOS-MSR                         & GMV-SR                          & MKW-SR                          &  & MSR                             & OOS-MSR                         & GMV-SR                          & MKW-SR                          &  & MSR                             & OOS-MSR                         & GMV-SR                          & MKW-SR                          &  & MSR                             & OOS-MSR                         & GMV-SR                          & MKW-SR                          \\ \cline{2-5} \cline{7-10} \cline{12-15} \cline{17-20} \cline{22-25} \cline{27-30} 
NW-GIC   & 0.517                           & 1.062                           & 0.680                           & 0.114                           &  & 0.520                           & 1.095                           & 0.251                           & 0.126                           &  & 0.331                           & 0.515                           & \textbf{0.211} & 0.073                           &  & 0.352                           & \textbf{0.545} & 0.140                           & 0.079                           &  & 0.208                           & 0.262                           & \textbf{0.091} & \textbf{0.051} &  & 0.219                           & \textbf{0.273} & 0.077                           & 0.053                           \\
NW-CV    & 0.517                           & 1.061                           & 0.634                           & 0.112                           &  & 0.521                           & 1.099                           & 0.251                           & 0.128                           &  & 0.331                           & 0.514                           & 0.212                           & \textbf{0.072} &  & 0.352                           & 0.546                           & 0.140                           & 0.079                           &  & 0.208                           & 0.262                           & 0.091                           & 0.051                           &  & 0.219                           & 0.273                           & 0.078                           & 0.053                           \\
POET     & 0.526                           & 1.055                           & 0.636                           & 0.167                           &  & 0.522                           & 1.095                           & 0.260                           & 0.144                           &  & 0.336                           & 0.511                           & 0.212                           & 0.102                           &  & 0.354                           & 0.548                           & 0.145                           & 0.089                           &  & 0.212                           & 0.263                           & 0.096                           & 0.066                           &  & 0.220                           & 0.276                           & 0.081                           & 0.058                           \\
NL-LW    & \textbf{0.487} & 1.705                           & \textbf{0.559} & 0.172                           &  & \textbf{0.480} & 2.249                           & 0.377                           & 0.333                           &  & \textbf{0.301} & 0.961                           & 0.322                           & 0.216                           &  & \textbf{0.300} & 1.350                           & 0.329                           & 0.391                           &  & \textbf{0.169} & 0.645                           & 0.265                           & 0.258                           &  & \textbf{0.163} & 0.931                           & 0.333                           & 0.416                           \\
SF-NL-LW & 0.516                           & 1.069                           & 0.689                           & \textbf{0.110} &  & 0.517                           & \textbf{1.094} & \textbf{0.249} & \textbf{0.120} &  & 0.330                           & 0.515                           & 0.215                           & 0.072                           &  & 0.350                           & 0.545                           & \textbf{0.139} & \textbf{0.076} &  & 0.207                           & 0.263                           & 0.091                           & 0.051                           &  & 0.217                           & 0.273                           & \textbf{0.076} & \textbf{0.051} \\
MAXSER   &                                 & \textbf{0.359} &                                 &                                 &  &                                 &                                 &                                 &                                 &  &                                 & \textbf{0.152} &                                 &                                 &  &                                 &                                 &                                 &                                 &  &                                 & \textbf{0.098} &                                 &                                 &  &                                 &                                 &                                 &                                 \\
         &                                 &                                 &                                 &                                 &  &                                 &                                 &                                 &                                 &  &                                 &                                 &                                 &                                 &  &                                 &                                 &                                 &                                 &  &                                 &                                 &                                 &                                 &  &                                 &                                 &                                 &                                 \\
         &                                 &                                 &                                 &                                 &  &                                 &                                 &                                 &                                 &  & \multicolumn{9}{c}{{\ul \textbf{Toeplitz $\rho = 0.5$}}}                                                                                                                                                                                                                         &  &                                 &                                 &                                 &                                 &  &                                 &                                 &                                 &                                 \\
NW-GIC   & 0.525                           & 1.067                           & 0.829                           & 0.134                           &  & 0.529                           & \textbf{1.095} & \textbf{0.266} & 0.157                           &  & 0.342                           & 0.521                           & \textbf{0.220} & 0.100                           &  & 0.365                           & \textbf{0.552} & 0.161                           & 0.113                           &  & 0.222                           & 0.271                           & 0.108                           & 0.083                           &  & 0.233                           & \textbf{0.283} & 0.108                           & 0.089                           \\
NW-CV    & 0.526                           & 1.067                           & 0.726                           & 0.132                           &  & 0.531                           & 1.099                           & 0.267                           & 0.159                           &  & 0.342                           & 0.521                           & 0.221                           & 0.100                           &  & 0.365                           & 0.553                           & 0.161                           & 0.113                           &  & 0.222                           & 0.271                           & 0.108                           & 0.084                           &  & 0.233                           & 0.283                           & 0.108                           & 0.089                           \\
POET     & 0.535                           & 1.061                           & 0.721                           & 0.190                           &  & 0.531                           & 1.096                           & 0.279                           & 0.175                           &  & 0.348                           & 0.518                           & 0.226                           & 0.133                           &  & 0.367                           & 0.556                           & 0.167                           & 0.124                           &  & 0.226                           & 0.273                           & 0.118                           & 0.100                           &  & 0.235                           & 0.286                           & 0.113                           & 0.095                           \\
NL-LW    & \textbf{0.495} & 1.694                           & \textbf{0.558} & 0.151                           &  & \textbf{0.489} & 2.231                           & 0.360                           & 0.290                           &  & \textbf{0.306} & 0.954                           & 0.316                           & 0.192                           &  & \textbf{0.313} & 1.340                           & 0.289                           & 0.342                           &  & \textbf{0.175} & 0.641                           & 0.233                           & 0.231                           &  & \textbf{0.177} & 0.926                           & 0.281                           & 0.365                           \\
SF-NL-LW & 0.523                           & 1.076                           & 0.819                           & \textbf{0.130} &  & 0.526                           & 1.095                           & 0.266                           & \textbf{0.151} &  & 0.340                           & 0.523                           & 0.225                           & \textbf{0.099} &  & 0.363                           & 0.553                           & \textbf{0.160} & \textbf{0.109} &  & 0.220                           & 0.273                           & \textbf{0.106} & \textbf{0.082} &  & 0.232                           & 0.284                           & \textbf{0.106} & \textbf{0.087} \\
MAXSER   &                                 & \textbf{0.363} &                                 &                                 &  &                                 &                                 &                                 &                                 &  &                                 & \textbf{0.158} &                                 &                                 &  &                                 &                                 &                                 &                                 &  &                                 & \textbf{0.091} &                                 &                                 &  &                                 &                                 &                                 &                                 \\
         &                                 &                                 &                                 &                                 &  &                                 &                                 &                                 &                                 &  &                                 &                                 &                                 &                                 &  &                                 &                                 &                                 &                                 &  &                                 &                                 &                                 &                                 &  &                                 &                                 &                                 &                                 \\
         &                                 &                                 &                                 &                                 &  &                                 &                                 &                                 &                                 &  & \multicolumn{9}{c}{{\ul \textbf{Toeplitz $\rho = 0.75$}}}                                                                                                                                                                                                                        &  &                                 &                                 &                                 &                                 &  &                                 &                                 &                                 &                                 \\
NW-GIC   & 0.542                           & 1.105                           & 1.300                           & 0.183                           &  & 0.549                           & \textbf{1.131} & 0.318                           & 0.227                           &  & 0.366                           & 0.558                           & 0.248                           & 0.166                           &  & 0.390                           & \textbf{0.593} & 0.224                           & 0.189                           &  & 0.250                           & 0.309                           & 0.174                           & 0.156                           &  & 0.264                           & \textbf{0.326} & 0.193                           & 0.170                           \\
NW-CV    & 0.542                           & 1.108                           & 1.131                           & 0.183                           &  & 0.551                           & 1.144                           & 0.319                           & 0.231                           &  & 0.366                           & 0.560                           & 0.248                           & 0.167                           &  & 0.390                           & 0.595                           & 0.223                           & 0.189                           &  & 0.250                           & 0.311                           & 0.174                           & 0.157                           &  & 0.264                           & 0.327                           & 0.193                           & 0.171                           \\
POET     & 0.553                           & 1.104                           & 1.086                           & 0.248                           &  & 0.552                           & 1.136                           & 0.333                           & 0.249                           &  & 0.373                           & 0.561                           & 0.261                           & 0.204                           &  & 0.393                           & 0.601                           & 0.235                           & 0.203                           &  & 0.256                           & 0.319                           & 0.192                           & 0.179                           &  & 0.267                           & 0.335                           & 0.201                           & 0.180                           \\
NL-LW    & \textbf{0.510} & 1.703                           & \textbf{0.548} & \textbf{0.109} &  & \textbf{0.510} & 2.215                           & 0.334                           & \textbf{0.187} &  & \textbf{0.324} & 0.956                           & 0.298                           & \textbf{0.132} &  & \textbf{0.337} & 1.339                           & 0.226                           & 0.232                           &  & \textbf{0.196} & 0.647                           & 0.194                           & 0.151                           &  & \textbf{0.202} & 0.931                           & \textbf{0.186} & 0.252                           \\
SF-NL-LW & 0.537                           & 1.115                           & 0.922                           & 0.176                           &  & 0.545                           & 1.136                           & \textbf{0.315} & 0.220                           &  & 0.361                           & 0.563                           & \textbf{0.245} & 0.157                           &  & 0.387                           & 0.597                           & \textbf{0.218} & \textbf{0.183} &  & 0.244                           & 0.313                           & \textbf{0.160} & \textbf{0.149} &  & 0.261                           & 0.330                           & 0.187                           & \textbf{0.167} \\
MAXSER   &                                 & \textbf{0.371} &                                 &                                 &  &                                 &                                 &                                 &                                 &  &                                 & \textbf{0.169} &                                 &                                 &  &                                 &                                 &                                 &                                 &  &                                 & \textbf{0.082} &                                 &                                 &  &                                 &                                 &                                 &                                 \\ \hline
\end{tabular}%
\begin{tablenotes}
\item The table shows the simulation results for the Toeplitz DGP. Each simulation was done with 100 iterations. We used sample sizes $n$ of 100, 200 and 400, and the number of stocks was either $n/2$ or $1.5n$ for the low-dimensional and the high-dimensional case, respectively. Each block of rows shows the results for a different value of $\rho$ in the Toeplitz DGP. The values in each cell show the average absolute estimation error for estimating the square of the Sharpe Ratio. 
\end{tablenotes}
\end{threeparttable}
\end{adjustbox}
\end{table}
\end{landscape}

\begin{landscape}
\begin{table}[] \label{tab2}
\caption{Simulation Results -- 3 Factor Toeplitz DGP with Real Factors}
\label{tab:sim-toe-fact-real}
\begin{adjustbox}{max width=1.45\textwidth}
\begin{threeparttable}
\begin{tabular}{lccccccccccccccccccccccccccccc}
\hline
{\ul }   & {\ul }                          & {\ul }                          & {\ul }                          & {\ul }                          & {\ul } & {\ul }                          & {\ul }                          & {\ul }                          & {\ul }                          & {\ul } & \multicolumn{9}{c}{{\ul Toeplitz $\rho = 0.25$}}                                                                                                                                                                                                                                 & {\ul } & {\ul }                          & {\ul }                          & {\ul }                          & {\ul }                          & {\ul } & {\ul }                          & {\ul }                          & {\ul }                          & {\ul }                          \\
         & \multicolumn{9}{c}{n = 100}                                                                                                                                                                                                                                                            &        & \multicolumn{9}{c}{n = 200}                                                                                                                                                                                                                                                      &        & \multicolumn{9}{c}{n = 400}                                                                                                                                                                                                                                                            \\ \cline{2-10} \cline{12-20} \cline{22-30}
         & \multicolumn{4}{c}{p = n/2}                                                                                                           &        & \multicolumn{4}{c}{p = 3n/2}                                                                                                          &        & \multicolumn{4}{c}{p = n/2}                                                                                                           &  & \multicolumn{4}{c}{p=3n/2}                                                                                                            &        & \multicolumn{4}{c}{p = n/2}                                                                                                           &        & \multicolumn{4}{c}{p = 3n/2}                                                                                                         \\ \cline{2-5} \cline{7-10} \cline{12-15} \cline{17-20} \cline{22-25} \cline{27-30}
         & MSR                             & OOS-MSR                         & GMV-SR                          & MKW-SR                          &        & MSR                             & OOS-MSR                         & GMV-SR                          & MKW-SR                          &        & MSR                             & OOS-MSR                         & GMV-SR                          & MKW-SR                          &  & MSR                             & OOS-MSR                         & GMV-SR                          & MKW-SR                          &        & MSR                             & OOS-MSR                         & GMV-SR                          & MKW-SR                          &        & MSR                             & OOS-MSR                         & GMV-SR                          & MKW-SR                          \\
NW-GIC   & 0.544                           & 1.237                           & 0.739                           & 0.152                           &        & 0.579                           & \textbf{1.269} & 0.343                           & 0.220                           &        & 0.369                           & 0.578                           & \textbf{0.235} & 0.117                           &  & 0.391                           & \textbf{0.658} & 0.200                           & 0.158                           &        & 0.242                           & 0.311                           & \textbf{0.148} & \textbf{0.085} &        & 0.254                           & \textbf{0.350} & \textbf{0.122} & 0.100                           \\
NW-CV    & 0.543                           & 1.237                           & \textbf{0.724} & 0.151                           &        & 0.580                           & 1.309                           & \textbf{0.341} & 0.224                           &        & 0.369                           & 0.578                           & 0.235                           & 0.117                           &  & 0.391                           & 0.658                           & \textbf{0.199} & 0.158                           &        & 0.242                           & 0.311                           & 0.148                           & 0.085                           &        & 0.254                           & 0.350                           & 0.122                           & 0.100                           \\
POET     & 0.558                           & 1.137                           & 1.041                           & 0.308                           &        & 0.590                           & 1.641                           & 0.463                           & 0.391                           &        & 0.418                           & 0.852                           & 0.372                           & 0.346                           &  & 0.445                           & 2.078                           & 0.459                           & 0.411                           &        & 0.338                           & 1.279                           & 0.436                           & 0.371                           &        & 0.351                           & 3.682                           & 0.484                           & 0.410                           \\
NL-LW    & \textbf{0.511} & 1.726                           & 0.906                           & \textbf{0.110} &        & \textbf{0.535} & 2.251                           & 0.481                           & \textbf{0.086} &        & \textbf{0.348} & 0.988                           & 0.289                           & \textbf{0.084} &  & \textbf{0.358} & 1.370                           & 0.260                           & \textbf{0.118} &        & \textbf{0.216} & 0.642                           & 0.206                           & 0.092                           &        & \textbf{0.223} & 0.953                           & 0.206                           & 0.169                           \\
SF-NL-LW & 0.540                           & 1.168                           & 0.736                           & 0.199                           &        & 0.567                           & 1.292                           & 0.349                           & 0.238                           &        & 0.366                           & 0.581                           & 0.235                           & 0.152                           &  & 0.383                           & 0.679                           & 0.203                           & 0.160                           &        & 0.240                           & 0.321                           & 0.157                           & 0.090                           &        & 0.251                           & 0.361                           & 0.126                           & \textbf{0.096} \\
MAXSER   &                                 & \textbf{0.375} &                                 &                                 &        &                                 &                                 &                                 &                                 &        &                                 & \textbf{0.166} &                                 &                                 &  &                                 &                                 &                                 &                                 &        &                                 & \textbf{0.081} &                                 &                                 &        &                                 &                                 &                                 &                                 \\
         &                                 &                                 &                                 &                                 &        &                                 &                                 &                                 &                                 &        &                                 &                                 &                                 &                                 &  &                                 &                                 &                                 &                                 &        &                                 &                                 &                                 &                                 &        &                                 &                                 &                                 &                                 \\
{\ul }   & {\ul }                          & {\ul }                          & {\ul }                          & {\ul }                          & {\ul } & {\ul }                          & {\ul }                          & {\ul }                          & {\ul }                          & {\ul } & \multicolumn{9}{c}{{\ul Toeplitz $\rho = 0.50$}}                                                                                                                                                                                                                                 & {\ul } & {\ul }                          & {\ul }                          & {\ul }                          & {\ul }                          & {\ul } & {\ul }                          & {\ul }                          & {\ul }                          & {\ul }                          \\
NW-GIC   & 0.548                           & 1.247                           & 0.760                           & 0.163                           &        & 0.584                           & \textbf{1.273} & 0.352                           & 0.232                           &        & 0.376                           & 0.585                           & 0.239                           & 0.132                           &  & 0.398                           & \textbf{0.666} & 0.213                           & 0.172                           &        & 0.251                           & 0.321                           & \textbf{0.159} & 0.100                           &        & 0.263                           & \textbf{0.360} & \textbf{0.143} & 0.115                           \\
NW-CV    & 0.548                           & 1.248                           & \textbf{0.743} & 0.162                           &        & 0.585                           & 1.314                           & \textbf{0.351} & 0.236                           &        & 0.376                           & 0.585                           & 0.239                           & 0.132                           &  & 0.398                           & 0.666                           & \textbf{0.212} & 0.173                           &        & 0.251                           & 0.321                           & 0.160                           & 0.100                           &        & 0.263                           & 0.360                           & 0.143                           & 0.115                           \\
POET     & 0.563                           & 1.148                           & 1.203                           & 0.320                           &        & 0.595                           & 1.642                           & 0.471                           & 0.402                           &        & 0.424                           & 0.855                           & 0.382                           & 0.358                           &  & 0.451                           & 2.065                           & 0.472                           & 0.422                           &        & 0.346                           & 1.276                           & 0.451                           & 0.383                           &        & 0.359                           & 3.648                           & 0.498                           & 0.420                           \\
NL-LW    & \textbf{0.516} & 1.732                           & 0.903                           & \textbf{0.107} &        & \textbf{0.540} & 2.244                           & 0.475                           & \textbf{0.079} &        & \textbf{0.349} & 0.990                           & 0.293                           & \textbf{0.077} &  & \textbf{0.366} & 1.372                           & 0.253                           & \textbf{0.101} &        & \textbf{0.226} & 0.648                           & 0.206                           & \textbf{0.083} &        & \textbf{0.229} & 0.955                           & 0.185                           & 0.151                           \\
SF-NL-LW & 0.543                           & 1.181                           & 0.753                           & 0.206                           &        & 0.572                           & 1.297                           & 0.358                           & 0.249                           &        & 0.371                           & 0.589                           & \textbf{0.237} & 0.160                           &  & 0.390                           & 0.687                           & 0.214                           & 0.173                           &        & 0.248                           & 0.331                           & 0.167                           & 0.102                           &        & 0.260                           & 0.371                           & 0.146                           & \textbf{0.110} \\
MAXSER   &                                 & \textbf{0.379} &                                 &                                 &        &                                 &                                 &                                 &                                 &        &                                 & \textbf{0.168} &                                 &                                 &  &                                 &                                 &                                 &                                 &        &                                 & \textbf{0.080} &                                 &                                 &        &                                 &                                 &                                 &                                 \\
         &                                 &                                 &                                 &                                 &        &                                 &                                 &                                 &                                 &        &                                 &                                 &                                 &                                 &  &                                 &                                 &                                 &                                 &        &                                 &                                 &                                 &                                 &        &                                 &                                 &                                 &                                 \\
{\ul }   & {\ul }                          & {\ul }                          & {\ul }                          & {\ul }                          & {\ul } & {\ul }                          & {\ul }                          & {\ul }                          & {\ul }                          & {\ul } & \multicolumn{9}{c}{{\ul Toeplitz $\rho = 0.75$}}                                                                                                                                                                                                                                 & {\ul } & {\ul }                          & {\ul }                          & {\ul }                          & {\ul }                          & {\ul } & {\ul }                          & {\ul }                          & {\ul }                          & {\ul }                          \\
NW-GIC   & 0.556                           & 1.295                           & 0.812                           & 0.189                           &        & 0.593                           & \textbf{1.318} & \textbf{0.377} & 0.263                           &        & 0.387                           & 0.623                           & 0.260                           & 0.164                           &  & 0.411                           & \textbf{0.708} & 0.246                           & 0.206                           &        & 0.267                           & 0.360                           & 0.198                           & 0.136                           &        & 0.281                           & \textbf{0.404} & 0.192                           & 0.150                           \\
NW-CV    & 0.556                           & 1.301                           & 0.802                           & 0.187                           &        & 0.595                           & 1.340                           & 0.383                           & 0.268                           &        & 0.388                           & 0.624                           & 0.260                           & 0.165                           &  & 0.412                           & 0.709                           & 0.246                           & 0.207                           &        & 0.268                           & 0.362                           & 0.198                           & 0.137                           &        & 0.281                           & 0.405                           & 0.192                           & 0.150                           \\
POET     & 0.571                           & 1.200                           & 2.071                           & 0.347                           &        & 0.605                           & 1.679                           & 0.490                           & 0.428                           &        & 0.436                           & 0.886                           & 0.409                           & 0.385                           &  & 0.464                           & 2.068                           & 0.501                           & 0.448                           &        & 0.361                           & 1.292                           & 0.483                           & 0.410                           &        & 0.375                           & 3.607                           & 0.529                           & 0.445                           \\
NL-LW    & \textbf{0.523} & 1.727                           & 0.947                           & \textbf{0.106} &        & \textbf{0.550} & 2.257                           & 0.468                           & \textbf{0.071} &        & \textbf{0.358} & 1.009                           & 0.287                           & \textbf{0.066} &  & \textbf{0.377} & 1.397                           & 0.242                           & \textbf{0.068} &        & \textbf{0.234} & 0.659                           & \textbf{0.195} & \textbf{0.062} &        & \textbf{0.242} & 0.979                           & \textbf{0.155} & \textbf{0.109} \\
SF-NL-LW & 0.549                           & 1.228                           & \textbf{0.783} & 0.222                           &        & 0.581                           & 1.346                           & 0.382                           & 0.272                           &        & 0.381                           & 0.631                           & \textbf{0.246} & 0.179                           &  & 0.402                           & 0.731                           & \textbf{0.241} & 0.200                           &        & 0.260                           & 0.371                           & 0.195                           & 0.127                           &        & 0.275                           & 0.417                           & 0.188                           & 0.141                           \\
MAXSER   &                                 & \textbf{0.385} &                                 &                                 &        &                                 &                                 &                                 &                                 &        &                                 & \textbf{0.173} &                                 &                                 &  &                                 &                                 &                                 &                                 &        &                                 & \textbf{0.079} &                                 &                                 &        &                                 &                                 &                                 &                                 \\ \hline
\end{tabular}%
\begin{tablenotes}
\item The table shows the simulation results for the Toeplitz DGP. Each simulation was done with 100 iterations. We used sample sizes $n$ of 100, 200 and 400, and the number of stocks was either $n/2$ or $1.5n$ for the low-dimensional and the high-dimensional case, respectively. Each block of rows shows the results for a different value of $\rho$ in the Toeplitz DGP. The values in each cell show the average absolute estimation error for estimating the square of the Sharpe Ratio. \end{tablenotes}
\end{threeparttable}

\end{adjustbox}
\end{table}
\end{landscape}

\section{Empirical Application}\label{emp}

For the empirical application, we use two subsamples. The first subsample uses data from January 1995 to December 2019 with an out-of-sample period from January 2005 to December 2019. We selected all stocks  in the S\&P 500 index for at least one month in the out-of-sample period and have data for the entire 1995-2019 period  resulting in 382 stocks. The second subsample starts in January 1990 and ends in December 2019 with an out-of-sample period from January 2000 to December 2019. Using the same criterion as the first subsample, the number of stocks was 321, which is around 15\% fewer than the first subsample. The objective is to have an out-of-sample competition between models, and we only estimated GMV and Markowitz portfolios for the plug-in estimators.
The first out-of-sample period includes only the recession of 2008. The second out-of-sample period includes the recessions of 2000 and 2008, and the out-of-sample periods reflect recent history.

The Markowitz return constraint $\rho_1$ is 0.8\% per month, and the MAXSER risk constraint is 4\%. In the low-dimensional experiment, we randomly select 50 stocks from the pool to estimate the models with the same stocks for all windows.  We also experimented with 25 stocks but did not report them. That table is available from the authors on demand. In the high-dimensional case, we use all available stocks.

We use a rolling window setup for the out-of-sample estimation of the Sharpe Ratio following \citet{caner2019}. Specifically, samples of size $n$ are divided into in-sample $(1:n_{I})$ and out-of-sample $(n_{I}+1:n)$. We start by estimating the portfolio $\widehat{\b w}_{n_I}$ in the in-sample period and the out-of-sample portfolio returns $\widehat{\b w}'_{n_I}y_{n_{I+1}}$. Then, we roll the window by one element $(2:n_{I}+1)$ and form a new in-sample portfolio $\widehat{\b w}_{n_{I+1}}$ and out-of-sample portfolio returns $\widehat{\b w}'_{n_{I+1}}y_{n_{I+2}}$. This procedure is repeated until the end of the sample.

The out-of-sample average return and variance without transaction costs are
\begin{equation*}
\widehat{\b \mu}_{os} = \frac{1}{n-n_{I}} \sum_{t=n_I}^{n-1} {\widehat{\b w}'_t}\b y_{t+1},\quad \widehat{\b\Sigma}_{y,os}^2 =  \frac{1}{n-n_{I}-1} \sum_{t=n_I}^{n-1} ({\widehat{\b w}'_t}\b y_{t+1} - \widehat{\b\mu}_{os})^2.
\end{equation*}

We estimate the Sharpe Ratios with and without transaction costs. The transaction cost, $c$, is defined as 50 basis points following \citet{demiguel2007optimal}. Let $y_{P,t+1} = {\widehat{\b w}'_t}\b y_{t+1}$ be the return of the portfolio in period $t+1$; in the presence of transaction costs, the returns will be defined as
\begin{equation*}
y_{P,t+1}^{Net} =  y_{P,t+1} - c(1+y_{P,t+1})\sum_{j=1}^p|{\widehat{w}_{t+1,j}}-{\widehat{w}^+_{t,j}}|,
\end{equation*}
where ${\widehat{w}^+_{t,j}} = {\widehat{w}_{t,j}}(1+y_{t+1,j})/(1+y_{t+1,P})$ and $y_{t,j}$ and $y_{t,P}$ are the excess returns of asset $j$ and the portfolio $P$ added to the risk-free rate. The adjustment made in ${\widehat{w}^+_{t,j}}$ is because the portfolio at the end of the period has changed compared to the portfolio at the beginning of the period.

The Sharpe Ratio is calculated from the average return and the variance of the portfolio in the out-of-sample period
\begin{equation*}
    SR = \frac{\widehat{\b\mu}_{os}}{\widehat{\b\Sigma}_{y,os}}.
\end{equation*}
The portfolio returns are replaced by the returns with transaction costs when we calculate the Sharpe Ratio with transaction costs.

We use the same test as \citet{ao2019} to compare the models. Specifically,
\begin{equation}
\label{eq:sharpe}
    H_0 ~ : ~ SR_{NW}\leq SR_0 ~~ vs ~~ H_a ~ : ~ SR_{NW}> SR_0,
\end{equation}
where $SR_{NW}$ is the Sharpe Ratio of our feasible nodewise model, which is tested against all remaining models. This is the \citet{jobson1981performance} test with \citet{memmel2003performance} correction.
We also considered the method of \citet{ledoit2008robust} for testing the significance of the winner and using the equally weighted portfolio as a benchmark; the results were very similar and hence are not reported.

We also include an equally weighted portfolio (EW). GMV-NW-GIC and GMV-NW-CV denote the nodewise method with GIC and cross validation tuning parameter choices, respectively, in the global minimum-variance portfolio (GMV). 

In each of our feasible nodewise models with GIC, CV,  we either use a single-factor model (market as the only factor) or three-factor model. They are denoted GMV-NW-GIC-SF, GMV-NW-GIC-3F for the global minimum variance portfolio analyzed with feasible nodewise method and GIC criterion for tuning parameter choice and single and three-factor models, respectively. In the same way, we define GMV-NW-CV-SF, GMV-NW-CV-3F. We take GMV-NW-GIC-SF as the benchmark to test against all other methods since it generally does well in different preliminary forecasts.

GMV-POET, GMV-NL-LW, and GMV-SF-NL-LW denote the POET, nonlinear shrinkage, and single-factor nonlinear shrinkage methods, respectively, which are described in the simulation section and also used in the global minimum-variance portfolio. The MAXSER is also used and explained in the simulation section. MW denotes the Markowitz mean-variance portfolio, and MW-NW-GIC-SF denotes the feasible nodewise method with GIC tuning parameter selection in the Markowitz portfolio with  a single factor. All the other methods with MW headers are analogous and thus self-explanatory.

The results are presented in Tables \ref{tab:emp_sp1} and \ref{tab:emp_sp2}. Table \ref{tab:emp_sp1} shows the results for the 2005-2019 out-of-sample period. Feasible nodewise methods do well  in terms of the Sharpe Ratio in Table \ref{tab:emp_sp1}.
For example, with transaction costs in the low-dimensional portfolio category, in terms of Sharpe Ratio (SR) (averaged over the out-of-sample time period), GMV-NW-GIC-SF is the best model. It has an SR of 0.210. In the case of high dimensional case with transaction costs in the same table, GMV-POET and our GMV-NW-GIC-SF virtually tie (difference in favor of POET in fourth decimal) at 0.214 for the Sharpe Ratio.

If we were to analyze only the Markowitz portfolio in Table \ref{tab:emp_sp1}, with transaction costs in high dimensions, MW-NW-GIC-SF has the highest SR of 0.211. Therefore, even in other subcategories of Markowitz portfolio, the feasible nodewise method dominates. Although statistical significance is not established, it is unclear that these significance tests have high power in our high-dimensional cases.

Table \ref{tab:emp_sp2} shows the results for the out-of-sample January 2000-2019 subsample.
We see that feasible nodewise methods dominate all scenarios except for the low-dimensional case with transaction costs. In  high dimensionality with transaction costs, GMV-NW-GIC-SF (Markowitz-nodewise-GIC) has an SR of 0.225, and the closest is GMV-POET with 0.204. Also, we experimented with two other out-sample periods of 2005-2017, 2000-2017, and the results are slightly better for our methods, and these can be shared on demand. 

\begin{table}[htb]
\caption{Empirical Results -- Out-of-Sample Period from Jan. 2005 to Dec. 2019}
\label{tab:emp_sp1}
\begin{adjustbox}{max width=\textwidth}
\begin{threeparttable}
\begin{tabular}{lccccccccccccccccccc}
\hline
              & \multicolumn{9}{c}{{\ul Without TC}}                                              & {\ul } & \multicolumn{9}{c}{{\ul With TC}}                                                 \\
              & \multicolumn{4}{c}{{\ul Low Dim.}} & {\ul } & \multicolumn{4}{c}{{\ul High Dim.}} & {\ul } & \multicolumn{4}{c}{{\ul Low Dim.}} & {\ul } & \multicolumn{4}{c}{{\ul High Dim.}} \\
              & SR     & AVG    & SD     & p-value &        & SR     & AVG    & SD     & p-value  &        & SR      & AVG    & SD    & p-value &        & SR     & AVG    & SD     & p-value  \\ \cline{1-5} \cline{7-10} \cline{12-15} \cline{17-20} 
EW            & 0.196          & 0.010   & 0.052   & 0.730   &  & 0.197          & 0.010   & 0.048   & 0.644   &  & 0.191          & 0.010   & 0.052   & 0.802   &  & 0.191          & 0.009   & 0.048   & 0.792   \\
GMV-NW-GIC-SF & \textbf{0.229} & 0.008   & 0.036   &         &  & 0.236          & 0.008   & 0.032   &         &  & \textbf{0.210} & 0.008   & 0.036   &         &  & 0.214          & 0.007   & 0.032   &         \\
GMV-NW-CV-SF  & 0.226          & 0.008   & 0.036   & 0.590   &  & 0.240          & 0.008   & 0.032   & 0.398   &  & 0.203          & 0.007   & 0.036   & 0.132   &  & 0.192          & 0.006   & 0.032   & 0.002   \\
GMV-NW-GIC-3F    & 0.215          & 0.007   & 0.034   & 0.576   &  & 0.214          & 0.007   & 0.033   & 0.520   &  & 0.191          & 0.007   & 0.034   & 0.424   &  & 0.183          & 0.006   & 0.033   & 0.398   \\
GMV-NW-CV-3F     & 0.212          & 0.007   & 0.034   & 0.474   &  & 0.226          & 0.007   & 0.032   & 0.790   &  & 0.183          & 0.006   & 0.034   & 0.278   &  & 0.132          & 0.004   & 0.032   & 0.032   \\
GMV-POET      & 0.218          & 0.007   & 0.034   & 0.682   &  & 0.232          & 0.007   & 0.030   & 0.914   &  & 0.203          & 0.007   & 0.034   & 0.822   &  & \textbf{0.214} & 0.006   & 0.030   & 0.996   \\
GMV-NL-LW     & 0.236          & 0.008   & 0.034   & 0.834   &  & 0.236          & 0.007   & 0.030   & 0.998   &  & 0.205          & 0.007   & 0.034   & 0.908   &  & 0.179          & 0.005   & 0.031   & 0.490   \\
GMV-SF-NL-LW  & 0.216          & 0.007   & 0.034   & 0.684   &  & \textbf{0.245} & 0.007   & 0.030   & 0.886   &  & 0.190          & 0.007   & 0.034   & 0.546   &  & 0.184          & 0.006   & 0.030   & 0.600   \\
MW-NW-GIC-SF  & 0.229          & 0.008   & 0.034   & 0.970   &  & 0.236          & 0.008   & 0.032   & 0.966   &  & 0.205          & 0.007   & 0.034   & 0.786   &  & 0.211          & 0.007   & 0.032   & 0.706   \\
MW-NW-CV-SF   & 0.228          & 0.008   & 0.034   & 0.942   &  & 0.242          & 0.008   & 0.032   & 0.620   &  & 0.197          & 0.007   & 0.034   & 0.482   &  & 0.190          & 0.006   & 0.032   & 0.056   \\
MW-NW-GIC-3F     & 0.214          & 0.007   & 0.033   & 0.628   &  & 0.217          & 0.007   & 0.033   & 0.606   &  & 0.185          & 0.006   & 0.033   & 0.444   &  & 0.183          & 0.006   & 0.033   & 0.416   \\
MW-NW-CV-3F      & 0.212          & 0.007   & 0.033   & 0.574   &  & 0.225          & 0.007   & 0.032   & 0.790   &  & 0.177          & 0.006   & 0.033   & 0.302   &  & 0.125          & 0.004   & 0.032   & 0.032   \\
MW-POET       & 0.223          & 0.007   & 0.032   & 0.880   &  & 0.229          & 0.007   & 0.030   & 0.844   &  & 0.200          & 0.006   & 0.032   & 0.794   &  & 0.207          & 0.006   & 0.030   & 0.840   \\
MW-NL-LW      & 0.220          & 0.008   & 0.034   & 0.860   &  & 0.235          & 0.007   & 0.030   & 0.980   &  & 0.186          & 0.006   & 0.034   & 0.636   &  & 0.177          & 0.005   & 0.030   & 0.540   \\
MW-SF-NL-LW   & 0.204          & 0.007   & 0.034   & 0.574   &  & 0.241          & 0.007   & 0.030   & 0.920   &  & 0.175          & 0.006   & 0.034   & 0.482   &  & 0.180          & 0.005   & 0.030   & 0.554   \\
MAXSER        & 0.161          & 0.010   & 0.065   & 0.510   &  &                &         &         &         &  & 0.024          & 0.002   & 0.066   & 0.116   &  &                &         &         &         \\ \hline
\end{tabular}%
\begin{tablenotes}
\item The table shows the Sharpe Ratio (SR), average returns (Avg), standard deviation (SD) and p-value of the \citet{jobson1981performance} test with \citet{memmel2003performance} correction. We also applied the
\citet{ledoit2008robust} test with circular bootstrap, and the results were very similar; therefore we only report those of the first test in this table.  The statistics were calculated from 180 rolling windows covering the period from Jan. 2005 to Dec. 2019, and the size of the estimation window was 120 observations.
\end{tablenotes}
\end{threeparttable}

\end{adjustbox}
\end{table}

\begin{table}[htb]
\caption{Empirical Results -- Out-of-Sample Period from Jan. 2000 to Dec. 2019}
\label{tab:emp_sp2}
\begin{adjustbox}{max width=\textwidth}
\begin{threeparttable}
\begin{tabular}{lccccccccccccccccccc}
\hline
{\ul }        & \multicolumn{9}{c}{{\ul Without TC}}                                              & {\ul } & \multicolumn{9}{c}{{\ul With TC}}                                                 \\
{\ul }        & \multicolumn{4}{c}{{\ul Low Dim.}} & {\ul } & \multicolumn{4}{c}{{\ul High Dim.}} & {\ul } & \multicolumn{4}{c}{{\ul Low Dim.}} & {\ul } & \multicolumn{4}{c}{{\ul High Dim.}} \\ \cline{1-5} \cline{7-10} \cline{12-15} \cline{17-20} 
{\ul }        & SR     & AVG    & SD     & p-value &        & SR     & AVG    & SD     & p-value  &        & SR      & AVG    & SD    & p-value &        & SR     & AVG    & SD     & p-value  \\ \cline{1-5} \cline{7-10} \cline{12-15} \cline{17-20} 
EW            & 0.201          & 0.010   & 0.047   & 0.874   &  & 0.210          & 0.010   & 0.047   & 0.546   &  & \textbf{0.195} & 0.009  & 0.047   & 0.998   &  & 0.203          & 0.010   & 0.047 & 0.758   \\
GMV-NW-GIC-SF & \textbf{0.213} & 0.008   & 0.035   &         &  & 0.245          & 0.008   & 0.034   &         &  & 0.195          & 0.007  & 0.035   &         &  & \textbf{0.225} & 0.008   & 0.034 &         \\
GMV-NW-CV-SF  & 0.212          & 0.008   & 0.036   & 0.940   &  & 0.249          & 0.008   & 0.034   & 0.374   &  & 0.191          & 0.007  & 0.036   & 0.454   &  & 0.206          & 0.007   & 0.033 & 0.006   \\
GMV-NW-GIC-3F    & 0.193          & 0.007   & 0.034   & 0.424   &  & 0.224          & 0.007   & 0.031   & 0.498   &  & 0.171          & 0.006  & 0.034   & 0.382   &  & 0.192          & 0.006   & 0.032 & 0.260   \\
GMV-NW-CV-3F     & 0.188          & 0.006   & 0.034   & 0.348   &  & 0.231          & 0.007   & 0.031   & 0.700   &  & 0.161          & 0.006  & 0.034   & 0.196   &  & 0.139          & 0.004   & 0.031 & 0.016   \\
GMV-POET      & 0.185          & 0.006   & 0.033   & 0.282   &  & 0.222          & 0.007   & 0.032   & 0.416   &  & 0.169          & 0.006  & 0.033   & 0.316   &  & 0.204          & 0.007   & 0.032 & 0.430   \\
GMV-NL-LW     & 0.160          & 0.006   & 0.035   & 0.172   &  & 0.232          & 0.007   & 0.029   & 0.838   &  & 0.131          & 0.005  & 0.035   & 0.120   &  & 0.175          & 0.005   & 0.029 & 0.398   \\
GMV-SF-NL-LW  & 0.172          & 0.006   & 0.034   & 0.252   &  & 0.242          & 0.007   & 0.028   & 0.934   &  & 0.145          & 0.005  & 0.034   & 0.196   &  & 0.184          & 0.005   & 0.028 & 0.398   \\
MW-NW-GIC-SF  & 0.211          & 0.007   & 0.034   & 0.872   &  & 0.243          & 0.008   & 0.032   & 0.868   &  & 0.189          & 0.006  & 0.034   & 0.644   &  & 0.219          & 0.007   & 0.032 & 0.602   \\
MW-NW-CV-SF   & 0.210          & 0.007   & 0.034   & 0.834   &  & \textbf{0.249} & 0.008   & 0.032   & 0.656   &  & 0.185          & 0.006  & 0.034   & 0.504   &  & 0.202          & 0.006   & 0.032 & 0.028   \\
MW-NW-GIC-3F     & 0.191          & 0.006   & 0.034   & 0.442   &  & 0.226          & 0.007   & 0.031   & 0.584   &  & 0.165          & 0.006  & 0.034   & 0.338   &  & 0.190          & 0.006   & 0.031 & 0.326   \\
MW-NW-CV-3F      & 0.184          & 0.006   & 0.034   & 0.324   &  & 0.228          & 0.007   & 0.031   & 0.652   &  & 0.153          & 0.005  & 0.034   & 0.162   &  & 0.132          & 0.004   & 0.031 & 0.038   \\
MW-POET       & 0.181          & 0.006   & 0.032   & 0.282   &  & 0.216          & 0.007   & 0.031   & 0.408   &  & 0.161          & 0.005  & 0.033   & 0.240   &  & 0.195          & 0.006   & 0.031 & 0.402   \\
MW-NL-LW      & 0.151          & 0.005   & 0.036   & 0.172   &  & 0.229          & 0.007   & 0.029   & 0.782   &  & 0.120          & 0.004  & 0.036   & 0.092   &  & 0.172          & 0.005   & 0.029 & 0.352   \\
MW-SF-NL-LW   & 0.161          & 0.006   & 0.035   & 0.248   &  & 0.237          & 0.007   & 0.028   & 0.886   &  & 0.131          & 0.005  & 0.035   & 0.152   &  & 0.178          & 0.005   & 0.028 & 0.398   \\
MAXSER        & 0.040          & 0.004   & 0.088   & 0.294   &  &                &         &         &         &  & -0.039         & -0.004 & 0.099   & 0.364   &  &                &         &       &         \\ \hline
\end{tabular}%
\begin{tablenotes}
\item The table shows the Sharpe Ratio (SR), average returns (Avg), standard deviation (SD) and p-value of the \citet{jobson1981performance} test with \citet{memmel2003performance} correction. We also applied the
\citet{ledoit2008robust} test with circular bootstrap, and the results were very similar; therefore we only report those of the first test in this table. The statistics were calculated from 240 rolling windows covering the period from Jan. 2005 to Dec. 2019, and the size of the estimation window was 120 observations.
\end{tablenotes}
\end{threeparttable}

\end{adjustbox}
\end{table}


In Table \ref{tab:stat}, we analyze turnover, leverage and maximum leverage (equations (\ref{eq:tn}), (\ref{eq:l}) and (\ref{eq:maxl}), respectively) of the portfolios in Tables \ref{tab:emp_sp1}-\ref{tab:emp_sp2}.

The definitions are as follows for turnover:
\begin{equation}
\label{eq:tn}
  \mbox{turnover} = \sum_{j=1}^p|{\widehat{w}_{t+1,j}}-{\widehat{w}^+_{t,j}}|,
\end{equation}
\noindent and leverage
\begin{equation}
\label{eq:l}
 \mbox{leverage} =  \left|\sum_{j = 1}^p \min\{\widehat{w}_{t+1,j},0\}\right|,
\end{equation}
\noindent and maximum leverage
\begin{equation}
\label{eq:maxl}
 \mbox{max leverage} =  \max_j\{\left| \min\{\widehat{w}_{t+1,j},0\} \right|\}.
\end{equation}

It is clear that in Table \ref{tab:stat}  in terms of turnover, leverage, maximum leverage, GMV-POET and GMV-NW-GIC-SF do well, with the best and close to best respectively if we discount EW portfolios.

\subsection{Time Series of Sharpe Ratios and Turnover}

\begin{figure}
     \centering
     \includegraphics[width=0.8\textwidth]{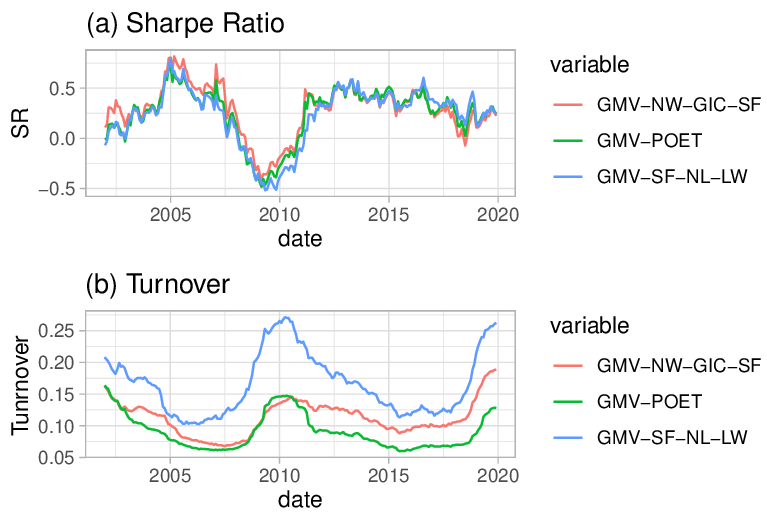}
\caption{24 months rolling Sharpe Ratio and turnover - Low Dimension with transaction costs}
     \label{fig:TS_LD}
\end{figure}

\begin{figure}[ht]
     \centering
     \includegraphics[width=0.8\textwidth]{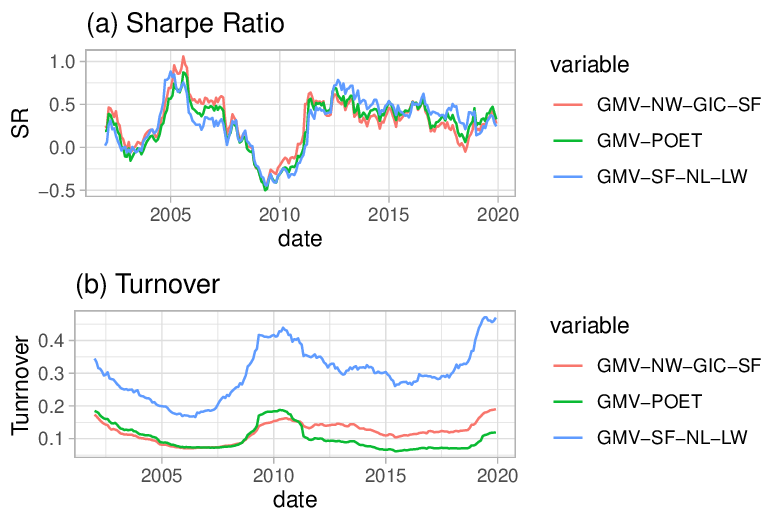}
\caption{24 months rolling Sharpe Ratio and turnover - High Dimension with transaction costs}
     \label{fig:TS_HD}
\end{figure}

Figures \ref{fig:TS_LD} and \ref{fig:TS_HD} shows Global Minimum Variance results of the NW-GIF-SF, the POET and the SF-NL-LW models with transaction costs. The results were obtained through a 24 months rolling window with the out-of-sample returns from the 2000-2019 experiment, which yields time-series that start in 2002 and end in 2019 for the Sharpe Ratio and the turnover. The main conclusion from the figures is that Nodewise works better in terms of the Sharpe Ratio in deep recessions like the 2008 crisis, but Nonlinear Shrinkage and POET are \textnormal{sup}erior when we have long periods of normality in the markets. Nodewise also delivers better Sharpe Ratios during the recovery of the crisis. On the turnover side, Nodewise and POET consistently have lower turnover than Nonlinear Shrinkage with POET being the overall lowest. However, during the 2008 crisis, especially in the high dimension setup, POET had a higher turnover than Nodewise. 

\newpage 

\begin{table}[htb]
\centering
\caption{Turnover and Leverage}
\label{tab:stat}
\begin{adjustbox}{max width=0.65\textwidth}
\begin{threeparttable}
\begin{tabular}{lccccccc}
\hline
             & \multicolumn{7}{c}{{\ul \textbf{2005-2019 Subsample}}}                                \\
             & \multicolumn{3}{c}{{\ul Low Dimension}} &  & \multicolumn{3}{c}{{\ul High Dimension}} \\
             & Turnover   & Leverage   & Max Leverage  &  & Turnover   & Leverage   & Max Leverage   \\ \cline{2-4} \cline{6-8}
EW            & 0.053    & 0.000    & 0.000        &  & 0.054    & 0.000    & 0.000        \\
GMV-NW-GIC-SF & 0.125    & 0.312    & 0.042        &  & 0.130    & 0.376    & 0.009        \\
GMV-NW-CV-SF  & 0.160    & 0.311    & 0.040        &  & 0.302    & 0.395    & 0.014        \\
GMV-NW-GIC-3F & 0.148    & 0.380    & 0.048        &  & 0.186    & 0.528    & 0.013        \\
GMV-NW-CV-3F  & 0.190    & 0.382    & 0.049        &  & 0.593    & 0.567    & 0.030        \\
GMV-POET      & 0.096    & 0.288    & 0.043        &  & 0.096    & 0.299    & 0.007        \\
GMV-NL-LW     & 0.198    & 0.420    & 0.057        &  & 0.325    & 0.807    & 0.024        \\
GMV-SF-NL-LW  & 0.163    & 0.383    & 0.050        &  & 0.341    & 0.904    & 0.025        \\
MW-NW-GIC-SF  & 0.154    & 0.331    & 0.046        &  & 0.150    & 0.382    & 0.009        \\
MW-NW-CV-SF   & 0.191    & 0.329    & 0.044        &  & 0.322    & 0.402    & 0.014        \\
MW-NW-GIC-3F  & 0.179    & 0.401    & 0.052        &  & 0.207    & 0.539    & 0.013        \\
MW-NW-CV-3F   & 0.220    & 0.401    & 0.051        &  & 0.626    & 0.582    & 0.030        \\
MW-POET       & 0.128    & 0.306    & 0.046        &  & 0.117    & 0.307    & 0.008        \\
MW-NL-LW      & 0.220    & 0.440    & 0.064        &  & 0.327    & 0.814    & 0.024        \\
MW-SF-NL-LW   & 0.184    & 0.400    & 0.052        &  & 0.344    & 0.912    & 0.025        \\
MAXSER        & 1.766    & 0.421    & 0.200        &  &          &          &              \\
              &          &          &              &  &          &          &              \\
              & \multicolumn{7}{c}{{\ul \textbf{2000-2019 Sub Sample}}}                    \\ \cline{2-4} \cline{6-8} 
EW            & 0.056    & 0.000    & 0.000        &  & 0.056    & 0.000    & 0.000        \\
GMV-NW-GIC-SF & 0.120    & 0.283    & 0.049        &  & 0.127    & 0.342    & 0.011        \\
GMV-NW-CV-SF  & 0.142    & 0.279    & 0.048        &  & 0.278    & 0.361    & 0.014        \\
GMV-NW-GIC-3F & 0.144    & 0.355    & 0.053        &  & 0.192    & 0.541    & 0.016        \\
GMV-NW-CV-3F  & 0.181    & 0.353    & 0.053        &  & 0.557    & 0.572    & 0.030        \\
GMV-POET      & 0.097    & 0.290    & 0.038        &  & 0.107    & 0.322    & 0.009        \\
GMV-NL-LW     & 0.196    & 0.396    & 0.068        &  & 0.311    & 0.782    & 0.027        \\
GMV-SF-NL-LW  & 0.173    & 0.383    & 0.062        &  & 0.310    & 0.849    & 0.026        \\
MW-NW-GIC-SF  & 0.142    & 0.296    & 0.050        &  & 0.148    & 0.351    & 0.011        \\
MW-NW-CV-SF   & 0.165    & 0.292    & 0.048        &  & 0.299    & 0.369    & 0.014        \\
MW-NW-GIC-3F  & 0.165    & 0.368    & 0.054        &  & 0.209    & 0.548    & 0.016        \\
MW-NW-CV-3F   & 0.203    & 0.364    & 0.054        &  & 0.582    & 0.581    & 0.030        \\
MW-POET       & 0.121    & 0.301    & 0.041        &  & 0.126    & 0.333    & 0.009        \\
MW-NL-LW      & 0.214    & 0.409    & 0.071        &  & 0.313    & 0.787    & 0.027        \\
MW-SF-NL-LW   & 0.197    & 0.395    & 0.067        &  & 0.314    & 0.855    & 0.025        \\
MAXSER        & 1.860    & 0.371    & 0.201        &  &          &          &              \\ \hline
\end{tabular}
\begin{tablenotes}
\item The table shows the average turnover, average leverage and average max leverage for all portfolios across all out-of-sample windows. The top panel shows the results for the 2000-2019 out-of-sample period, and the second panel shows the results for the 2005-2019 out-of-sample period.
\end{tablenotes}
\end{threeparttable}
\end{adjustbox}
\end{table}

\newpage

\section{Conclusion}
We provide a hybrid factor model combined with nodewise regression method that can control for risk and obtain the maximum expected return of a large portfolio. Our result is novel and holds even when $p>n$. We allow for an increasing number of factors, with possible unbounded largest eigenvalue of the covariance matrix of errors. Sparsity is assumed on the precision matrix of errors rather than the covariance matrix of errors. We also show that the maximum out-of-sample Sharpe Ratio can be estimated consistently. Furthermore, we also develop a formula for the maximum Sharpe Ratio when the sum of the weights of the portfolio  is one. A consistent estimate for the constrained case is also shown. Then, we extended our results to the consistent estimation of the Sharpe Ratios in two widely used portfolios in the literature. It will be essential to extend our results to more restrictions on portfolios.

\bibliographystyle{chicagoa}
\bibliography{maxsrsep14-mc}

\newpage

\appendix

\setcounter{page}{1}

\setcounter{equation}{0}\setcounter{lemma}{0}\setcounter{assum}{0}\setcounter{thm}{0}\renewcommand{\theequation}{A.%
\arabic{equation}}\renewcommand{\thelemma}{A.\arabic{lemma}}%
\renewcommand{\theassum}{A.\arabic{assum}}%
\renewcommand{\thethm}{A.\arabic{thm}}%
\renewcommand{\baselinestretch}{1}\baselineskip=15pt

\begin{center}
\begin{huge}
Supplementary Material to

\vspace{0.4cm}
Sharpe Ratio  Analysis in High Dimensions:

\vspace{0.4cm}
Residual-Based Nodewise Regression in Factor Models
\end{huge}

\vspace{0.8cm}
\begin{Large}
Mehmet Caner \qquad Marcelo C. Medeiros \qquad Gabriel F. R. Vasconcelos
\end{Large}
\end{center}

\section*{Supplement A}

Supplement A is divided into several parts. The first part has preliminary proofs, norm inequalities, definitions, and a maximal inequality that is extended in a very minor form from the existing literature. The second part has the proofs of lemmata that lead to proof of Theorem \ref{rthm1}. The first two parts relate only to the proof of Theorem \ref{rthm1}. The third part is only related to the proof of Theorem \ref{rthm2}. Part 4 is related to all the remaining proofs of the theorems in this paper. 

\subsection*{Part 1}

We start with a lemma that provides norm inequalities. Let $\b A_1: p \times K$, $\b B_1: K \times K$ matrices and $x: K \times 1$ vector.

\begin{lemma}\label{la1}
(i). \[ \| \b A_1 \b B_1 \b x \|_{\infty} \le K^2 \| \b A_1 \|_{\infty} \| \b B_1 \|_{\infty} \| \b x \|_{\infty}.\]
(ii). \[ \| \b A_1 \b B_1 \b A_1' \|_{\infty} \le K^2 \| \b A_1 \|_{\infty}^2 \| \b B_1 \|_{\infty}.\]
\end{lemma}

\noindent
{\bf Proof of Lemma \ref{la1}}. (i). Set $\b B_1 \b x = \b x_1$, and let $\b a_j'$ be the $1 \times K$ row vector of $\b A_1$
\begin{eqnarray}
\| \b A_1 \b x_1 \|_{\infty} &=& \max_{1 \le j \le p} | \b a_j' \b x_1 | \le \left[ \max_{1 \le j \le p} \|\b a_j \|_1
\right] \| \b x_1 \|_{\infty} \label{pla10} \\
& \le & \left[ K \| \b A_1 \|_{\infty}
\right] \| \b x_1 \|_{\infty} 
 =  \left[ K \| \b A_1 \|_{\infty} 
\right] \| \b B_1 x \|_{\infty} \label{pla11} \\
& \le & K^2 \| \b A_1 \|_{\infty} \| \b B_1 \|_{\infty} \| x\|_{\infty},
\end{eqnarray}
where we use H{\"o}lder's inequality for the first inequality, and the relation between $l_1, l_{\infty}$ norms for the second inequality, and to get the last inequality we repeat the first two  inequalities.

(ii). Use Section 4.3 of \cite{vdg2016}
\begin{equation}
\| \b A_1 \b B_1 \b A_1' \|_{\infty} \le \| \b A_1 \|_{\infty} [ \| \b B_1 \b A_1' \|_{l_1}],\label{pla12}
\end{equation}
where $\|.\|_{l_1}$ is the maximum absolute column sum norm of $\b B_1\b A_1'$ matrix (i.e. $l_1$ induced matrix norm). Let $\b b_l$' be $1 \times K$ row vector of $\b B_1$, and $\b a_j$ is the $j$th column of $\b A_1'$ matrix.
\begin{eqnarray}
\| \b B_1 \b A_1' \|_{l_1} &= &\max_{1 \le j \le p} \sum_{l=1}^K \left|\b b_l'\b a_j\right| 
 \le \left[\max_{1 \le j \le p} \left\|\b a_j \right\|_{\infty}\right] \left[ \sum_{l=1}^K \left\|\b b_l \right\|_1\right] \nonumber \\
& \le & \left\| \b A_1 \right\|_{\infty} \left[ K \max_{1 \le l \le K} \left\|\b b_l \right\|_1\right] 
 \le  \left\| \b A_1 \right\|_{\infty} \left[ K^2 \left\| \b B_1 \right\|_{\infty} \right]\label{pla13}
\end{eqnarray}
where we use H{\"o}lder's inequality for the first inequality, and $l_1, l_{\infty}$ norm relation for the other inequalities. Next use (\ref{pla13}) in (\ref{pla12}) to get
\[ \| \b A_1 \b B_1 \b A_1' \|_{\infty} \le K^2 \| \b A_1 \|_{\infty}^2 \| \b B_1 \|_{\infty}.\]
\begin{flushright}
{\bf Q.E.D.}
\end{flushright}

Next we provide a lemma that is directly from Lemma A.2 of \cite{fan2011}.

\begin{lemma}\label{la2}(\cite{fan2011}).
Suppose that two random variables $Z_1$, $Z_2$ satisfy the following exponential type tail condition. There exist $r_{z_1}, r_{z_2} \in (0,1)$ and $b_{z_1}, b_{z_2} > 0$ constant such that for all $ s >0$
\[ 
\P [ | Z_l | > s] \le \exp\left[1 - (s/b_{z_l})^{r_{z_l}}\right], \quad l=1,2,\]
Then, for some $r_{z_3} >0$, and $b_{z_3} >0$
\[ 
\P \left[| Z_1 Z_2 | > s \right] \le \exp\left[1 - (s/b_{z_3})^{r_{z_3}}\right].
\]
\end{lemma}
We provide now the following maximal inequality due to Theorem 1 of \cite{merl2011}, and used in the proof of Lemma A.3(i) and proof of Lemma B.1(ii) in \cite{fan2011}.
To that effect, we provide a general assumption on data, and then show the theorem and its proof.

\begin{assumL}
{\it (i). $\b X_t, \b Y_t$ are vectors of dimension $d_x$ and $d_y$, respectively, for $t=1,\cdots, n$.  They are both stationary and ergodic. Also $\{\b X_t,\b Y_t\}$ are strong mixing with strong mixing coefficients are satisfying
\[ \alpha(t) \le \exp(-C t^{r_{xy}}),\]
with $t$, a positive integer, and $r_{xy} > 0$ a positive constant. (ii). We also let $\b X_t, \b Y_t$ satisfy the exponential tail condition for $j_1=1, \cdots, d_x, j_2=1,\cdots, d_y$
\[ 
\P [ | X_{j_1,t} | > s ] \le \exp\left[-(s/b_x)^{r_x}\right],
\]
for positive constants $b_x, r_x >0$, and
\[ \P [ |Y_{j_2,t} | > s] \le \exp\left[-(s/b_y)^{r_y}\right],\]
with positive constants $b_y, r_y >0$. Also, we need to assume $3 r_x^{-1} + r_{xy}^{-1} > 1, 3 r_y^{-1} + r_{xy}^{-1} > 1$.}
\end{assumL}

\begin{thm}\label{thma1}
Under Assumption L1, (\cite{fan2011}).
\begin{eqnarray*}
\P\left[\max_{1 \le j_1 \le d_x} \max_{1 \le j_2 \le d_y} \frac{ \left|\sum_{t=1}^n\left(X_{j_1,t} Y_{j_2,t} - \E[X_{j_1,t} Y_{j_2,t}]\right)\right|}{n} >s
 \right] & \le & d_x d_y \{ n \exp \left[ \frac{- (n s)^{\gamma}}{C_1}
 \right]\\
 & +& \exp \left[\frac{-  (n^2 s^2)}{C_2 ( 1 + n C_3)}
 \right] \\
 & + & \exp \left[\frac{-(n s)^2}{C_4 n } \exp \left( \frac{(ns)^{\gamma(1- \gamma)}}{C_5 [\ln(ns)]^{\gamma}}
 \right)
 \right]\}
\end{eqnarray*}
with $0 < \gamma < 1$, and $\gamma$ is defined as $\gamma^{-1}:= 1.5 r_x^{-1} + 1.5 r_y^{-1} + r_{xy}^{-1}$.
\end{thm}

\noindent
{\bf Proof of Theorem \ref{thma1}}. This is a simple application of Lemma A.2 above with Assumption L1 for Theorem 1 of \cite{merl2011}, and Bonferroni union bound.
\begin{flushright}
{\bf Q.E.D.}
\end{flushright}

\subsection*{Part 2}

We start with an important maximal inequality applied to factor models in nodewise regression setting. Some of the results are already in Lemma A.3, Lemma B.1 of \cite{fan2011}. We show them so that readers can see all results without referral to other literature. We also provide two new results Lemma \ref{la3}(ii), (v) due to nodewise regression interaction with factor models.

\begin{lemma}\label{la3}
Under Assumptions \ref{as1}-\ref{as3}, for $C > C_m>0$, with $m=1,2,3,4,5$ with $C_m$ that is used in Theorem \ref{thma1}.

(i). \[ \P \left[ \max_{1 \le j \le p} \max_{1 \le l \le p} \left| \frac{1}{n} \sum_{t=1}^n u_{l,t} u_{j,t} - \E[u_{l,t} u_{j,t}]\right| > C \sqrt{\ln(p)/n}
\right] = O\left(\frac{1}{p^2}\right).\]

(ii). Denote $\b U_{-j}$ as the  $(p-1)\times n$ matrix in (\ref{-yj}), and let the $l$ th row and $t$ th column element $U_{-j,l,t}$ and $\b\eta_j$ as $n \times 1 $ vector, and the $t$ th element as $\eta_{j,t}$
\[
\P\left[ \max_{1 \le j \le p} \max_{1 \le l \le p}\left| \frac{1}{n} \sum_{t=1}^n U_{-j,l,t} \eta_{j,t}\right| > C \sqrt{\ln(p)/n}
\right] = O\left(\frac{1}{p^2}\right).
\]

(iii). \[\P \left[\max_{1 \le k \le K} \max_{1 \le j \le p}\left| \frac{1}{n} \sum_{t=1}^n f_{k,t} u_{j,t}\right| > C \sqrt{\ln(p)/n}
\right] = O\left(\frac{1}{p^2}\right).\]

(iv). Let $f_{k_1,t}, f_{k_2,t}$ represent $k_1, k_2$ factors (elements) of the vector $f_t$
\[
\P \left[ \max_{1 \le k_1 \le K} \max_{1 \le k_2 \le K}\left| \frac{1}{n} \sum_{t=1}^n f_{k_1,t} f_{k_2,t} - \E[f_{k_1,t} f_{k_2,t}]\right| > C \sqrt{\ln(n)/n}
\right] = O\left(\frac{1}{n^2}\right).
\]

(v).
\[
\P \left[ \max_{1 \le k \le K} \max_{1 \le j \le p}\left| \frac{1}{n} \sum_{t=1}^n f_{k,t} \eta_{j,t}\right| > C \sqrt{ \bar{s}\ln(p)/n}
\right] = O\left(\frac{1}{p^2}\right).
\]
\end{lemma}

\noindent
{\bf Proof of Lemma \ref{la3}}.
(i). This is Lemma \ref{la3}(i) of \cite{fan2011}.

(ii). The proof follows from Theorem \ref{thma1} and Assumption \ref{as3} provides the tail probability through the same algebra as in p.3346 of \cite{fan2011}.

(iii). This is Lemma B1(ii) of \cite{fan2011}.

(iv). This is Lemma B1(i) of \cite{fan2011}.

(v). The proof will involve several steps and this is due to interaction of factor models ($f_{k,t}$) and nodewise error ($\eta_{j,t}$). Start with the definition
of
\begin{eqnarray}
\eta_{j,t} & = & u_{j,t}- \b u_{-j,t}'\b\gamma_j = [u_{j,t}, \b u_{-j,t}']
\begin{bmatrix}
1 \\
-\b\gamma_j
\end{bmatrix}\\
& = & \b u_t' \b C_j,\label{pla31}
\end{eqnarray}
where $\b C_j:= \begin{bmatrix}
1 \\
-\b\gamma_j \end{bmatrix}$ is a $p \times 1$ vector, and $\b u_t':=[u_{j,t}, \b u_{-j,t}']$, which is $1 \times p$ row vector.
Next,
\begin{eqnarray}
\max_{1\le j \le p} \max_{1 \le k \le K} \left| \frac{1}{n} \sum_{t=1}^n f_{k,t} \eta_{j,t} \right| & = &
\max_{1 \le j \le p} \max_{1 \le k \le K} \left| \frac{1}{n} \sum_{t=1}^n f_{k,t}\b u_t' \b C_j\right| \nonumber \\
& \le & \max_{1 \le k \le K} \left\| \frac{1}{n} \sum_{t=1}^n f_{k,t} \b u_t'\right\|_{\infty} \max_{1 \le j \le p} \left\| \b C_j \right\|_1 \nonumber \\
& = & \max_{1 \le j \le p} \max_{1 \le k \le K} \left| \frac{1}{n} \sum_{t=1}^n f_{k,t} \b u_{j,t} \right| \max_{1 \le j \le p} \left\| \b C_j \right\|_1 \label{pla32}
\end{eqnarray}
where we use (\ref{pla31}) for the first equality and H{\"o}lder's inequality for the first inequality. Consider
\begin{equation}
\max_{1 \le j \le p} \| \b C_j \|_1 \le 1 + \max_{1 \le j \le p} \| \b \gamma_j \|_1,\label{pla33}
\end{equation}
where we use $\b C_j$ definition. Noting that $\b\Sigma_{n,-j,-j}$ is $p-1 \times p-1$ submatrix of $\b\Sigma_n$ consisting all rows and columns of $\b\Sigma_n$ except the $j$th one.
See that
\[ 
\frac{\b\gamma_j'\b\Sigma_{n,-j,-j}\b\gamma_j}{\b\gamma_j' \b\gamma_j} \ge \textnormal{Eigmin} (\b\Sigma_{n,-j,-j}) \ge \textnormal{Eigmin} (\b\Sigma_n) \ge c > 0.\]
 Then,
\begin{equation}
\|\b\gamma_j \|_2^2 =\b\gamma_j'\b\gamma_j \le \frac{\b\gamma_j' \b\Sigma_{n,-j,-j}\b\gamma_j}{\textnormal{Eigmin}(\b\Sigma_n)} \le C < \infty,\label{pla34}
\end{equation}
where we use the last inequality above for the first inequality in (\ref{pla34}) and (B.48) of \cite{canerkock2018} for the second inequality in (\ref{pla34}), given our Assumption
$\max_{1 \le j \le p} \E[u_{j,t}^2] \le C < \infty$. Hence, 
\begin{equation}
\max_{1 \le j \le p} \|\b\gamma_j \|_1 \le \sqrt{\bar{s}} \max_{1 \le j \le p } \|\b\gamma_j \|_2 \le \sqrt{\bar{s}} C = O ( \sqrt{\bar{s}}), \label{pla35}
\end{equation}
by (\ref{pla34}). Clearly, by (\ref{pla33})-(\ref{pla35})
\[\max_{1 \le j \le p} \| \b C_j \|_1 = O (\sqrt{\bar{s}}).\]
Then since $\b \Omega_j:= \frac{\b C_j}{\tau_j^2}$ and by (\ref{pt11}) 
\begin{equation}
\max_{1 \le j \le p} \| \b\Omega_j \|_1 = O (\sqrt{\bar{s}}).\label{pla36}
\end{equation}
Next, use Lemma \ref{la3}(iii) and (\ref{pla36}) in (\ref{pla32}) to show
\[
\P \left[ \max_{1 \le j \le p} \max_{1 \le k \le K} \left| \frac{1}{n} \sum_{t=1}^n f_{k,t} \eta_{j,t} \right| \ge C \frac{\sqrt{\bar{s} \ln(p)}}{\sqrt{n}}\right] = O(1/p^2).
\]

This also implies that, since $\b X:=(\b f_1, \cdots, \b f_n):  K \times n$ matrix, and $\b \eta_j:=(\eta_{j,1}, \cdots, \eta_{j,n})': n \times 1$
\begin{equation}
\max_{1 \le j \le p} \| \b X \b\eta_j/n\|_{\infty} = O_p \left( \frac{\sqrt{\bar{s} \ln(p)}}{\sqrt{n}}\right).\label{pla37}
\end{equation}

\begin{flushright}
{\bf Q.E.D.}
\end{flushright}

Now we start defining two events, and we condition the next lemma, which is $l_1$ bound on nodewise regression estimates, on these two events. Then we relax this restriction, and show that an unconditional result for $l_1$ norm of the nodewise regression estimates after finding that these two events converge in probability to one.
Define
\begin{equation}
 {\cal A}_1:= \left\{ 2 \max_{1 \le j \le p} \|\b\eta_{xj}' \widehat{\b U}_{-j}'/n\|_{\infty} \le \lambda_n\right\},\label{eve1}
 \end{equation}
and define the population adaptive restricted eigenvalue condition, as in \cite{canerkock2018}, for $j=1,\cdots, p$, and let $\b\delta_{S_j}$ represent the vector
with $S_j$ indices in $\b\delta_j$, and all the other elements  than $S_j$ indices in $\b\delta_j$ set to zero
\begin{equation}
\phi^2 (s_j):= \min_{\b\delta_j \in \R^{p-1}} \left\{ \frac{\b\delta_j' \b\Sigma_{n,-j,-j}\b\delta_j}{\|\b\delta_{S_j} \|_2^2}:\b\delta \in \R^p - \{ 0 \}, \|\b\delta_{S_j^c} \|_1 \le 3
\sqrt{s_j} \| \b \delta_{S_j} \|_2
\right\},\label{ev1}
\end{equation}
and the empirical version of the adaptive restricted eigenvalue condition is as follows, with $\widehat{\b U}_{-j}: p-1 \times n $ matrix
\begin{equation}
\widehat{\phi}^2 (s_j):= \min_{\b\delta_j \in \R^{p-1}} \left\{ \frac{\b\delta_j' (\widehat{\b U}_{-j} \widehat{\b U}_{-j}'/n) \b\delta_j}{\|\b\delta_{S_j} \|_2^2}: \b\delta \in \R^p - \{ 0 \}, \|\b\delta_{S_j^c} \|_1 \le 3
\sqrt{s_j} \|\b\delta_{S_j} \|_2
\right\},\label{ev2}
\end{equation}
and the event is for each $j=1,\cdots, p$
\[ {\cal A}_{2j}:= \{ \widehat{\phi}^2(s_j) \ge \phi^2(s_j)/2\}.\]

We have the following $l_1$ bound result.

\begin{lemma}\label{la4}
Under $\b A_1 \cap A_{2j}$, with $\lambda_n >0$ in (\ref{eve1})
\[ \max_{1 \le j \le p} \| \widehat{\b\gamma}_j - \b\gamma_j \|_1 \le \frac{24 \lambda_n \bar{s}}{\phi^2(\bar{s})} = O_p ( \lambda_n \bar{s}),\]
\end{lemma}
We specify the formula and the rate for $\lambda_n$ in the next Lemma.

\noindent
{\bf Proof of Lemma \ref{la4}}. Start with $\widehat{\b\gamma}_j$ definition in (\ref{rv15})

\begin{equation}
\| \widehat{\b u}_j - \widehat{\b U}_{-j}' \widehat{\b \gamma}_j \|_n^2 + 2 \lambda_n \sum_{j=1}^{p-1} | \widehat{\b \gamma}_j |\le
\| \widehat{\b u}_j - \widehat{\b U}_{-j}'\b\gamma_j \|_n^2 + 2 \lambda_n \sum_{j=1}^{p-1} |\b\gamma_j|.\label{pla41}
\end{equation}
Use (\ref{rv14}) to have $\widehat{\b u}_j - \widehat{\b U}_{-j}' \widehat{\b\gamma}_j = \b\eta_{xj} - \widehat{\b U}_{-j}' ( \widehat{\b\gamma}_j - \b\gamma_j)$ and this last equation can be substituted into first left side term and first right side term in (\ref{pla41}) to have
\begin{equation}
\|\b\eta_{xj} - \widehat{\b U}_{-j}' ( \widehat{\b\gamma}_j - \b\gamma_j) \|_n^2 + 2 \lambda_n \sum_{j=1}^{p-1} | \widehat{\b\gamma}_j | \le
\|\b\eta_{xj} \|_n^2 + 2 \lambda_n \sum_{j=1}^{p-1} |\b\gamma_j |.\label{pla42}
\end{equation}
\noindent Simplify  the first term on the left and the first term on the right side of (\ref{pla42}),
\begin{equation}
\|\widehat{\b U}_{-j}' (\widehat{\b\gamma}_j - \b\gamma_j) \|_n^2 + 2 \lambda_n \sum_{j=1}^{p-1} | \widehat{\b\gamma}_j | \le
2 \left| \frac{\b\eta_{xj}' \widehat{\b U}_{-j}'}{n} (\widehat{\b \gamma}_j - \b \gamma_j ) \right| + 2 \lambda_n \sum_{j=1}^{p-1} | \b \gamma_j |.\label{pla43}
\end{equation}
Since we use ${\cal A}_1$ and then H{\"o}lder's inequality
\begin{equation}
\| \widehat{\b U}_{-j}' (\widehat{\b\gamma}_j - \b\gamma_j )\|_n^2 + 2 \lambda_n \sum_{j=1}^{p-1} | \widehat{\b\gamma}_j | \le \lambda_n
\| \widehat{\b\gamma}_j - \b\gamma_j \|_1 + 2 \lambda_n \sum_{j=1}^{p-1} |\b\gamma_j |.\label{pla44}
\end{equation}
Use $\|\widehat{\b\gamma}_j \|_1 = \|\widehat{\b\gamma}_{S_j} \|_1 + \|\widehat{\b\gamma}_{S_j^c}\|_1$, on the second term on the left side of (\ref{pla44})
($S_j$ represents the indices of nonzero cells in row $j$ of the precision matrix, and $S_j^c$ represents the indices of zero cells in row $j$ of the precision matrix).
\begin{equation}
\| \widehat{\b U}_{-j}' (\widehat{\b\gamma}_j - \b\gamma_j) \|_n^2 + 2 \lambda_n \sum_{j \in S_j^c} | \widehat{\b\gamma}_j| \le
\lambda_n \| \widehat{\b\gamma}_j  - \b\gamma_j \|_1 + 2 \lambda_n \sum_{j=1}^{p-1} |\b\gamma_j | - 2 \lambda_n \sum_{j \in S_j} | \widehat{\b\gamma}_j |.\label{pla45}
\end{equation}

Now use $\sum_{ j \in S_j^c} |\b\gamma_j | = 0$ in the second term on the right side of (\ref{pla45}) and use reverse triangle inequality:
\begin{equation}
\|\widehat{\b U}_{-j}' (\widehat{\b \gamma}_j - \b \gamma_j) \|_n^2 + 2 \lambda_n \sum_{j \in S_j^c} | \widehat{\b \gamma}_j| \le
\lambda_n \| \widehat{\b\gamma}_j  - \b\gamma_j \|_1 + 2 \lambda_n \sum_{j \in S_j} |\widehat{\b\gamma}_j - \b\gamma_j |.\label{pla46}
\end{equation}
Next by $ \|\widehat{\b\gamma}_j - \b\gamma_j \|_1 = \| \widehat{\b\gamma}_{S_j} - \b\gamma_{S_j} \|_1 + \| \widehat{\b\gamma}_{S_j^c} \|_1$ for the first term on the right side of (\ref{pla46})
\begin{equation}
\| \widehat{\b U}_{-j}' (\widehat{\b \gamma}_j - \b \gamma_j ) \|_n^2 + \lambda_n  \sum_{j \in S_j^c} | \widehat{\b \gamma}_j | \le
3 \lambda_n \sum_{j \in S_j} | \widehat{\b \gamma}_j - \b \gamma_j|.\label{pla47}
\end{equation}
Use the norm inequality $\| \widehat{\b\gamma}_{S_j} - \b\gamma_{S_j} \|_1 \le \sqrt{s_j} \| \widehat{\b\gamma}_{S_j} - \b\gamma_{S_j} \|_2$
\begin{equation}
\|\widehat{\b U}_{-j}' (\widehat{\b\gamma}_j - \b\gamma_j ) \|_n^2 + \lambda_n \sum_{j \in S_j^c} | \widehat{\b\gamma}_j| \le 3 \lambda_n \sqrt{s_j} \| \widehat{\b\gamma}_{S_j} - \b\gamma_{S_j} \|_2.\label{pla48}
\end{equation}
Now ignoring the first term above and dividing the rest by $\lambda_n >0$, provides the restricted set condition (cone condition) in adaptive restricted eigenvalue condition
\begin{equation}
\| \widehat{\b\gamma}_{S_j^c} \|_1  \le 3 \sqrt{s_j} \| \widehat{\b\gamma}_{S_j} - \b\gamma_{S_j} \|_2.\label{pla48a}
\end{equation}
Set $\b\delta_j= \widehat{\b\gamma}_j - \b\gamma_j$ in the empirical adaptive restricted set condition in (\ref{ev2}), then use the empirical adaptive restricted eigenvalue condition in (\ref{pla48})
\[ 
\|\widehat{\b U}_{-j}' (\widehat{\b \gamma}_j - \b\gamma_j) \|_n^2 + \lambda_n \sum_{j \in S_j^c} | \widehat{\b \gamma}_j |\le
3 \lambda_n \sqrt{s_j} \frac{ \| \widehat{\b U}_{-j}' (\widehat{\b \gamma}_j - \b\gamma_j) \|_n}{\widehat{\phi} (s_j)}.
\]
Then use $3ab \le a^2/2 + 9 b^2/2$ with $b = \frac{\lambda_n \sqrt{s_j}}{\widehat{\phi}(s_j)}$, $a= \| \widehat{\b U}_{-j}' (\widehat{\b\gamma}_j - \b\gamma_j)\|_n$.
\[ 
\| \widehat{\b U}_{-j}' (\widehat{\b\gamma}_j - \b\gamma_j) \|_n^2 + \lambda_n \sum_{j \in S_j^c} | \widehat{\b\gamma}_j |\le  \frac{\| \widehat{\b U}_{-j}' (\widehat{\b\gamma}_j - \b\gamma_j) \|_n^2}{2} +
\frac{9 \lambda_n^2 s_j}{2 \widehat{\phi}^2 (s_j)}.\]
Use ${\cal A}_{2j}$ in the first term on the right side and simplify
\[ \| \widehat{\b U}_{-j}' (\widehat{\b\gamma}_j - \b\gamma_j) \|_n^2 + 2 \lambda_n \sum_{j \in S_j^c} | \widehat{\b\gamma}_j| \le
\frac{18 \lambda_n^2 s_j}{\phi^2 (s_j)}.\]
This implies
\begin{equation}
 \| \widehat{\b U}_{-j}' (\widehat{\b\gamma}_j - \b\gamma_j) \|_n^2 \le
\frac{18 \lambda_n^2 s_j}{\phi^2 (s_j)}.\label{pla49}
\end{equation}
Now to get $l_1$ bound, ignore the first term in (\ref{pla48}) and add both sides $\lambda_n \| \widehat{\b\gamma}_{S_j} - \b\gamma_{S_j} \|_1$
\begin{equation}
\lambda_n \sum_{j \in S_j^c} | \widehat{\b\gamma}_j | + \lambda_n \sum_{j \in S_j} | \widehat{\b\gamma}_j - \b\gamma_j | = \lambda_n \| \widehat{\b\gamma}_j - \b\gamma_j \|_1
\le \lambda_n \| \widehat{\b\gamma}_{S_j} - \b\gamma_{S_j} \|_1 + 3 \lambda_n \sqrt{s_j} \| \widehat{\b\gamma}_{S_j} - \b\gamma_{S_j} \|_2.\label{pla410}
\end{equation}
Use the norm inequality $\| \widehat{\b\gamma}_{S_j} - \b\gamma_{S_j} \|_1 \le \sqrt{s_j} \| \widehat{\b\gamma}_{S_j} - \b\gamma_{S_j}\|_2$ for the first term on the right side of (\ref{pla410})
\[ \lambda_n \| \widehat{\b\gamma}_j - \b\gamma_j \|_1 \le 4 \lambda_n \sqrt{s_j} \| \widehat{\b\gamma}_{S_j} - \b\gamma_{S_j} \|_2.\]
This can be simplified as
\[  \| \widehat{\b\gamma}_j - \b\gamma_j \|_1 \le 4 \sqrt{s_j} \| \widehat{\b\gamma}_{S_j} - \b\gamma_{S_j} \|_2.\]
and can use the empirical adaptive restricted eigenvalue condition in (\ref{ev2})
\[ \| \widehat{\b\gamma}_j - \b\gamma_j \|_1 \le 4 \sqrt{s_j} \frac{ \| \widehat{\b U}_{-j}' (\widehat{\b\gamma}_j - \b\gamma_j)\|_n}{\widehat{\phi} (s_j)}.\]
Next, use (\ref{pla49}) and ${\cal A}_{2j}$ to have
\[ \| \widehat{\b\gamma}_j - \b\gamma_j \|_1 \le \frac{24 \lambda_n s_j}{\phi^2 (s_j)} \le \frac{24 \lambda_n \bar{s}}{\phi^2 (\bar{s})}.\]
Last inequality above is true by noticing $s_j \le \bar{s}$ by $\bar{s}$ definition, and then by definition of population adaptive restricted eigenvalue condition
$\phi^2 (s_j) \ge \phi^2 (\bar{s})$.
\begin{flushright}
{\bf Q.E.D.}
\end{flushright}

Now we evaluate two events, in the next two lemmata.
\begin{lemma}\label{la5}
Under Assumptions \ref{as1}-\ref{as4}
\[ \P ( {\cal A}_1 ) \ge 1 - O\left(\frac{1}{p^2}\right) - O\left(\frac{1}{n^2}\right),\]
and
\[ \lambda_n := C \left[ K^2 \frac{\sqrt{\bar{s}} \ln(p)}{n} + \sqrt{\frac{\ln(p)}{n}}\right],\] 
with $C>0$.
\end{lemma}

\noindent
{\bf Proof of Lemma \ref{la5}}.  Start with ${\cal A}_1$ definition in (\ref{eve1}).
Use (\ref{rv13})-(\ref{rv14a}) and $\b M_X:= \b I_n - \b X'(\b X \b X')^{-1} \b X$ is idempotent such that
\begin{equation}
\frac{\widehat{\b U}_{-j} \b\eta_{xj}}{n} = \frac{\b U_{-j} \b\eta_j}{n} - \left(\frac{\b U_{-j}\b X'}{n}
\right) \left( \frac{\b X \b X'}{n}
\right)^{-1} \left( \frac{\b X \eta_j}{n}
\right)\label{pla51}
\end{equation}
Next, we use the triangle inequality in $l_{\infty}$ norm such that
\begin{equation}
\left\| \frac{\b\eta_{xj}' \widehat{\b U}_{-j}' }{n}\right\|_{\infty} =
\left\| \frac{\widehat{\b U}_{-j} \b\eta_{xj}}{n}\right\|_{\infty} \le \left\| \frac{\b U_{-j} \b\eta_j}{n} \right\|_{\infty} + \left\| \left( \frac{\b U_{-j} \b X'}{n}
\right) \left( \frac{\b X \b X'}{n}
\right)^{-1} \left(\frac{\b X \b \eta_j}{n}
\right) \right\|_{\infty}
\label{pla52}
\end{equation}
Note that $\b U$ is a $p \times n$ matrix and $\b U_{-j}$ is the $p-1 \times n$ submatrix, which is $\b U$ without the $j$th row. As a consequence,
\begin{equation}
\max_{1 \le j \le p} \left\| \frac{\b U_{-j} \b\eta_j}{n} \right\|_{\infty} \le C \sqrt{\ln(p)/n},\label{pla53}
\end{equation}
with probability at  least $1 - O(1/p^2)$ by Lemma \ref{la3}(ii). Next, for the second right side term in (\ref{pla52}) we have that
\begin{equation}
\max_{1 \le j \le p} \left\| \left(\frac{\b U_{-j} X'}{n}
\right) \left( \frac{\b X \b X'}{n}
\right)^{-1} \left(\frac{\b X \b\eta_j}{n}
\right) \right\|_{\infty} \le K^2 \max_{1 \le j \le p} \left\| \frac{\b U_{-j} \b X'}{n} \right\|_{\infty} \left\| \left(\frac{\b X \b X'}{n} \right)^{-1}
\right\|_{\infty} \max_{1 \le j \le p}\left\| \frac{\b X \b \eta_j}{n} \right\|_{\infty}.\label{pla54}
\end{equation}
by Lemma \ref{la1}(i). We evaluate each term in (\ref{pla54}). Note that $\b X=(\b f_1, \cdots, \b f_n):K \times n$, $\b U: p \times n$
\[ 
\max_{1 \le j \le p} \left\| \frac{\b U_{-j}\b X'}{n} \right\|_{\infty} \le \left\| \b U \b X'/n \right\|_{\infty} = \max_{1 \le j \le p} \max_{1 \le k \le K} \left| \frac{ \sum_{t=1}^n u_{j,t} f_{k,t}}{n}\right|.
\]
Then, by Lemma \ref{la3}(iii),
\begin{equation}
\P\left[\max_{1 \le j \le p} \max_{1 \le k \le K} \left| \frac{1}{n} \sum_{t=1}^n u_{j,t} f_{k,t}\right| \le C \sqrt{\ln(p)/n}\right] \ge 1- O(1/p^2).\label{pla55}
\end{equation}
Then, by Assumption \ref{as4} and Lemma \ref{la3}(iv),
\begin{equation}
\left\|(\b X \b X'/n)^{-1} \right\|_{\infty} \le C,\label{pla56}
\end{equation}
with probability at least $1 - O(1/n^2)$. Next, since $\b\eta_j: n \times 1$, and $\b\eta_{j,t}$ is the $t$th element
\begin{equation}
\max_{1 \le j \le p} \left\| \frac{\b X \b\eta_j}{n} \right\|_{\infty} = \max_{1 \le j \le p} \max_{1 \le k \le K} \left| \frac{1}{n} \sum_{t=1}^n f_{k,t} \eta_{j,t}\right|.\label{pla56a}
\end{equation}
Then, by Lemma \ref{la3}(v),
\begin{equation}
\P\left[\max_{1 \le j \le p} \max_{1 \le k \le K} \left| \frac{1}{n} \sum_{t=1}^n f_{k,t} \eta_{j,t} \right| \le C \sqrt{\frac{\bar{s} \ln(p)}{n}}\right] \ge 1- O(1/p^2).\label{pla57}
\end{equation}
Combine (\ref{pla53})-(\ref{pla57}) in (\ref{pla52}) in order to form
\begin{equation}
\max_{ 1 \le j \le p} \| \frac{\widehat{\b U}_{-j} \b\eta_{xj}}{n} \|_{\infty} \le C \left[ K^2 \frac{\sqrt{\bar{s}} \ln(p)}{n} + \sqrt{\frac{\ln(p)}{n}} \right],\label{pla57a}
\end{equation}
with probability at least $ 1 - O(1/p^2) - O(1/n^2)$. Now use (\ref{eve1}) to get $\lambda_n$. 
\begin{flushright}
{\bf Q.E.D.}
\end{flushright}

\begin{lemma}\label{la6}
Under Assumptions \ref{as1}-\ref{as5}, for $j=1,\cdots, p$, we have that
\[
\P\left({\cal A}_{2j}\right) \ge 1 - O(1/p^2) - O(1/n^2).
\]
\end{lemma}

\noindent
{\bf Proof of Lemma \ref{la6}}.
For each $j=1,\cdots, p$, add and subtract $\b\delta_j' (\b U_{-j} \b U_{-j}'/n) \b\delta_j$
\begin{eqnarray*}
 \frac{\b\delta_j' \widehat{\b U}_{-j} \widehat{\b U}_{-j}' \b\delta_j}{n} & = & \frac{\b\delta_j' \widehat{\b U}_{-j} \widehat{\b U}_{-j}' \b\delta_j}{n} -
 \frac{\b\delta_j' \b U_{-j} \b U_{-j}' \b\delta_j}{n} +\frac{\b\delta_j' \b U_{-j} \b U_{-j}' \b\delta_j}{n}  \\
 & \ge &  \frac{\b\delta_j' \b U_{-j} \b U_{-j}' \b\delta_j}{n} - \left|\frac{\b \delta_j' \widehat{\b U}_{-j} \widehat{\b U}_{-j}' \b\delta_j}{n}
-\frac{\b\delta_j' \b U_{-j} \b U_{-j}' \b\delta_j}{n}\right| \end{eqnarray*}
Next, add and subtract $\frac{\b\delta_j' \b\Sigma_{n,-j,-j} \b\delta_j}{n}$ from the right side of the previous inequality and get
\begin{eqnarray}
\frac{\b\delta_j' \widehat{\b U}_{-j} \widehat{\b U}_{-j}' \b\delta_j}{n}
& \ge & \frac{\b\delta_j' \b\Sigma_{n,-j,-j}\b\delta_j}{n} - \left| \frac{\b\delta_j' \b U_{-j} \b U_{-j}' \b\delta_j }{n} - \frac{\b\delta_j' \b\Sigma_{n,-j,-j} \b\delta_j}{n}\right|
\nonumber \\
& - & \left|\frac{\b\delta_j' \widehat{\b U}_{-j} \widehat{\b U}_{-j}'\b\delta_j}{n} - \frac{\b\delta_j' \b U_{-j} \b U_{-j}' \b\delta_j }{n}\right|\label{pla60}
\end{eqnarray}
Note that second right side term with absolute value in (\ref{pla60}) can be bounded by using H{\"o}lder's inequality twice
\[ 
\left|\frac{\b\delta_j' \widehat{\b U}_{-j} \widehat{\b U}_{-j}' \b\delta_j}{n} - \frac{\b\delta_j' \b U_{-j} \b U_{-j}' \b\delta_j }{n} \right|
\le \| \b\delta_j \|_1^2 
\left\| \frac{\widehat{\b U}_{-j}\widehat{\b U}_{-j}'}{n} - \frac{\b U_{-j} \b U_{-j}'}{n} \right\|_{\infty}.
 \]
 By the same analysis applied to the first right side term with absolute value in (\ref{pla60}) and simplifying
\begin{eqnarray}
\frac{\b\delta_j' \widehat{\b U}_{-j} \widehat{\b U}_{-j}' \b\delta_j}{n}
& \ge & \frac{\b\delta_j' \b\Sigma_{n,-j,-j} \b\delta_j}{n} \nonumber \\
& - & \left\|\b\delta_j \right\|_1^2
\left[\left\| \frac{\widehat{\b U}_{-j} \widehat{\b U}_{-j}'}{n} - \frac{\b U_{-j} \b U_{-j}'}{n}\right \|_{\infty} +
\left\| \frac{\b U_{-j} \b U_{-j}'}{n} - \b\Sigma_{n,-j,-j} \right\|_{\infty}
\right]\label{pla61}
\end{eqnarray}
In (\ref{pla61}) we start considering  by (\ref{rv13})
\begin{equation}
\frac{\widehat{\b U}_{-j} \widehat{\b U}_{-j}'}{n} - \frac{\b U_{-j} \b U_{-j}'}{n} = - \frac{\b U_{-j} \b X'}{n} \left(\frac{\b X \b X'}{n}
\right)^{-1} \frac{\b X \b U_{-j}'}{n}.\label{pla62}
\end{equation}
Using (\ref{pla62}), Lemma \ref{la1}(ii),  (\ref{pla55}), and (\ref{pla56}) we get that
\begin{eqnarray}
\left\| \frac{\widehat{\b U}_{-j} \widehat{\b U}_{-j}'}{n} - \frac{\b U_{-j} \b U_{-j}'}{n} \right\|_{\infty} &\le & K^2 \left\| \frac{\b U_{-j} \b X' }{n} \right\|_{\infty}^2\left \| (\frac{\b X \b X'}{n})^{-1}\right\|_{\infty}
\nonumber \\
& \le & C \frac{K^2 \ln(p)}{n}, \label{pla63}
\end{eqnarray}
with probability at least $1 - O(1/p^2) - O(1/n^2)$. Next, in (\ref{pla61}) see that $\b\Sigma_{n,-j,-j}$ is a submatrix of $\b\Sigma_n$, and $\b U_{-j}$ is a submatrix of $\b U$ as described above and 
\begin{equation}
\left\|\frac{\b U_{-j} \b U_{-j}'}{n} - \b\Sigma_{n,-j,-j} \right\|_{\infty} \le
\left\| \frac{\b U \b U'}{n} - \b\Sigma_n \right\|_{\infty} \le C \sqrt{\ln(p)/n},\label{pla64}
\end{equation}
with probability at least $1 - O(1/p^2)$ by Lemma \ref{la3}(i). We need to provide some simplification for $\|\b\delta_j\|_1^2$ term in (\ref{pla61}).
Next, since the  cone condition in adaptive restricted eigenvalue condition is satisfied in (\ref{pla48a})
\[ 
\|\b\delta_{S_j^c} \|_1 \le 3 \sqrt{s_j} \|\b\delta_{S_j} \|_2.
\]
Then, add $\|\b\delta_{S_j} \|_1 $ to the left side and right side and use the norm inequality that puts an upper bound on the $l_1$ norm in terms of the $l_2$ norm. Hence,
\begin{eqnarray*}
\|\b\delta_{S_j^c} \|_1 + \|\b\delta_{S_j} \|_1 & = & \| \b\delta_j \|_1 \le \|\b\delta_{S_j} \|_1 + 3 \sqrt{s_j} \| \b\delta_{S_j} \|_2 \\
& \le & 4 \sqrt{s_j} \|\b\delta_{S_j} \|_2.
\end{eqnarray*}
So, we have that
\begin{equation}
\frac{\|\b\delta_j \|_1^2}{\|\b\delta_{S_j} \|_2^2} \le 16 s_j.\label{pla65}
\end{equation}
Now, divide (\ref{pla61}) by $ \|\b\delta_{S_j} \|_2^2 >0$ and use (\ref{pla63}) and (\ref{pla65}):
\begin{eqnarray}
\frac{\b\delta_j' \widehat{\b U}_{-j} \widehat{\b U}_{-j}' \b\delta_j/n}{\|\b\delta_{S_j}\|_2^2} & \ge &
\frac{\b\delta_j' \b\Sigma_{n,-j,-j} \b\delta_j}{\left\|\b\delta_{S_j} \right\|_2^2 }- 16 s_j K^2 \left\| \frac{\b U_{-j} \b X'}{n}\right\|_{\infty}^2 \left\| (\frac{\b X \b X'}{n})^{-1}\right\|_{\infty}
\nonumber \\
& - & 16 s_j \left\| \frac{\b U_{-j} \b U_{-j}'}{n} - \b\Sigma_{n,-j,-j} \right\|_{\infty}.\label{pla66}
\end{eqnarray}
Next, using the empirical and population adaptive restricted eigenvalue definitions and minimizing over $\b\delta_j$ we have that
\begin{eqnarray}
\widehat{\phi}^2 (s_j) & \ge  & \phi^2 (s_j) - 16 s_j K^2\left\| \frac{\b U_{-j} \b X'}{n}\right\|_{\infty}^2\left\|\left(\frac{\b X \b X'}{n}\right)^{-1}\right\|_{\infty}
\nonumber \\
& - & 16 s_j \left\| \frac{\b U_{-j} \b U_{-j}'}{n} - \b\Sigma_{n,-j,-j} \right\|_{\infty}.\label{pla67}
\end{eqnarray}
Note that, if we have with probability approaching one (wpa1 from now on)
\begin{equation}
16 s_j \left[ K^2 \left\| \frac{\b U_{-j} \b X'}{n}\|_{\infty}^2 \| (\frac{\b X \b X'}{n})^{-1}\right\|_{\infty}
+ 16 s_j \left\| \frac{\b U_{-j} \b U_{-j}'}{n} - \b\Sigma_{n,-j,-j} \right\|_{\infty}\right] \le \phi^2 (s_j)/2,\label{pla67a}
\end{equation}
we have with wpa1
\[ 
\widehat{\phi}^2 (s_j)/2 \ge \phi^2 (s_j)/2.
\]
Thus, we need to show that following probability goes to zero
\begin{equation}
\begin{split}
\P[\widehat{\phi}^2 (s_j)&<\phi^2 (s_j)/2] \\&\le
\P\left[16 s_j\left(K^2\left\|\frac{\b U_{-j} \b X'}{n}\right\|_{\infty}^2\left\|\left(\frac{\b X \b X'}{n}\right)^{-1}\right\|_{\infty}
+ \left\|\frac{\b U_{-j} \b U_{-j}'}{n} - \b\Sigma_{n,-j,-j} \right\|_{\infty}\right) > \phi^2 (s_j)/2\right]\label{pla68}
\end{split}
\end{equation}

Set $\epsilon_n:= 16 s_j\left[K^2 \ln(p)/n+ \sqrt{\ln(p)/n}\right]$. Clearly, by (\ref{pla63}) and (\ref{pla64}) we have that
\begin{equation}
\begin{split}
&\P\left[16 s_j\left(K^2\left\|\frac{\b U_{-j} \b X'}{n}\right\|_{\infty}^2 \left\|\left(\frac{\b X \b X'}{n}\right)^{-1}\right\|_{\infty}
+ \left\|\frac{\b U_{-j}\b U_{-j}'}{n} - \b\Sigma_{n,-j,-j} \right\|_{\infty}\right) > \epsilon_n\right] \\ &\qquad\qquad\qquad\le O(1/p^2) + O(1/n^2).\label{pla69}
\end{split}
\end{equation}
Since $\epsilon_n \to 0$ by Assumption \ref{as5}, by (\ref{pla68})(\ref{pla69}) $P (\widehat{\phi}^2 ( s_j) < \phi^2 (s_j)/2) \to 0$.
\begin{flushright}
{\bf Q.E.D.}
\end{flushright}

One crucial point is that we need to get a low bound for $\cap_{j=1}^p {\cal A}_{2j}$. In that respect, from (\ref{pla67a})
\[
\begin{split}
&\left(16 s_j \left[ K^2 \left\| \frac{\b U_{-j} \b X' }{n} \right\|_{\infty}^2 \left\| (\frac{\b X \b X'}{n})^{-1} \right\|_{\infty}  + \left\| \frac{\b U_{-j} \b U_{-j}'}{n} - \b\Sigma_{n,-j,-j} \right\|_{\infty}\right]
\le \phi^2 (s_j)/2 \right) \\&\qquad\qquad\subseteq \{ \widehat{\phi}^2 (s_j) \ge \phi^2 (s_j)/2\} = {\cal A}_{2j}.
\end{split}
\]

Clearly by the definitions of $\b\Sigma_n$ and $\b\Sigma_{n,-j,-j}$ and population adaptive restricted eigenvalue condition, we have that
\begin{eqnarray}
&16s_j&\left[K^2\left\|\frac{\b U_{-j} \b X'}{n} \right\|_{\infty}^2 \left\|\left(\frac{\b X \b X'}{n}\right)^{-1}\right\|_{\infty} + \left\|\frac{\b U_{-j} \b U_{-j}'}{n} - \b\Sigma_{n,-j,-j} \right\|_{\infty}\right] \nonumber \\
& \le & 16 \bar{s}
\left[K^2\left\|\frac{\b U \b X'}{n} \right\|_{\infty}^2 \left\|\left(\frac{\b X \b X'}{n}\right)^{-1}\right\|_{\infty}  + \left\| \frac{\b U \b U'}{n} - \b\Sigma_n \right\|_{\infty} \right]
\nonumber \\
& \le & \phi^2 (\bar{s})/2 \le \phi^2 (s_j)/2.\label{pla610}
\end{eqnarray}

So (\ref{pla610}) implies that,  for $j=1,\cdots, p$,
\[ 
\left[16\bar{s}\left[K^2\left\|b U \b X'/n\right\|_{\infty}^2  \left\|\left(\b X \b X'/n\right)^{-1} \right\|_{\infty}  + \left\| \b U \b U'/n - \b\Sigma_n \right\|_{\infty}\right] \le \phi^2 (\bar{s})/2\right] \subseteq {\cal A}_{2j}.
\]
This means that
\[ 
\left[16 \bar{s} \left[ K^2 \left\| \b U \b X'/n\right\|_{\infty}^2  \left\| \left(\b X \b X'/n\right)^{-1} \right\|_{\infty}  + \left\| \b U \b U'/n - \b\Sigma_n \right\|_{\infty}\right] \le \phi^2 (\bar{s})/2\right] \subseteq \cap_{j=1}^p {\cal A}_{2j}.
\]
Next, by (\ref{pla56}) and (\ref{pla64}), via Lemma \ref{la3}(iii), we have that
\begin{eqnarray}
\P\left[\left(\cap_{j=1}^p {\cal A}_{2,j}\right)^c\right] & \le & \P\left(16 \bar{s}\left[K^2\left\|\b U \b X'/n\right\|_{\infty}^2 \left\|\left(\b X \b X'/n\right)^{-1} \right\|_{\infty} + \left\|\b U \b U'/n -\b\Sigma_n \right\|_{\infty}\right]> \phi^2 (\bar{s})/2\right) \nonumber \\
& \le & O(1/p^2) + O(1/n^2).\label{pla611}
\end{eqnarray}


We provide the main consistency result for residual based nodewise regression result. 

\begin{lemma}\label{la7}
Under Assumptions \ref{as1}-\ref{as5}
\[ 
\max_{1 \le j \le p} \| \widehat{\b\gamma}_j - \b\gamma_j \|_1 = O_p\left(\lambda_n \bar{s}\right)=
o_p(1).
\]
\end{lemma}

\noindent
{\bf Proof of Lemma \ref{la7}}. Use Lemmata \ref{la5}-\ref{la6} and (\ref{pla611}) to have
\[ 
\P\left({\cal A}_1 \cap \{\cap_{j=1}^p {\cal A}_{2j} \}\right) \ge 1 - O(1/p^2) - O(1/n^2).
\]
Then, combine above with Lemma \ref{la4} to have the desired result via Assumption \ref{as5} and Lemma \ref{la5} to have $\lambda_n \bar{s} = o(1)$. 

\begin{flushright}
{\bf Q.E.D.}
\end{flushright}

Next, we provide proof of consistency for the estimates of the reciprocal of the main diagonal elements  of the precision matrix.

\begin{lemma}\label{la8}
Under Assumptions \ref{as1}-\ref{as5}
\[
\max_{1 \le j \le p}\left|\widehat{\tau}_j^2 - \tau_j^2\right| = O_p\left(\bar{s}^{1/2} \lambda_n \right) = o_p (1).
\]
\end{lemma}

\noindent
{\bf Proof of Lemma \ref{la8}}. Start with $\widehat{\tau}_j^2$ and the definition in (\ref{rv16}). For all $j=1,\cdots, p$,
\[
\widehat{\tau}_j^2:= \widehat{\b u}_j' (\widehat{\b u}_j - \widehat{\b U}_{-j}' \widehat{\b \gamma}_j)/n
\]
and $\tau_j^2:= \E\left[\eta_{j,t}^2\right]$, with
$ \eta_{j,t}:= u_{j,t}- \b u_{-j,t}'\b\gamma_j$, and $\b\eta_{j}:=(\eta_{j,1}, \cdots, \eta_{j,n})': n \times 1$ vector $\b\eta_{xj}= \b M_X \b \eta_j$. Using (\ref{rv14}) for $\widehat{\b u}_j$ in $\widehat{\tau}_j^2$ definition we have
\begin{eqnarray*}
\widehat{\tau}_j^2 & = & (\widehat{\b U}_{-j}' \b\gamma_j + \b\eta_{xj})'(\b\eta_{xj} - \widehat{\b U}_{-j}' (\widehat{\b \gamma}_j - \b\gamma_j))/n \\
& = & \frac{\b \eta_{xj}' \b \eta_{xj}}{n} - \frac{\b \eta_{xj}' \widehat{\b U}_{-j}' (\widehat{\b \gamma}_j - \b \gamma_j)}{n} \\
& + & \frac{\b \gamma_j' \widehat{\b U}_{-j} \b\eta_{xj}}{n}  - \frac{\b\gamma_j' \widehat{\b U}_{-j} \widehat{\b U}_{-j}' (\widehat{\b \gamma}_j - \b\gamma_j)}{n}.
\end{eqnarray*}
By the triangle inequality we get
\begin{eqnarray}
\max_{1 \le j \le p} \left| \widehat{\tau}_j^2 - \tau_j^2\right| & \le & \max_{1 \le j \le p} \left| \frac{\b\eta_{xj}' \b\eta_{xj}}{n} - \tau_j^2\right| + \max_{1 \le j \le p}
\left|\frac{\eta_{xj}' \widehat{U}_{-j}'\left(\widehat{\b \gamma}_j - \b \gamma_j\right)}{n}\right|\nonumber  \\
& + & \max_{1 \le j \le p} \left| \frac{\b\gamma_j' \widehat{\b U}_{-j}\b\eta_{xj}}{n}\right| + \max_{1 \le j \le p} \left| \frac{\b\gamma_j' \widehat{\b U}_{-j} \widehat{\b U}_{-j}' (\widehat{\b\gamma}_j - \b\gamma_j)}{n} \right|.\label{pla81}
\end{eqnarray}

Consider each term in (\ref{pla81}) carefully. Start with definition; $\b M_X:= \b I_n - \b X' (\b X \b X')^{-1} \b X$, and $\b M_X$ being idempotent.
\begin{eqnarray}
\max_{1\le j \le p}\left|\frac{\b\eta_{xj}'\b\eta_{xj}}{n} - \tau_j^2\right| & \le & \max_{1 \le j \le p}\left|\frac{\b\eta_j'\b\eta_j}{n} - \tau_j^2\right| \nonumber \\
& + & \max_{1 \le j \le p }\left| \frac{\b\eta_j' \b X' }{n} \left(\frac{\b X \b X'}{n}
\right)^{-1} \frac{\b X \b \eta_j}{n}\right|.\label{pla82}
\end{eqnarray}
First, exactly as in Lemma \ref{la3}(i) with Assumption \ref{as2}(ii)(iv), $ 3 r_2^{-1} + r_0^{-1} > 1$ we have by Theorem \ref{thma1} that
\begin{equation}
\max_{1 \le j \le p} \left| \frac{\b\eta_j' \b\eta_j}{n} - \tau_j^2\right| = O_p\left(\sqrt{\ln(p)/n}\right).\label{pla83}
\end{equation}
Then note that $ \b X \b \eta_j: K \times 1$ vector, and $\b X \b X': K \times K$ matrix. Therefore, 
\begin{eqnarray}
\max_{1 \le j \le p}\left|\frac{\b\eta_j' \b X' }{n} \left( \frac{\b X \b X'}{n}
\right)^{-1} \frac{\b X \b \eta_j}{n}\right| & \le & \max_{1 \le j \le p}  \left\| \frac{\b\eta_j' \b X'}{n}\right\|_1 \max_{1 \le j \le p} \left\| \left( \frac{\b X\b X'}{n}
\right)^{-1} \frac{\b X \b\eta_j}{n}\right\|_{\infty} \nonumber \\
& \le &  \left[\max_{1 \le j \le p} \left\| \frac{\b\eta_j' \b X'}{n} \right\|_1\right] K \left\| \left( \frac{\b X \b X'}{n}
\right)^{-1} \right\|_{\infty} \max_{1 \le j \le p} \left\| \frac{\b X \b \eta_j}{n} \right\|_{\infty} \nonumber \\
& \le & \left[  K \max_{1 \le j \le p} \left\| \frac{\eta_j' X'}{n} \right\|_{\infty} \right] \left[ K \left\| \left( \frac{XX'}{n}\right)^{-1} \right\|_{\infty} \max_{1 \le j \le p} \| \frac{X \eta_j}{n} \|_{\infty}
\right] \nonumber \\
& = & \left[ K \max_{1 \le j \le p} \left\| \frac{\b\eta_j' \b X'}{n} \right\|_{\infty}
\right]^2 \left\| \left(\frac{\b X \b X'}{n} \right)^{-1} \right\|_{\infty},\label{pla84}
\end{eqnarray}
where we use H{\"o}lder's inequality for the first inequality, and (\ref{pla10}) and (\ref{pla11}) for the second inequality, and the norm inequality between $l_1$ and $l_{\infty}$ norms for the third inequality
(i.e. $\| \b x \|_1 \le \textnormal{dim}(\b x) \|\b x\|_{\infty}, \textnormal{dim}(\b x):$ dimension of the vector x). Next by (\ref{pla56}), (\ref{pla56a}), and (\ref{pla57}), we have by (\ref{pla84}) that
\begin{equation}
\max_{1 \le j \le p} \left| \frac{\b\eta_j' \b X'}{n} \left( \frac{\b X \b X'}{n}
\right)^{-1} \frac{\b X \b \eta_j}{n}\right| = O_p\left(K^2 \bar{s} \frac{\ln(p)}{n}\right).\label{pla85}
\end{equation}
Combine (\ref{pla83}) and (\ref{pla85}) in (\ref{pla82}) to have the first term on the right side of (\ref{pla81}) by Assumption \ref{as5} to get the last equality in (\ref{pla86})
\begin{equation}
\max_{1 \le j \le p} \left| \frac{\b\eta_{xj}' \b\eta_{xj}}{n} - \tau_j^2\right| = O_p\left( K^2 \bar{s} \frac{\ln(p)}{n}\right) + O_p\left(\sqrt{\frac{\ln(p)}{n}}\right)
.\label{pla86}
\end{equation}

See that by Lemma \ref{la5} (with probability approaching one) and (\ref{eve1}) that
\begin{equation}
\left\|\b\eta_{xj}' \widehat{\b U}_{-j}'/n \right\|_{\infty} \le \lambda_n/2 .\label{lr0}
\end{equation}
 In (\ref{pla81}) consider the second term on the right side by (\ref{lr0}), Lemma \ref{la7}
\begin{eqnarray}
\max_{1 \le j \le p}\left|\frac{\b \eta_{xj}' \widehat{\b U}_{-j}'}{n} (\widehat{\b\gamma}_j - \b\gamma_j)\right| & \le & \max_{1 \le j \le p} \left\| \frac{\b\eta_{xj}' \widehat{\b U}_{-j}'}{n}\right\|_{\infty}
\max_{1 \le j \le p}\left\| \widehat{\b \gamma}_j - \b \gamma_j\right\|_1 \nonumber \\
& = & O_p\left(\lambda_n\right)O_p\left(  \bar{s} \lambda_n \right) 
 = O_p \left(\bar{s} \lambda_n^2  \right).\label{pla87}
\end{eqnarray}

Consider the third term on the right side of (\ref{pla81}), where we use H{\"o}lder's inequality to get 
\begin{eqnarray}
\max_{1 \le j \le p}\left|\b\gamma_j' \frac{\widehat{\b U}_{-j} \b\eta_{xj}}{n}\right| & \le & \left[\max_{1 \le j \le p} \left\| \b \gamma_j \right\|_1\right]
\left[\max_{1 \le j \le p} \left\| \frac{\widehat{\b U}_{-j} \b\eta_{xj}}{n} \right\|_{\infty}\right] \nonumber \\
& = & O\left(\bar{s}^{1/2}\right)O_p\left( \lambda_n\right)= O_p\left(\sqrt{\bar{s}} \lambda_n \right),\label{pla88}
\end{eqnarray}
for the rates we use (\ref{pla35}), (\ref{lr0}). Last we consider the fourth term on the right side of (\ref{pla81}). To get a better rate, we start with the Karush-Kuhn-Tucker (KKT) conditions in (\ref{rv15}). The following $p-1$ equations form the KKT
\[ 
\lambda_n \widehat{\b\kappa}_j + \frac{\widehat{\b U}_{-j} \widehat{\b U}_{-j}'}{n} \widehat{\b \gamma}_j - \frac{\widehat{\b U}_{-j} \widehat{\b u}_j}{n} = \b 0_{p-1},
\]
where $\widehat{\b\kappa}_j$ is the sub-differrential and explained in more detail in p.160 of Caner and Kock (2018) which replaces the gradient in non-differential penalties. Also
for all $j=1\cdots,p$
$\| \widehat{\b\kappa}_j \|_{\infty} \le 1.$
Use (\ref{rv14}) for $\widehat{\b u}_j$
and rewrite KKT as
\[
\frac{\widehat{\b U}_{-j} \widehat{\b U}_{-j}'}{n} (\widehat{\b\gamma}_j - \b\gamma_j) = \frac{\widehat{\b U}_{-j} \b\eta_{xj}}{n} - \lambda_n \widehat{\b\kappa}_j.\]
Then, by the triangle inequality, we have that
\begin{eqnarray}
\left\|\frac{\widehat{\b U}_{-j} \widehat{\b U}_{-j}'}{n} \left(\widehat{\b\gamma}_j - \b\gamma_j\right)\right\|_{\infty} & \le & \left\| \frac{\widehat{\b U}_{-j} \b\eta_{xj}}{n} \right\|_{\infty} + \lambda_n \left\| \widehat{\b\kappa}_j \right\|_{\infty} \nonumber \\
& \le & \left\| \frac{\widehat{\b U}_{-j} \b\eta_{xj}}{n} \right\|_{\infty} + \lambda_n \nonumber \\
& = & O_p\left(\lambda_n \right).\label{pla810}
\end{eqnarray}

Then the fourth term on the right side of (\ref{pla81})
\begin{eqnarray}
\max_{1 \le j \le p}\left|\b\gamma_j'\frac{\widehat{\b U}_{-j} \widehat{\b U}_{-j}'}{n}\left(\widehat{\b\gamma}_j - \b\gamma_j\right)\right|
& \le & \max_{1 \le j \le p} \left\|\b\gamma_j \right\|_1 \max_{1 \le j \le p}\left\|\frac{\widehat{\b U}_{-j} \widehat{\b U}_{-j}'}{n} (\widehat{\b\gamma}_j - \b\gamma_j)\right\|_{\infty}
\nonumber \\
& = & O_p\left(\sqrt{\bar{s}} \lambda_n\right),\label{pla811}
\end{eqnarray}
where we use H{\"o}lder's inequality, (\ref{pla35}) and (\ref{pla810}). Clearly, (\ref{pla88}) and (\ref{pla811}) are the slowest among the four terms on the right side of (\ref{pla81}), and we use Assumption \ref{as5} to get the desired result.

\begin{flushright}
{\bf Q.E.D.}
\end{flushright}

\noindent
{\bf Proof of Theorem \ref{rthm1}}. First, we derive some of the key results. By definition of $\tau_j^2$, for $j=1,\cdots,p$, and since $\b\Omega:=\b\Sigma_n^{-1}$, with Assumption
\ref{as1}
\begin{equation}
\tau_j^2:= \frac{1}{\Omega_{j,j}} \ge \frac{1}{\textnormal{Eigmax} (\b\Omega)} = \textnormal{Eigmin}(\b\Sigma_n) \ge c > 0.\label{pt11}
\end{equation}
Note that $\min_{1 \le j \le p} \tau_j^2$ is bounded away from zero. Next
\begin{equation}
\min_{1 \le j \le p} \widehat{\tau}_j^2 = \min_{1 \le j \le p} | \widehat{\tau}_j^2 - \tau_j^2 + \tau_j^2| \ge \min_{1 \le j \le p} \tau_j^2 -
\max_{1 \le j \le p} | \widehat{\tau}_j^2 - \tau_j^2|.\label{pt12}
\end{equation}
is bounded away from zero wpa1 by Lemma \ref{la8}. Then
\begin{equation}
\max_{1 \le j \le p} \left|\frac{1}{\widehat{\tau}_j^2} - \frac{1}{\tau_j^2}\right| = \max_{1 \le j \le p} \frac{\left| \widehat{\tau}_j^2 - \tau_j^2\right|}{\widehat{\tau}_j^2 \tau_j^2}
= O_p\left(\sqrt{\bar{s}} \lambda_n \right) = o_p (1).\label{pt13}
\end{equation}
by Lemma \ref{la8}, (\ref{pt11}), and (\ref{pt12}). Now we complete the proof by using the formula for $\widehat{\b\Omega}_j, \b\Omega_j$.

\begin{eqnarray*}
\max_{1 \le j \le p}\left\|\widehat{\b\Omega}_j - \b\Omega_j \right\|_1 & = & \max_{1 \le j \le p} \left\| \frac{\widehat{\b C}_j}{\widehat{\tau}_j^2} - \frac{\b C_j}{\tau_j^2} \right\|_1 \nonumber \\
& \le & \max_{1 \le j \le p} \left| \frac{1}{\widehat{\tau}_j^2} - \frac{1}{\tau_j^2} \right| + \max_{1 \le j \le p} \left\| \frac{\widehat{\b\gamma}_j}{\widehat{\tau}_j^2} - \frac{\b\gamma_j}{\tau_j^2} \right\|_1
\nonumber \\
& \le & \max_{1 \le j \le p} \left| \frac{1}{\widehat{\tau}_j^2} - \frac{1}{\tau_j^2} \right|
+ \max_{1 \le j \le p} \frac{\left\|\widehat{\b\gamma}_j - \b\gamma_j\right\|_1}{\widehat{\tau}_j^2} \\
& + & \max_{1 \le j \le p} \left\|\b\gamma_j\right\|_1 \max_{1 \le j \le p}\left| \frac{1}{\widehat{\tau}_j^2} - \frac{1}{\tau_j^2}\right|\\
& = & O_p\left(\bar{s} \lambda_n \right) = o_p (1),
\end{eqnarray*}
where we use (\ref{pt13}), Lemma \ref{la7}, (\ref{pla35}) for the rates, and the last equality is by Assumption \ref{as5}.

\begin{flushright}
{\bf Q.E.D.}
\end{flushright}

\subsection*{Part 3}

After the proof of Theorem \ref{rthm1} we provide lemmata that lead to proof of Theorem \ref{rthm2}. We start with a lemma that is related to norm inequalities.
First define generic matrices, $\b A_1: p \times K, \b A_2: K \times p$, $\b D_1: K \times K, \b D_2: p \times p$, also define a row vector $\b x':1 \times p$, and also define $p \times p$ matrices $\b A_3, \b D_3$.

\begin{lemma}\label{la9}

(i). \[ \| \b A_1 \b D_1 \b A_2\|_{l_{\infty}} \le p K^{1/2}  \| \b A_1 \|_{l_{\infty}} \| \b A_2\|_{\infty} \| \b D_1 \|_{l_2} .\]

(ii). \[ \| \b A_2 \b D_2 \b A_1 \|_{\infty} \le p \| \b A_2\|_{\infty} \| \b A_1 \|_{\infty}  \| \b D_2 \|_{l_{\infty}}.\]

(iii).\[ \| \b x' \b A_3 \b D_3 \|_1 \le \| \b x \|_1 \| \b D_3 \|_{l_{\infty}} \| \b A_3 \|_{l_{\infty}}.\]

(iv). \[\| \b A_2 \b D_2 \b A_1 \|_{\infty} \le p \| \b A_2 \|_{\infty} \| \b A_1 \|_{\infty} \| \b D_2 \|_{l_1}.\]
\end{lemma}

\noindent
{\bf Proof of Lemma \ref{la9}}.

(i). \begin{eqnarray*}
\| \b A_1 \b D_1 \b A_2 \|_{l_{\infty}} & \le & \| \b A_1 \|_{l_{\infty}} \| \b D_1 \b A_2 \|_{l_{\infty}} 
 \le  \| \b A_1 \|_{l_{\infty}} [p \| \b A_2\|_{\infty}] \| \b D_1  \|_{l_{\infty}}\\
& \le & p \| \b A_1 \|_{l_{\infty}} \| \b A_2 \|_{\infty} [ K^{1/2} \| \b D_1\|_{l_2} ] \\
\end{eqnarray*}
where we use submultiplicativity of matrix norms for the first inequality, and submultiplicativity of matrix norms and the following for the second inequality,
\[ \| \b A_2 \|_{l_{\infty}}:= \max_{1 \le k \le K} \|\b A_{2,k}' \|_1\le p \max_{1 \le k \le K} \max_{1 \le j \le p} | A_{2,kj} |,\]
where $\b A_{2,k}'$ and $\b A_{2,kj}$ are the $k$th row of $\b A_2$, and $k,j$ element of $\b A_2$ respectively.
Then, for the last inequality, we use a matrix norm inequality that provides an upper bound for $l_{\infty}$ matrix norm in terms of spectral norm in p.365 of \cite{hj2013}.

(ii). \begin{eqnarray*}
\| \b A_2 \b D_2 \b A_1 \|_{\infty} & \le & \| \b A_2 \|_{\infty} \| \b D_2 \b A_1\|_{l_1} 
 \le  p \| \b A_2 \|_{\infty} \| \b A_1 \|_{\infty} \| \b D_2 \|_{l_{\infty}}
\end{eqnarray*}
where we use section 4.3 of \cite{vdg2016} for the first inequality, and  the second inequality can be seen by defining $\b D_{2,j}'$ as the $j$th row of $\b D_2$, and $\b A_{1,k}$ as the $k$th column of $\b A_1$ and using H{\"o}lder's inequality
\begin{eqnarray*}
\| \b D_2 \b A_1 \|_{l_1}& := &\max_{1 \le k \le K} \sum_{j=1}^p |\b D_{2,j}'\b A_{1,k}| \le \max_{1 \le k \le K} \sum_{j=1}^p \| \b D_{2,j}\|_1 \|\b A_{1,k} \|_{\infty} \\
& \le & p \max_{1 \le j \le p} \| \b D_{2,j} \|_1 \left( \max_{1 \le k \le K}  \| \b A_{1,k} \|_{\infty} \right) 
 =  p \| \b D_2 \|_{l_{\infty}} \| \b A_1 \|_{\infty}.
\end{eqnarray*}

(iii).
\begin{eqnarray*}
\| \b x' \b A_3 \b D_3 \|_1 = \| \b D_3' \b A_3' x \|_1 & \le & \| \b x \|_1 \| \b D_3' \b A_3'\|_{l_1} \\
& \le & \| \b x \|_1 \| \b D_3' \|_{l_1} \| \b A_3 ' \|_{l_1} 
 =  \| \b x \|_1 \| \b D_3 \|_{l_{\infty}} \| \b A_3 \|_{l_{\infty}},
\end{eqnarray*}
where we use p.345 of \cite{hj2013} for the first inequality, and $l_1$ matrix norm submultiplicativity for the second inequality, and the last equality
is by seeing that transpose of $l_1$ matrix norm is $l_{\infty}$ matrix norm.

(iv).
\begin{eqnarray*}
\| \b A_2 \b D_2 \b A_1 \|_{\infty} & \le & \| \b A_2 \|_{\infty} \| \b D_2 \b A_1 \|_{l_1} \\
& \le & \| \b A_2 \|_{\infty} \| \b D_2 \|_{l_1} \| \b A_1 \|_{l_1} 
 \le  p \| \b A_2 \|_{\infty} \| \b D_2 \|_{l_1} \| \b A_1 \|_{\infty},
\end{eqnarray*}
where we use p.44 \cite{vdg2016} dual norm inequality for the first inequality, then for the second inequality we use submultiplicativity property of matrix norms,and for the last inequality we use $\| \b A_1 \|_{l_1}:= \max_{1 \le k \le K} \sum_{j=1}^p | A_{1,jk}| \le p \| \b A_1 \|_{\infty}$, where $A_{1,jk}$ is the $j,k$ th cell in $\b A_1$.

\begin{flushright}
{\bf Q.E.D.}
\end{flushright}

\begin{lemma}\label{la10}
Under Assumptions \ref{as1}-\ref{as4}, \ref{as6}-\ref{as7}

(i).
\[ 
\| \widehat{\b B} - \b B \|_{l_{\infty}} = O_p\left(K^{3/2} \sqrt{\ln(p)/n}\right),
\]

(ii).
\[ 
\| \widehat{\b B} \|_{l_{\infty}} = O_p (K),
\]

(iii). 
\[ 
\|\b B\|_{l_{\infty}} = O (K).
\]

(iv). 
\[ 
\| \b B' \|_{l_{\infty}} = \| \b B \|_{l_1} = O(p).
\]
\end{lemma}

\noindent
{\bf Proof of Lemma \ref{la10}}. 

(i). By (\ref{rv11})
\begin{eqnarray*}
\| \widehat{\b B} - \b B \|_{l_{\infty}} &:= & \max_{1 \le j \le p} \| \widehat{\b b}_j' - \b b_j' \|_1 = \max_{1 \le j \le p} \|\widehat{\b b}_j - \b b_j \|_1 
 =  \max_{1 \le j \le p} \left\| \left( \frac{\b X \b X'}{n}\right)^{-1} \frac{\b X \b u_j}{n}\right\|_1 \\
& \le & \left\|\left(\frac{\b X \b X'}{n}\right)^{-1}\right\|_{l_1}  \max_{1 \le j \le p} \left\| \frac{\b X \b u_j}{n}\right\|_1 
 \le  \left\| \left(\frac{\b X \b X'}{n}\right)^{-1} \right\|_{l_1} \left[K \max_{1 \le j \le p} \left\| \frac{\b X \b u_j}{n}\right\|_{\infty}\right] \\
& \le & K \left[ K^{1/2} \left\|\left(\frac{\b X \b X'}{n}\right)^{-1}\right\|_{l_2}\right] \max_{1 \le j \le p}\left\|\frac{\b X \b u_j}{n}\right\|_{\infty}\\
& = & O\left(K^{3/2}\right) O_p\left(\sqrt{\ln(p)/n}\right),
\end{eqnarray*}
where we use $l_{\infty}$ norm definition for the first equality, and for the first inequality we use p.345 of \cite{hj2013}, which is $ \| \b A \b x \|_1 \le \| \b A \|_{l_1} \| \b x \|_1$ for a generic matrix $\b A$, and generic vector $\b x$, for the third inequality we use the upper bound of $l_1$ induced matrix norm in terms of spectral norm, as in p.365 of \cite{hj2013}. The rates are  from (B.3)(B.4) of \cite{fan2011}, and  Lemma \ref{la3}(iii).

(ii). See that
\begin{equation}
\| \b B \|_{l_{\infty}} = \max_{1 \le j \le p} \| \b b_j' \|_1 = O (K),\label{pla10-1}
\end{equation}
by Assumption \ref{as6} that $| b_{jk} | \le C $ for a positive constant $C$ and uniformly over $j=1,\cdots, p$, $k=1, \cdots, K$. Next, using the results above with Assumption \ref{as7}, we have
\[ 
\| \widehat{\b B} \|_{l_{\infty}} = O_p (K).
\]

(iii). This is proved in (\ref{pla10-1}).

(iv). The proof of (iv) is the same as in (ii) above except, with $\b b_k$ as the $k$th column of matrix $\b B$.
\[ \| \b B \|_{l_1} = \max_{1 \le k \le K } \| \b b_k \|_1 = O (p),\]
by Assumption \ref{as6}.

\begin{flushright}
{\bf Q.E.D.}
\end{flushright}

Before the next lemma, we extend two following results which is Lemma B.4 in \cite{fan2011} to the case of increasing maximal eigenvalue of errors.
\begin{lemma}\label{la10a}
Under Assumptions \ref{as4},\ref{as6}, and \ref{as7}(i),
with $c>0, C>0$, and positive finite constants

(i). 
\[
\textnormal{Eigmin} (\b B' \b\Omega \b B ) \ge \frac{c p}{C r_n}.
\]

(ii). 
\[
\left\|\left(\textnormal{cov}(\b f_t)^{-1} + \b B' \b\Omega \b B\right)^{-1}\right\|_{l_2} = O\left(\frac{r_n}{p}\right).\]
\end{lemma}

\noindent
{\bf Proof of Lemma \ref{la10a}}. We follow the proof of Lemma B.4 in \cite{fan2011}.
(i). Since $\b\Omega:= \b\Sigma_n^{-1}$,
\begin{eqnarray*}
\textnormal{Eigmin}(\b B' \b\Omega \b B) & \ge & \textnormal{Eigmin}(\b\Omega) \textnormal{Eigmin} (\b B' \b B) 
 =  [\textnormal{Eigmax} (\b\Sigma_n)]^{-1} \textnormal{Eigmin} (\b B' \b B) \\
& \ge & \frac{cp }{C r_n},
\end{eqnarray*}
by Assumption \ref{as6}, \ref{as7}(i).

(ii). Using Assumption \ref{as4}
\begin{equation}
\textnormal{Eigmin} [([\textnormal{cov} (\b f_t)]^{-1} + \b B' \b\Omega \b B )] \ge \textnormal{Eigmin}(\b B' \b\Omega \b B)\ge \frac{c p}{C r_n}.\label{lb42}
\end{equation}
We have the desired result by (\ref{lb42}), and since for an invertible matrix A, $\textnormal{Eigmax}\left(\b A^{-1}\right)= 1/ \textnormal{Eigmin}(\b A)$.

\begin{flushright}
{\bf Q.E.D.}
\end{flushright}

As described above in the main text, we form the symmetrized version of our feasible nodewise regression estimator for this part of the paper: $\widehat{\b\Omega}_{sym}:= \frac{\widehat{\b\Omega} + \widehat{\b\Omega}'}{2}$.

\begin{lemma}\label{la11}
(i). Under Assumptions \ref{as1}-\ref{as4},\ref{as6}-\ref{as7}
\[ \| \widehat{\b B}' \widehat{\b\Omega}_{sym} \widehat{\b B} - \b B' \b\Omega \b B \|_{l_2} = O_p \left(p  K \max(\bar{s} \lambda_n, \bar{s}^{1/2} K^{1/2} \sqrt{\frac{\ln(p)}{n}})\right).\]

(ii). Under Assumptions \ref{as1}-\ref{as4}, \ref{as6}-\ref{as7}
\[ \| \widehat{\b G} \|_{l_2} = O_p \left(\frac{r_n}{p}\right),\]
with $\widehat{\b G}:= \left([\widehat{\textnormal{cov}(\b f_t)}]^{-1} + \widehat{\b B}' \widehat{\b\Omega}_{sym} \widehat{\b B}\right)^{-1}$.
\end{lemma}

\noindent
{\bf Proof of Lemma \ref{la11}}.
(i). We  start with simple adding and subtracting($ \widehat{\b B} = (\widehat{\b B} -\b B ) + \b B)$, $\widehat{\b\Omega}_{sym}=(\widehat{\b\Omega}_{sym} - \b\Omega ) + \b\Omega)$ and the triangle inequality.
Hence, 
\begin{eqnarray}
\| \widehat{\b B}' \widehat{\b\Omega}_{sym} \widehat{\b B} - \b B' \b\Omega \b B\|_{\infty} & \le & \| (\widehat{\b B} - \b B)' (\widehat{\b\Omega}_{sym} - \b\Omega ) (\widehat{\b B} - \b B )\|_{\infty} \nonumber \\
& + & 2 \| (\widehat{\b B} - \b B)' (\widehat{\b\Omega}_{sym} - \b\Omega )  \b B \|_{\infty} + \| (\widehat{\b B} - \b B )' \b\Omega (\widehat{\b B} - \b B ) \|_{\infty} \nonumber \\
& + & \|\b B' (\widehat{\b\Omega}_{sym} - \b\Omega )\b B \|_{\infty} + 2 \| (\widehat{\b B} - \b B)' \b\Omega \b B \|_{\infty}.\label{pla11-1}
\end{eqnarray}

Analyze each term in (\ref{pla11-1}), and by Lemma \ref{la9}(ii)(iv)
\begin{eqnarray}
\| (\widehat{\b B} - \b B)' (\widehat{\b\Omega}_{sym} - \b\Omega ) (\widehat{\b B} - \b B )\|_{\infty} & \le  &\frac{1}{2}  \| (\widehat{\b B} - \b B )'  (\widehat{\b\Omega} - \b\Omega) (\widehat{\b B} - \b B )\|_{\infty} \nonumber \\
& + & \frac{1}{2}  \| (\widehat{\b B} - \b B )'  (\widehat{\b\Omega}' - \b\Omega) (\widehat{\b B} - \b B )\|_{\infty} \nonumber \\
& \le & \frac{p}{2} \| \widehat{\b B} - \b B \|_{\infty}^2 \| \widehat{\b\Omega} - \b\Omega \|_{l_{\infty}} + \frac{p}{2} \| \widehat{\b B} - \b B \|_{\infty}^2 \| \widehat{\b\Omega}' - \b\Omega \|_{l_1} \nonumber \\
& = & p O_p\left(\frac{K \ln(p)}{n}\right)O_p\left(\bar{s} \lambda_n\right),\label{pla11-2}
\end{eqnarray}
where we use (B.14) of Fan et al (2011) which is: $\| \widehat{\b B} - \b B \|_{\infty}= O_p\left(\sqrt{\frac{K\ln (p)}{n}}\right)$, and Theorem \ref{rthm1} with
\begin{equation}
\| \widehat{\b\Omega} - \b\Omega \|_{l_{\infty}}=
\| \widehat{\b\Omega}' - \b\Omega \|_{l_1}:= \max_{1 \le j \le p} \| \widehat{\b\Omega}_j - \b\Omega_j \|_1,\label{eqeq}
\end{equation}
since $l_1$ norm of transpose of $\widehat{\b\Omega}$ involves rows of $\widehat{\b\Omega}$ (hence columns of $\widehat{\b\Omega}'$).

For the second term in (\ref{pla11-1})
\begin{eqnarray}
\| (\widehat{\b B} - \b B )' (\widehat{\b\Omega}_{sym} - \b\Omega) \b B \|_{\infty} &\le&
\| (\widehat{\b B} - \b B)' (\widehat{\b\Omega} - \b\Omega) \b B \|_{\infty}/2  \nonumber \\
& + & \| (\widehat{\b B} - \b B )' (\widehat{\b\Omega}' - \b\Omega ) \b B \|_{\infty}/2 \nonumber \\
& \le & \frac{p}{2} \| \widehat{\b B} - \b B \|_{\infty} \| \b B \|_{\infty} \| \widehat{\b\Omega} - \b\Omega \|_{l_{\infty}} \nonumber \\
& + & \frac{p}{2} \| \widehat{\b B} - \b B \|_{\infty} \| \b B \|_{\infty} \| \widehat{\b\Omega}' - \b\Omega \|_{l_1} \nonumber \\
& = & p O_p\left(\sqrt{\frac{ K\ln(p)}{n}}\right)O(1) O_p\left( \bar{s} \lambda_n\right),\label{pla11-3}
\end{eqnarray}
where we use, $\widehat{\b\Omega}_{sym}$,  Lemma \ref{la9}(ii)(iv) for the first-second inequalities, (B.14) of \cite{fan2011}, Assumption \ref{as6}, and Theorem \ref{rthm1} and (\ref{eqeq}) for the rates.
Now consider the third term in (\ref{pla11-1})
\begin{equation}
\| (\widehat{\b B} - \b B)' \b\Omega (\widehat{\b B} - \b B )\|_{\infty}  \le  p \| \widehat{\b B} - \b B \|_{\infty}^2 \| \b\Omega \|_{l_{\infty}} 
 =  p\left[O_p\left(\sqrt{\frac{K \ln(p)}{n}}\right)\right]^2 O \left(\bar{s}^{1/2}\right),\label{pla11-4}
\end{equation}
where we use Lemma \ref{la9}(ii) for the first inequality, (B.14) of \cite{fan2011}, and $\| \b\Omega \|_{l_{\infty}}:=\max_{1 \le j \le p}
\| \b\Omega_j'  \|_1 = \max_{1 \le j \le p} \| \b\Omega_j  \|_1 = O\left(\bar{s}^{1/2}\right)$ as in (\ref{pla36}). We consider the fourth term in (\ref{pla11-1})
\begin{eqnarray}
\|\b B' (\widehat{\b\Omega}_{sym}- \b\Omega ) \b B \|_{\infty} & \le&
\frac{1}{2} \|\b B' (\widehat{\b\Omega} - \b\Omega ) \b B \|_{\infty} + \frac{1}{2} \| \b B' (\widehat{\b\Omega}' - \b\Omega ) \b B \|_{\infty} \nonumber \\
& \le & \frac{p}{2} \| \b B \|_{\infty}^2 \| \widehat{\b\Omega} - \b\Omega \|_{l_{\infty}} + \frac{p}{2}
 \| \b B \|_{\infty}^2 \| \widehat{\b\Omega}' - \b\Omega \|_{l_1} \nonumber \\
& = & p [O(1)]^2 O_p\left(\bar{s} \lambda_n \right),\label{pla11-5}
\end{eqnarray}
where we use symmetry of $\widehat{\b\Omega}_{sym}$, Lemma \ref{la9}(ii)(iv) for the first and second inequality, and Assumption \ref{as6}, and Theorem 1
(\ref{eqeq})
for the rates. Also analyze the fifth term in (\ref{pla11-1})
\begin{equation}
\| (\widehat{\b B} - \b B )' \b\Omega \b B \|_{\infty} \le p \| \widehat{\b B} - \b B \|_{\infty} \| \b B \|_{\infty} \| \b\Omega \|_{l_{\infty}} 
 =  p O_p\left(\sqrt{\frac{K \ln(p)}{n}}\right) O (1) O \left(\bar{s}^{1/2}\right),\label{pla11-6}
\end{equation}
where we use Lemma \ref{la9}(ii) for the inequality, and the rates are by (B.14) of \cite{fan2011}, Assumption \ref{as6}, and
$\| \b\Omega \|_{l_{\infty}}:=\max_{1 \le j \le p}
\| \b\Omega_j'  \|_1 = \max_{1 \le j \le p} \| \b\Omega_j  \|_1 = O\left(\bar{s}^{1/2}\right)$ as in (\ref{pla36}).  The slowest rate is the maximum of the rates (\ref{pla11-5}) and (\ref{pla11-6}) above. So,
\begin{equation}
\| \widehat{\b B}' \widehat{\b\Omega}_{sym} \widehat{\b B} - \b B' \b\Omega \b B \|_{\infty} = O_p \left( p \max(\bar{s} \lambda_n , \bar{s}^{1/2} K^{1/2} \sqrt{\frac{\ln(p)}{n}} ) \right).\label{pla11-7}
\end{equation}
Then, by norm inequality tying spectral norm to $\|.\|_{\infty}$ norm in p.365 of \cite{hj2013}, and since  $\widehat{\b B}' \widehat{\b\Omega}_{sym} \widehat{\b B} - \b B' \b\Omega \b B$
is $K \times K$ matrix
\begin{equation}
\| \widehat{\b B}' \widehat{\b\Omega}_{sym} \widehat{\b B} - \b B' \b\Omega \b B \|_{l_2} \le K \| \widehat{\b B}' \widehat{\b\Omega}_{sym} \widehat{\b B} - \b B' \b\Omega \b B \|_{\infty}
= O_p \left( p  K \max(\bar{s} \lambda_n, \bar{s}^{1/2} K^{1/2} \sqrt{\frac{\ln(p)}{n}})\right).\label{pla11-8}
\end{equation}

\begin{flushright}
{\bf Q.E.D.}
\end{flushright}

(ii). Since $\widehat{\textnormal{cov}(\b f_t)}^{-1}, (\textnormal{cov} (\b f_t))^{-1}$ does not involve the precision matrix estimator, we proceed as in \cite{fan2011}, Lemma B5(ii). Specifically (B.20) of \cite{fan2011} provide
\[ 
\|[\widehat{\textnormal{cov}(\b f_t)}]^{-1} - [\textnormal{cov}(\b f_t)]^{-1} \|_{l_2} = O_p\left(\sqrt{\ln (n)/n}\right).\]
Using (\ref{pla11-8})  and the equation above we develop a larger bound 
\begin{equation}
\| ([\widehat{\textnormal{cov}(\b f_t)}]^{-1} + \widehat{\b B}' \widehat{\b\Omega}_{sym} \widehat{\b B})- ([\textnormal{cov} (\b f_t)]^{-1} + \b B' \b\Omega \b B ) \|_{l_2} = O_p \left( 
p  K \max(\bar{s} \lambda_n , \bar{s}^{1/2} K^{1/2} \sqrt{\frac{\max[\ln(p),\ln(n)]}{n}})\right).\label{pla11-8a}
\end{equation}
Note that
\[  p  K \max\left(\bar{s} \lambda_n , \bar{s}^{1/2} K^{1/2} \sqrt{\frac{\max[\ln(p),\ln(n)]}{n}}\right)= o (p/r_n),\]
since by Assumption \ref{as7}, $r_n  K \max\left(\bar{s} \lambda_n, \bar{s}^{1/2} K^{1/2} \sqrt{\frac{\max[\ln(p),\ln(n)]}{n}}\right)= o(1)$. So (\ref{pla11-8a}) has the rate
\begin{equation}
\| ([\widehat{\textnormal{cov}(\b f_t)}]^{-1} + \widehat{\b B}' \widehat{\b\Omega}_{sym} \widehat{\b B})- ([\textnormal{cov} (\b f_t)]^{-1} + \b B' \b\Omega \b B ) \|_{l_2} = o_p (p/r_n).\label{pla11-9}
\end{equation}
Then using Lemma A.1(i) of \cite{fan2011}, with (\ref{lb42}) and (\ref{pla11-9})
\begin{equation}
\textnormal{Eigmin} ([\widehat{\textnormal{cov}(\b f_t)}]^{-1} + \widehat{\b B}' \widehat{\b\Omega}_{sym} \widehat{\b B}) \ge \frac{c p }{C r_n},\label{pla11-9a}
\end{equation}
wpa1 with $r_n <<p$ as in Assumption \ref{as7}. By (\ref{pla11-9a}), and seeing that for invertible matrix $\b A$, $\textnormal{Eigmax}(\b A^{-1}) = 1 / \textnormal{Eigmin}(\b A)$,
\[ \| ([\widehat{\textnormal{cov}(\b f_t)}]^{-1} + \widehat{\b B}' \widehat{\b\Omega}_{sym} \widehat{\b B})^{-1} \|_{l_2} = O_p \left(\frac{r_n}{p}\right).\]

\begin{flushright}
{\bf Q.E.D.}
\end{flushright}

We restate the definitions of major terms that are used.
\begin{equation}
\widehat{\b G}:= (\widehat{\textnormal{cov}(\b f_t)}^{-1} + \widehat{\b B}' \widehat{\b\Omega}_{sym} \widehat{\b B})^{-1}.\label{pla12-1}
\end{equation}
\begin{equation}
\b G:= (\textnormal{cov}(\b f_t)^{-1} + \b B' \b\Omega \b B)^{-1}.\label{pla12-2}
\end{equation}

Next, remembering
\begin{equation}
\widehat{\b L}:= \widehat{\b B} \widehat{\b G} \widehat{\b B}' \quad \b L:=\b B \b G \b B'.\label{pla11-10}
\end{equation}
and 
\begin{equation}
 l_n:=  r_n^2 K^{5/2} \max\left(\bar{s} \lambda_n, \bar{s}^{1/2} K^{1/2} \sqrt{\frac{\max[\ln(p),\ln(n)]}{n}}\right).\label{ln}
 \end{equation}
 
We have the next lemma which will be instrumental in proving Theorem 2.
\begin{lemma}\label{la12}
Under Assumptions \ref{as1}-\ref{as4}, \ref{as6}-\ref{as7}
\[\| \widehat{\b L} - \b L \|_{l_{\infty}} =O_p (l_n)  = o_p (1).\]
\end{lemma}

\noindent
{\bf Proof of Lemma \ref{la12}}. 
Start with, by adding and subtracting and triangle inequality
\begin{eqnarray}
\| \widehat{\b L} - \b L \|_{l_{\infty}} &\le & \| (\widehat{\b B} - \b B )  \widehat{\b G} (\widehat{\b B} - \b B )' \|_{l_{\infty}} + \| (\widehat{\b B} - \b B ) \widehat{\b G} \b B'\|_{l_{\infty}} \nonumber \\
& + & \| \b B \widehat{\b G} (\widehat{\b B} - \b B)'\|_{l_{\infty}} + \| \b B \widehat{G} \b B' - \b B \b G \b B' \|_{l_{\infty}}.\label{pla12-3}
\end{eqnarray}
Consider the first term in (\ref{pla12-3})
\begin{eqnarray}
\| (\widehat{\b B} - \b B ) \widehat{\b G} (\widehat{\b B} - \b B )' \|_{l_{\infty}} & \le & p K^{1/2} \| \widehat{\b B} - B \|_{l_{\infty}} \| \widehat{\b G} \|_{l_2} \| \widehat{\b B} - \b B \|_{\infty} \nonumber \\
& = & p K^{1/2} O_p \left(K^{3/2} \sqrt{\frac{\ln(p)}{n}}\right) O_p\left(\frac{r_n}{p}\right) O_p\left(\sqrt{\frac{K\ln (p)}{n}}\right) \nonumber \\
& = & O_p\left(r_n K^{5/2} \frac{\ln(p)}{n}\right),\label{pla12-4}
\end{eqnarray}
where we use Lemma \ref{la9}(i) for the first inequality, Lemma \ref{la10}-\ref{la11}, and (B.14) of  \cite{fan2011}: $\| \widehat{\b B} - \b B \|_{\infty} = O_p\left(\sqrt{\frac{K\ln (p)}{n}}\right)$ for the rates. Next in (\ref{pla12-3}), we consider the second term on the right side
\begin{eqnarray}
\| (\widehat{\b B} - \b B) \widehat{\b G} \b B' \|_{l_{\infty}} & \le & p K^{1/2} \| \widehat{\b B} - \b B \|_{l_{\infty}} \| \widehat{\b G} \|_{l_2} \| \b B \|_{\infty} \nonumber \\
& = & p K^{1/2} O_p\left(K^{3/2}\sqrt{\frac{\ln(p)}{n}}\right) O_p\left(\frac{r_n}{p}\right) O(1) = O_p\left(r_n K^2 \sqrt{\frac{\ln(p)}{n}}\right),\label{pla12-5}
\end{eqnarray}
where we use Lemma \ref{la9}(i) for the first inequality, and for the rates  use Lemma \ref{la10}-\ref{la11}, and Assumption \ref{as6} which shows that factor loadings are uniformly bounded away from infinity. Analyze the third term in (\ref{pla12-3}).
\begin{eqnarray}
\| \b B \widehat{\b G} (\widehat{\b B} - \b B)' \|_{l_{\infty}} & \le & p K^{1/2} \| \b B \|_{l_{\infty}} \| \widehat{\b G} \|_{l_2} \| \widehat{\b B} - \b B \|_{\infty} \nonumber \\
& = & p K^{1/2} O (K) O_p\left(\frac{r_n}{p}\right) O_p\left( \sqrt{\frac{K\ln(p)}{n}}\right) = O_p\left( r_n K^2 \sqrt{\frac{\ln(p)}{n}}\right),\label{pla12-6}
\end{eqnarray}
where we use Lemma \ref{la9}(i) for the first inequality, Lemma \ref{la10}-\ref{la11}, and (B.14) of  \cite{fan2011}: $\| \widehat{\b B} - \b B \|_{\infty} = O_p\left(\sqrt{\frac{K \ln (p)}{n}}\right)$ for the rates. Now we analyze the fourth term on the right side of (\ref{pla12-3}).
\begin{equation}
\|\b B (\widehat{\b G} - \b G)\b B' \|_{l_{\infty}} \le p K^{1/2} \| \b B \|_{l_{\infty}} \| \widehat{\b G} - \b G \|_{l_2} \| \b B \|_{\infty},\label{pla12-4}
\end{equation}
where we use Lemma \ref{la9}(i). We have  from (\ref{pla12-1})(\ref{pla12-2})
and by submultiplicativity of $l_2$ matrix norm (spectral norm)

\begin{eqnarray}
\| \widehat{\b G} - \b G \|_{l_2} & \le & \| \widehat{\b G} \|_{l_2} \| \b G \|_{l_2} \| (\widehat{\textnormal{cov}(\b f_t)}^{-1} + \widehat{\b B}' \widehat{\b\Omega}_{sym} \widehat{\b B}) - (\textnormal{cov}(\b f_t)^{-1} + \b B' \b\Omega \b B) \|_{l_2} \nonumber \\
& = & O_p\left(\frac{r_n}{p}\right)O\left(\frac{r_n}{p}\right) O_p\left(pK\max(\bar{s} \lambda_n, \bar{s}^{1/2} K^{1/2} \sqrt{\frac{\max[\ln(p),\ln(n)]}{n}})\right) \nonumber \\
& =& O_p \left( r_n^2 p^{-1} K \max(\bar{s} \lambda_n, \bar{s}^{1/2} K^{1/2} \sqrt{\frac{\max[\ln(p),\ln(n)]}{n}})\right),\label{pla12-5}
\end{eqnarray}
where we use Lemma \ref{la11}, and  $\| \b G \|_{l_2} = O (r_n/p)$ by Lemma \ref{la10a}, (\ref{pla11-8a}). Substitute (\ref{pla12-5}) into (\ref{pla12-4}) via Lemma \ref{la10}
\begin{equation}
\| \b B (\widehat{\b G} - \b G ) \b B' \|_{l_{\infty}} = O_p \left(r_n^2  K^{5/2} \max(\bar{s} \lambda_n, \bar{s}^{1/2} K^{1/2} \sqrt{\frac{\max[\ln(p),\ln(n)]}{n}})\right).\label{pla12-6}
\end{equation}
Since the last rate is the slowest among all  on the right side of (\ref{pla12-3}) we have the desired result.
\begin{flushright}
{\bf Q.E.D.}
\end{flushright}

\noindent
{\bf Proof of Theorem \ref{rthm2}}. From (\ref{sw4}), and using triangle inequality
\begin{equation}
\max_{1 \le j \le p} \| \widehat{\b\Gamma}_j - \Gamma_j \|_1=
\max_{1 \le j \le p } \| \widehat{\b\Gamma}_j' - \b\Gamma_j' \|_1\le \max_{1 \le j \le p} \| \widehat{\b\Omega}_j' - \b\Omega_j'\|_1 +
\max_{1 \le j \le p}
\|\widehat{\b\Omega}_j' \widehat{\b L} \widehat{\b\Omega} - \b\Omega_j' \b L \b\Omega\|_1.\label{pt2-1}
\end{equation}
We consider second right side term in (\ref{pt2-1}). Add and subtract $\b\Omega_j' \widehat{\b L} \widehat{\b\Omega}$ via triangle inequality
\begin{equation}
\max_{1 \le j \le p}
\|\widehat{\b\Omega}_j' \widehat{\b L} \widehat{\b\Omega} - \b\Omega_j' \b L \b\Omega\|_1 \le \max_{1 \le j \le p} \| (\widehat{\b\Omega}_j - \b\Omega_j)' \widehat{\b L} \widehat{\b\Omega} \|_1 +
\max_{1 \le j \le p} \| \b\Omega_j' (\widehat{\b L} \widehat{\b\Omega} - \b L \b\Omega)\|_1.\label{pt2-2}
\end{equation}
We analyze the first term on the right side of (\ref{pt2-2}) and try to simplify by adding and subtracting $(\widehat{\b\Omega}_j - \b\Omega_j)' \widehat{\b L} \b\Omega$, and triangle inequality
\begin{equation}
\max_{1 \le j \le p} \| (\widehat{\b\Omega}_j - \b\Omega_j)' \widehat{\b L} \widehat{\b\Omega} \|_1 \le \max_{1 \le j \le p} \| (\widehat{\b\Omega}_j - \b\Omega_j)' \widehat{\b L} (\widehat{\b\Omega}- \b\Omega)\|_1
+ \max_{1 \le j \le p} \| (\widehat{\b\Omega}_j - \b\Omega_j)' \widehat{\b L} \b\Omega \|_1.\label{pt2-3}
\end{equation}
Then on the first right side term in (\ref{pt2-3}) add and subtract $(\widehat{\b\Omega}_j - \b\Omega_j)' \b L (\widehat{\b\Omega} - \b\Omega)$ via triangle inequality
\[
\max_{1 \le j \le p} \| (\widehat{\b\Omega}_j - \b\Omega_j)' \widehat{\b L} (\widehat{\b\Omega} - \b\Omega) \|_1 \le \max_{1 \le j \le p} \| (\widehat{\b\Omega}_j - \b\Omega_j)' (\widehat{\b L} - \b L ) (\widehat{\b\Omega} - \b\Omega ) \|_1 + \max_{1 \le j \le p} \| (\widehat{\b\Omega}_j - \b\Omega_j)' \b L (\widehat{\b\Omega} - \b\Omega) \|_1.
\]
Now for the second right side term in (\ref{pt2-3}) add and subtract $(\widehat{\b\Omega}_j - \b\Omega_j)' \b L  \b\Omega$ via triangle inequality
\[
\max_{1 \le j \le p} \| (\widehat{\b\Omega}_j- \b\Omega_j)' \widehat{\b L} \b\Omega \|_1 \le  \max_{1 \le j \le p} \| (\widehat{\b\Omega}_j - \b\Omega_j)' (\widehat{\b L} - \b L ) \b\Omega \|_1 +
\max_{1 \le j \le p} \| (\widehat{\b\Omega}_j - \b\Omega_j)' \b L \b\Omega \|_1.
\]
Substitute the last two inequalities into (\ref{pt2-3})
\begin{eqnarray}
\max_{1 \le j \le p} \| (\widehat{\b\Omega}_j - \b\Omega_j)' \widehat{\b L} \widehat{\b\Omega} \|_1 &\le &
\max_{1 \le j \le p} \| (\widehat{\b\Omega}_j - \b\Omega_j)' (\widehat{\b L} - \b L ) (\widehat{\b\Omega} - \b\Omega ) \|_1 \nonumber \\
&  + & \max_{1 \le j \le p} \| (\widehat{\b\Omega}_j - \b\Omega_j)' \b L (\widehat{\b\Omega} - \b\Omega) \|_1 \nonumber \\
& + & \max_{1 \le j \le p} \| (\widehat{\b\Omega}_j - \b\Omega_j)' (\widehat{\b L} - \b L ) \b\Omega \|_1 \nonumber \\
&+&
\max_{1 \le j \le p} \| (\widehat{\b\Omega}_j - \b\Omega_j)' \b L \b\Omega \|_1.\label{pt2-4}
\end{eqnarray}
Now in (\ref{pt2-2}) we consider the second term on the right side, add and subtract $ \b\Omega_j' \b L \widehat{\b\Omega} $ via triangle inequality
\begin{equation}
\max_{1 \le j \le p} \| \b\Omega_j' \widehat{\b L} \widehat{\b\Omega} - \b\Omega_j' \b L \b\Omega \|_1 \le \max_{1 \le j \le p}
\| \b\Omega_j' (\widehat{\b L} - \b L ) \widehat{\b\Omega} \|_1 + \max_{1 \le j \le p} \| \b\Omega_j' \b L (\widehat{\b\Omega} - \b\Omega ) \|_1.\label{pt2-5}
\end{equation}
Also add and subtract $\b\Omega_j' (\widehat{\b L} - \b L ) \b\Omega$ to the first term on the right side of (\ref{pt2-5}) above, to have
\begin{eqnarray}
\max_{1 \le j \le p} \| \b\Omega_j' \widehat{\b L} \widehat{\b\Omega} - \b\Omega_j' \b L \b\Omega \|_1 &\le & \max_{1 \le j \le p} \| \b\Omega_j' (\widehat{\b L} - \b L ) (\widehat{\b\Omega} - \b\Omega ) \|_1 \nonumber \\
& + & \max_{1 \le j \le p} \| \b\Omega_j' (\widehat{\b L} - \b L ) \b\Omega \|_1 \nonumber \\
& + & \max_{1 \le j \le p} \| \b\Omega_j' \b L (\widehat{\b\Omega} - \b\Omega) \|_1.\label{pt2-6}
\end{eqnarray}
Combine (\ref{pt2-4})(\ref{pt2-6}) into (\ref{pt2-2}) right side to have
\begin{eqnarray}
\max_{1 \le j \le p}
\|\widehat{\b\Omega}_j' \widehat{\b L} \widehat{\b\Omega} - \b\Omega_j' \b L \b\Omega\|_1 &\le &
\max_{1 \le j \le p} \| (\widehat{\b\Omega}_j - \b\Omega_j)' (\widehat{\b L} - \b L ) (\widehat{\b\Omega} - \b\Omega ) \|_1 \nonumber \\
&  + & \max_{1 \le j \le p} \| (\widehat{\b\Omega}_j - \b\Omega_j)' \b L (\widehat{\b\Omega} - \b\Omega) \|_1 
 +  \max_{1 \le j \le p} \| (\widehat{\b\Omega}_j - \b\Omega_j)' (\widehat{\b L} - \b L ) \b\Omega \|_1 \nonumber \\
&+&
\max_{1 \le j \le p} \| (\widehat{\b\Omega}_j - \b\Omega_j)' \b L \b\Omega \|_1 
 + 
\max_{1 \le j \le p} \| \b\Omega_j' (\widehat{\b L} - \b L ) (\widehat{\b\Omega} - \b\Omega ) \|_1 \nonumber \\
& + & \max_{1 \le j \le p} \| \b\Omega_j' (\widehat{\b L} - \b L ) \b\Omega \|_1 
 +  \max_{1 \le j \le p} \| \b\Omega_j' \b L (\widehat{\b\Omega} - \b\Omega) \|_1.\label{pt2-7}
\end{eqnarray}

To consider all the terms in (\ref{pt2-7}) we need to find some rates about terms.  In that respect,
\begin{eqnarray}
\| \b L \|_{l_{\infty}} & :=& \| \b B \b G \b B' \|_{l_{\infty}} 
 \le  \| \b B \|_{l_{\infty}} \| \b B \|_{l_1} \| \b G \|_{l_{\infty}}  \nonumber \\
& \le & \| \b B \|_{l_{\infty}} \| \b B \|_{l_1}  [K^{1/2} \| \b G \|_{l_2}] \nonumber \\
& = & O	 (K)  O(p) K^{1/2} O\left(\frac{r_n}{p}\right) =  O\left(r_n K^{3/2}\right),\label{pt2-8}
\end{eqnarray}
where we use definition of $\b L$ for the first equality in (\ref{pla11-10}), $\b G$ is defined in (\ref{pla12-2}), and we use submultiplicativity of $l_{\infty}$ norm for the first inequality, and the relation between spectral norm and $l_{\infty}$ norm from p.365 of \cite{hj2013} for the second inequality, and the rates are from (\ref{pla10-1}), Lemma \ref{la10},  Lemma \ref{la10a} and $G$ definition. Next we need the following, by using the same analysis in (B.55) of \cite{canerkock2018} via strict stationary of the data, or (\ref{pla36}) here
\begin{equation}
\| \b\Omega \|_{l_{\infty}}: = \max_{1 \le j \le p} \| \b\Omega_j' \|_1= \max_{1 \le j \le p } \| \b\Omega_j \|_1 = O (\sqrt{\bar{s}}).\label{pt2-9}
\end{equation}

We consider each term on the right side of (\ref{pt2-7}).
\begin{eqnarray}
max_{1 \le j \le p} \| (\widehat{\b\Omega}_j - \b\Omega_j)' (\widehat{\b L} - \b L ) (\widehat{\b\Omega} - \b\Omega ) \|_1
& \le & [ \max_{1 \le j \le p} \| \widehat{\b\Omega}_j - \b\Omega_j\|_1^2] \| \widehat{\b L} - \b L \|_{l_{\infty}}\label{pt2-9a} \\
& = & [O_p\left(\bar{s} \lambda_n\right)]^2 
   O_p (l_n),\label{pt2-10}
\end{eqnarray}
where we use Lemma \ref{la9}(iii), and
\begin{equation}
\| \widehat{\b\Omega} - \b\Omega \|_{l_{\infty}} := \max_{1 \le j \le p} \| \widehat{\b\Omega}_j' - \b\Omega_j' \|_1 = \max_{1 \le j \le p} \|\widehat{\b\Omega}_j - \b\Omega_j \|_1,\label{pt2-11}
\end{equation}
for the inequality in (\ref{pt2-9a}) and use Lemma \ref{la12}, and Theorem \ref{rthm1} for the rates.

We consider the second term on the right side of (\ref{pt2-7}).
\begin{eqnarray}
max_{1 \le j \le p} \| (\widehat{\b\Omega}_j - \b\Omega_j)' \b L  (\widehat{\b\Omega} - \b\Omega ) \|_1
& \le & [ \max_{1 \le j \le p} \| \widehat{\b\Omega}_j - \b\Omega_j\|_1^2] \| \b L \|_{l_{\infty}} \nonumber \\
&=&  [O_p\left(\bar{s} \lambda_n \right)]^2 
    O_p (r_n K^{3/2}),\label{pt2-12}
\end{eqnarray}
where we use Lemma \ref{la9}(iii), and  (\ref{pt2-11})
for the inequality in (\ref{pt2-12}) and use (\ref{pt2-8}), and Theorem \ref{rthm1} for the rates. We analyze the third term on the right side of (\ref{pt2-7})
\begin{eqnarray}
\max_{1 \le j \le p} \| (\widehat{\b\Omega}_j - \b\Omega_j)' (\widehat{\b L} - \b L ) \b\Omega \|_1 & \le &[ \max_{1 \le j \le p} \| \widehat{\b\Omega}_j - \b\Omega_j \|_1] \, \| \widehat{\b L} - \b L \|_{l_{\infty}}
\, \| \b\Omega \|_{l_{\infty}} \nonumber \\
& = & O_p\left(\bar{s} \lambda_n\right) 
  O_p ( l_n) O (\bar{s}^{1/2}),\label{pt2-13}
\end{eqnarray}
where we use Lemma \ref{la9}(iii) for the first inequality, 
  and the rates are by (\ref{pt2-9}), Lemma \ref{la12}, Theorem 1. Now consider the fourth term on the right side of (\ref{pt2-7})
\begin{eqnarray}
\max_{1 \le j \le p} \| (\widehat{\b\Omega}_j - \b\Omega_j)' \b L \b\Omega \|_1 & \le & \max_{1 \le j \le p} \| \widehat{\b\Omega}_j - \b\Omega_j \|_1 \| \b L \|_{l_{\infty}}
\| \b\Omega \|_{l_{\infty}} \nonumber \\
& = & O_p\left(\bar{s} \lambda_n \right) O_p\left(r_n K^{3/2}\right)O\left(\bar{s}^{1/2}\right),\label{pt2-14}
\end{eqnarray}
where we use Lemma \ref{la9}(iii) for the inequality, and Theorem \ref{rthm1}, (\ref{pt2-8})(\ref{pt2-9}) for the rate. Now consider the fifth term on the right side of (\ref{pt2-7}).
\begin{eqnarray}
\max_{1 \le j \le p } \| \b\Omega_j' (\widehat{\b L} - L ) (\widehat{\b\Omega} - \b\Omega) \|_1 & \le & [ \max_{1 \le j \le p } \| \b\Omega_j \|_1]
\| \widehat{\b L} - \b L \|_{l_{\infty}} \| \widehat{\b\Omega} - \b\Omega \|_{l_{\infty}} \nonumber \\
& = & O (\bar{s}^{1/2}) O_p ( l_n) 
  O_p \left( \bar{s} \lambda_n \right),\label{pt2-15}
\end{eqnarray}
where we use lemma \ref{la9}(iii) for the first inequality, and Theorem \ref{rthm1}, Lemma \ref{la12}, (\ref{pt2-9}) for the rates. Consider the sixth term on the right side of (\ref{pt2-7})
\begin{equation}
\max_{1 \le j \le p } \| \b\Omega_j' (\widehat{\b L} - \b L ) \b\Omega \|_1  \le  [ \max_{1 \le j \le p} \| \b\Omega_j \|_1^2] \| \widehat{\b L} - \b L \|_{l_{\infty}} 
 =  O (\bar{s}) O_p ( l_n),\label{pt2-16}
\end{equation}
where we use Lemma \ref{la9}(iii) for the inequality, and use (\ref{pt2-9}), and Lemma \ref{la12} for the rates. Now analyze the seventh term on the right side of (\ref{pt2-7})
\begin{eqnarray}
\max_{1 \le j \le p} \| \b\Omega_j' \b L (\widehat{\b\Omega} - \b\Omega) \|_1 & \le & [ \max_{1 \le j \le p} \| \b\Omega_j \|_1] \| \b L \|_{l_{\infty}} \| \widehat{\b\Omega} - \b\Omega \|_{l_{\infty}} \nonumber \\
& = & O (\bar{s}^{1/2}) O_p ( r_n K^{3/2}) O_p\left(\bar{s} \lambda_n \right),\label{pt2-17}
\end{eqnarray}
where we use Lemma \ref{la9}(iii) for the inequality, and for the rates we use (\ref{pt2-8})(\ref{pt2-9}) Theorem 1. Note that among all (\ref{pt2-10})-(\ref{pt2-17}), the slowest rate is
by (\ref{pt2-16}) by the definition of  $l_n$ in (\ref{ln}) and by Assumption \ref{as7}, with $\bar{s} l_n \to 0$.

So we have by Assumption \ref{as7}
\begin{equation}
\max_{1 \le j \le p} \| \widehat{\b\Omega}_j' \widehat{\b L} \widehat{\b\Omega} - \b\Omega_j' L \b\Omega \|_1 = O_p ( \bar{s} l_n ) = o_p (1).\label{pt2-18}
\end{equation}
This ends the proof of  (i) with using Theorem 1 and (\ref{pt2-18}) in (\ref{pt2-1}).

(ii). Since
\[ 
\b y_t = \b B \b f_t + \b u_t,
\]
as in \cite{fan2011} with $y_t,u_t$ being $p \times 1$ vector of asset returns, and errors respectively at time $t=1,\cdots, n$.
\[ \widehat{\b \mu} - \b \mu = \b B \left[ \frac{1}{n} \sum_{t=1}^n (\b f_t - \E[\b f_t]) \right] + \frac{1}{n} \sum_{t=1}^n \b u_t,\]
by Assumption \ref{as1}. Consider

\begin{eqnarray*}
 \left\| \widehat{\b \mu} - \b \mu \right\|_{\infty} &\le& \left\| \b B \frac{1}{n} \sum_{t=1}^n (\b f_t - \E[\b f_t])\right\|_{\infty} + \left\| \frac{1}{n} \sum_{t=1}^n \b u_t \right\|_{\infty} \\
 & \le & \left\| \b B \right\|_{l_{\infty}} \left\| \frac{1}{n} \sum_{t=1}^n (\b f_t - \E[\b f_t]) \right\|_{\infty} + \left\| \frac{1}{n} \sum_{t=1}^n \b u_t \right\|_{\infty} \\
 & = & O (K) O_p\left(\sqrt{\ln(n)/n}\right) + O_p\left( \sqrt{\ln(p)/n}\right),
 \end{eqnarray*}
Clearly, by the proof of Lemma \ref{la1}(i) here we have $\|\b A \b x\|_{\infty} \le \| \b A \|_{l_{\infty}} \| \b x \|_{\infty}$ for a generic vector $\b x$, and a matrix $\b A$. Then,
by Lemma \ref{la10}(iii) and Theorem A.1 we get the rate.

\begin{flushright}
{\bf Q.E.D.}
\end{flushright}

\subsection*{Part 4}

First, we start with a maximal eigenvalue bound which will be used in the proof of Theorem 8.
Here, we provide the rate for maximal eigenvalue of covariance matrix of returns $\b\Sigma_y$. See that 
\[ \textnormal{Eigmax}(\b\Sigma_y) \le \textnormal{Eigmax}[\b B \textnormal{cov}(\b f) \b B'] + \textnormal{Eigmax}( \b\Sigma_n) \le \textnormal{Eigmax}(\textnormal{cov}(\b f)) \textnormal{Eigmax}(\b B \b B') + \textnormal{Eigmax}(\b\Sigma_n).\]
Since by Assumption \ref{as7}, $r_n/p \to 0$, and $\textnormal{Eigmax} (\b\Sigma_n) \le C r_n$, with the above inequality and specifically by Assumption \ref{as6}(ii), with Proposition 2.1 of \cite{fan2013}

\begin{equation}
 \textnormal{Eigmax} (\b\Sigma_y) = O (p).\label{maxe}
 \end{equation}
This is true for $\textnormal{cov}(f) = \b I_k$ in \cite{fan2013}. The result holds for general $\textnormal{cov}(\b f)$ as discussed in section 2.1 of \cite{fan2013}. 

Then we provide  the proofs of Theorems 3-7.

\noindent
{\bf Proof of Theorem \ref{gmv}}. First, we start with definitions of $\widehat{\b A}:=  \b 1_p' \widehat{\b \Gamma} \b 1_p/p$, $\widehat{\b F}:=1_p' \widehat{\b \Gamma} \widehat{\b \mu}/p$,
$\b A := \b 1_p' \b\Sigma_y^{-1} \b 1_p/p$, $\b F := \b 1_p' \b\Sigma_y^{-1} \b \mu/p$.

\begin{eqnarray}
\left| \frac{\widehat{SR}_{nw}^2}{SR^2} - 1
\right| & = & \left|
\frac{p(\b 1_p' \widehat{\b \Gamma} \widehat{\b \mu}/p)^2 (\b 1_p' \widehat{\b \Gamma} \b 1_p/p)^{-1}}{p(\b 1_p' \b\Sigma_y^{-1}\mu/p)^2 (\b 1_p' \b\Sigma_y^{-1} \b 1_p/p)^{-1}} - 1
\right| \nonumber \\
& = & \left|  \frac{\widehat{F}^2  A}{F^2 \widehat{A} } -1
\right| 
 = 
\left|  \frac{  (\widehat{F}^2 A - F^2 \widehat{A} )}{  F^2 \widehat{A} }
\right|. \label{psr1}
\end{eqnarray}


\noindent Now consider the numerator in (\ref{psr1}):
\begin{eqnarray}
| \widehat{F}^2  A - F^2 \widehat{A} | & = & | \widehat{F}^2  A - \widehat{F}^2 \widehat{A} + \widehat{F}^2 \widehat{A} - F^2 \widehat{A} | \nonumber \\
& \le & | \widehat{F}^2 ( \widehat{A} - A ) | + | (\widehat{F}^2 - F^2) \widehat{A} | 
 \le  | \widehat{F}^2 ( \widehat{A} - A )| + | \widehat{F} - F| | \widehat{F} + F| |\widehat{A}|.\label{psr3}
 \end{eqnarray}

\noindent Analyze the first term on the right side of (\ref{psr3}):
\begin{eqnarray}
 \widehat{F}^2 &=& | \widehat{F}^2 - F^2 + F^2|  \nonumber \\
 & \le & | \widehat{F}^2 - F^2 | + F^2 
  \le  | \widehat{F} - F | | \widehat{F} + F | + F^2.\label{psr4}
  \end{eqnarray}

\noindent Then, by Lemma \ref{tl3} in Supplement B, via Assumption \ref{as8}
\begin{equation}
| \widehat{F} - F | = O_p ( K \bar{s} l_n) = o_p(1).\label{psr5a}
\end{equation}
Then,
\begin{eqnarray}
| \widehat{F} + F |& \le& | \widehat{F} | + |F| 
 \le  | \widehat{F} - F | + 2 | F| \nonumber \\
& = & o_p (1) + O( K^{1/2}) 
 =  O_p (K^{1/2}),\label{psr5b}
\end{eqnarray}
where we use (\ref{psr5a}) and Lemma \ref{tl5} in Supplement B.  By (\ref{psr5a})(\ref{psr5b}) and Lemma \ref{tl5} in (\ref{psr4}), we have
\begin{equation}
\widehat{F}^2 = O_p (K).\label{psr6}
\end{equation}
Then, by Lemma \ref{tl2} in Supplement B and (\ref{psr6}),
\begin{equation}
| \widehat{F}^2 (\widehat{A} - A ) | \le \widehat{F}^2 | \widehat{A} - A | = O_p ( K \bar{s} l_n) = o_p (1).\label{psr7}
\end{equation}

\noindent Then, the second term on the right side of (\ref{psr3}) is
\begin{equation}
| \widehat{F} - F | | \widehat{F} + F |  |\widehat{A}| = O_p ( K \bar{s} l_n) O_p (K^{1/2}) O_p (1) = O_p (K^{3/2} \bar{s} l_n)= o_p (1),\label{psr8}
\end{equation}
by (\ref{psr5a})(\ref{psr5b}) and Lemma \ref{tl2}, Lemma \ref{tl5} in  Supplement B, and the last equality is by Assumption \ref{as8}.
Use (\ref{psr7})(\ref{psr8}) in (\ref{psr3}) with Assumption \ref{as8}
\begin{equation}
 | \widehat{F}^2  A - F^2 \widehat{A} | = O_p (K^{3/2} \bar{s} l_n) = o_p (1).\label{psr9}
\end{equation}
Now consider the denominator in (\ref{psr1}). Note that
\[   F^2 \widehat{A} =  F^2 (\widehat{A}- A) + F^2 A \ge  F^2 A -   | F^2 (\widehat{A} - A)|.\]
So by Assumption \ref{as8}(ii)
\begin{equation}
 F^2 A \ge C^2 c >0.\label{a60}
\end{equation}

Next
\begin{equation}
 | \widehat{F}^2 (\widehat{A} - A )| = O_p ( K \bar{s} l_n) = o_p (1).\label{a60a}
\end{equation}
by (\ref{psr7}) and Assumption \ref{as8}. Combine (\ref{psr9}) with (\ref{a60})(\ref{a60a}) in (\ref{psr1}) to obtain the desired result.
\begin{flushright}
{\bf Q.E.D.}
\end{flushright}

\noindent
{\bf Proof of Theorem \ref{mmv}}. To ease the notation in the proofs, set $AD - F^2 =z$, $A \rho_1^2 - 2 F \rho_1 + D =v$. The estimates will be $\widehat{z}= \widehat{A} \widehat{D} - \widehat{F}^2$, $\widehat{v}=\widehat{A} \rho_1^2 - 2 \widehat{F} \rho_1 + \widehat{D}$.
Then,
\begin{eqnarray}
 \left|  \frac{\widehat{SR}_{mv}^2}{SR_{mv}^2} - 1
 \right| & = & \left|
 \frac{ \widehat{z}/\widehat{v}}{z/ v} -1
  \right|
 \nonumber \\
 & = & \left| \frac{\widehat{z}}{\widehat{v}} \frac{v}{z} - 1
 \right| 
  =  \left|  \frac{ \widehat{z} v - \widehat{v} z }{\widehat{v} z }
 \right|.\label{ptmv1}
  \end{eqnarray}

\noindent First, analyze the denominator of (\ref{ptmv1}).
\begin{eqnarray}
 | \widehat{v} z | &=& |  (\widehat{v} - v) z + vz|. \nonumber \\
 & \ge & | v z | - | (\widehat{v} - v) z | 
 \ge  | v z | - | \widehat{v} - v | |z|. \label{ptmv3a}
 \end{eqnarray}
Then, by Lemma \ref{tl2}-\ref{tl4} in Supplement B, triangle inequality  and $\rho_1$ being bounded away from zero and finite, by Assumption \ref{as8},
\begin{equation}
  | \widehat{v} - v | = | (\widehat{A} - A ) \rho_1^2  - 2 (\widehat{F} - F ) \rho_1 + (\widehat{D} - D )| = O_p ( K^2 \bar{s} l_n)= o_p (1).\label{ptmv3b}
  \end{equation}
We also know that by the conditions in theorem statement $z= AD - F^2 \ge C_1 > 0$, and $ v= A \rho_1^2 - 2 F  \rho_1 + D \ge C_1 > 0$.
Then, see that by Lemma \ref{tl5} in Supplement B
\begin{equation}
|z| = | AD - F^2| \le A D = O(K).\label{a54a}
\end{equation}
Thus, by (\ref{ptmv3b})(\ref{a54a}) and $ z \ge C_1 > 0, v \ge C_1 >0$ with Assumption \ref{as8}: $K^3 \bar{s} l_n \to 0$ in (\ref{ptmv3a}), we have
\begin{equation}
| \widehat{v} z | = o_p (1) + C_1^2 > 0.\label{ptmv2}
\end{equation}

\noindent Consider the numerator in (\ref{ptmv1}):
\begin{equation}
| \widehat{z} v - \widehat{v} z  | = | \widehat{z} v - vz+ vz - \widehat{v} z| \le | \widehat{z} - z | |v| + | z | | \widehat{v} - v |.\label{ptmv0}
\end{equation}
By Lemma \ref{tl6} in Supplement B, and Assumption \ref{as8}
\begin{equation}
| \widehat{z} - z | = | (\widehat{A} \widehat{D} - \widehat{F}^2 ) - (AD - F^2)| = O_p ( K^2 \bar{s} l_n) = o_p (1).\label{ptmv3aa}
\end{equation}
Clearly, by Lemma \ref{tl5} in Supplement B and triangle inequality with $\rho_1$ being finite,
\begin{equation}
| v | = | A \rho_1 -2 F \rho_1 + D | = O(K).\label{ptm3g}
\end{equation}
Then, use (\ref{ptmv3b})(\ref{a54a})(\ref{ptmv3aa})(\ref{ptm3g}) in (\ref{ptmv0}) by Assumption \ref{as8}
\begin{equation}
| \widehat{z} v - \widehat{v} z | = O_p ( K^3 \bar{s} l_n) = o_p (1).\label{ptm3h}
\end{equation}
Use (\ref{ptmv2})(\ref{ptm3h}) in (\ref{ptmv1}) to obtain the desired result.
\begin{flushright}
{\bf Q.E.D.}
\end{flushright}

\noindent
{\bf Proof of Theorem \ref{nwsr}}. See that
\[ \left|  \frac{\widehat{MSR}^2/p}{MSR^2/p} - 1
\right| =
\left| \frac{\widehat{\b \mu}' \widehat{\b \Gamma} \widehat{\b \mu}/p}{{\b \mu'} {\b  \Gamma} {\b \mu} /p} - 1 \right| = | \frac{\widehat{\b \mu}' \widehat{\b \Gamma} \widehat{\b \mu}/p - {\b \mu'} {\b \Gamma} {\b  \mu}/p|}{{\b \mu}' {\b \Gamma}{\b  \mu}/p }
.\]

\noindent Lemma \ref{tl4} in Supplement B  shows that
\begin{equation}
| \widehat{\b \mu}' \widehat{\b \Gamma} \widehat{\b \mu}/p - \widehat{\b \mu'}\widehat{\b  \Gamma}\widehat{\b \mu}/p| = O_p( K^2 \bar{s} l_n).\label{pt2.3}
\end{equation}
Combining (\ref{d4}),(\ref{pt2.3}) with Assumption \ref{as8},
\[  \left| \frac{\widehat{\b \mu}' \widehat{\b \Gamma} \widehat{\b \mu}/p}{\widehat{\b \mu'}\widehat{\b  \Gamma}\widehat{\b  \mu}/p } - 1 \right| = O_p ( K^2 \bar{s} l_n) = o_p (1).\]

\begin{flushright}
{\bf Q.E.D.}
\end{flushright}

\noindent
{\bf Proof of Theorem \ref{tme1}}. Note that by the definition of $MSR_c$ in (\ref{e1}) and $A,F,D$ terms,
\[ \frac{MSR_c^2}{p} = D - (F^2/ A),\]
and the estimate is
\[ \frac{\widehat{MSR}_c^2}{p} = \widehat{D} - (\widehat{F}^2/\widehat{A}),\]
where $\widehat{A} = \b 1_p' \widehat{\b \Gamma} \b 1_p/p$, $\widehat{F} = \b 1_p' \widehat{\b \Gamma} \widehat{\b \mu}/p$, $\widehat{D} = \widehat{\b \mu}' \widehat{\b \Gamma} \widehat{\b \mu}/p$.
Then, clearly
\begin{equation}
\frac{\frac{\widehat{MSR}_c^2}{p}}{\frac{MSR_c^2}{p}} = \left[ \frac{\widehat{A} \widehat{D} - \widehat{F}^2}{AD - F^2}
\right] \left[  \frac{A}{\widehat{A}}
\right].\label{ptme11}
\end{equation}
We start with
\begin{equation}
| \widehat{A} - A | = O_p ( \bar{s} l_n)=o_p (1),\label{a6a}
\end{equation}
by Lemma \ref{tl2} in Supplement B.
Then by Assumption \ref{as8}
\begin{equation}
A \ge c > 0.\label{a32a}
\end{equation}
Thus, clearly we obtain, since $|\widehat{A}| \ge A - | \widehat{A}- A|$, 
\[ \left| \frac{A}{\widehat{A}} - 1
\right| = \left| \frac{A - \widehat{A}}{\widehat{A}}
\right| \le \frac{ | \widehat{A} - A |}{ [A - | \widehat{A} - A |]} = \frac{O_p ( \bar{s} l_n )}{1/C - O_p ( \bar{s} l_n)} 
\]
which implies  for the denominator 
\begin{equation}
| \frac{A}{\widehat{A}} - 1| =  O_p ( \bar{s} l_n)=o_p (1).\label{ptme12}
\end{equation}

Next, Lemma \ref{tl6} in  Supplement B  establishes that
\[ |(\widehat{A} \widehat{D} - \widehat{F}^2 ) - ( A D - F^2)| = O_p ( K^2 \bar{s} l_n) = o_p (1).\]
We can use the condition that $AD - F^2 \ge C_1 > 0$, and thus we combine the results above to obtain
\begin{equation}
\left| \frac{ \widehat{A} \widehat{D} - \widehat{F}^2}{AD - F^2} - 1\right| = O_p ( K^2 \bar{s} l_n)=o_p (1).\label{ptme13}
\end{equation}
Since
\[
\frac{\frac{\widehat{MSR}_c^2}{p}}{\frac{MSR_c^2}{p}} =
\left[ \left( \frac{\widehat{A} \widehat{D} - \widehat{F}^2}{AD - F^2} - 1\right) + 1 \right]
\left[ \left( \frac{A}{\widehat{A}} -1 \right) + 1 \right]
\]
Combine (\ref{ptme12})(\ref{ptme13}) in (\ref{ptme11}) to obtain
\begin{eqnarray}
\left|  \frac{\widehat{MSR}_c^2/p}{MSR_c^2/p} - 1
\right| &\le& \left| \frac{ \widehat{A} \widehat{D} - \widehat{F}^2}{AD - F^2} - 1\right|
\left| \frac{A}{\widehat{A}} - 1\right|  \nonumber \\
&+& \left| \frac{A}{\widehat{A}} - 1\right| + \left| \frac{ \widehat{A} \widehat{D} - \widehat{F}^2}{AD - F^2} - 1\right| \\
& = & O_p ( K^2 \bar{s} l_n) = o_p (1),\label{ptme14}
\end{eqnarray}
where the rate is the slowest among the three right-hand-side terms. 

\begin{flushright}
{\bf Q.E. D}
\end{flushright}

\noindent
{\bf Proof of Theorem \ref{tme2}}. Note that we define $\b\Gamma: \b\Sigma_y^{-1}$.
We need to start with
\begin{equation}
\left|  \frac{(\widehat{MSR}^*)^2/p}{(MSR^*)^2/p} - 1
\right| = \frac{\left| (\widehat{MSR}^*)^2/p - (MSR^*)^2/p
\right|}{(MSR^*)^2/p}\label{ptme21}
\end{equation}

Define the event $E_1 = \{|\b 1_p' \widehat{\b \Gamma} \widehat{\b \mu}/p - \b 1_p'  \b \Gamma \b \mu /p | \le \epsilon \}$, where $\epsilon > 0$. We condition the proof on event $E_1$, then at the end of the proof we show that $\P(E_1)\to 1$. Start with the condition $\b 1_p' \Gamma \mu / p \ge  C > 2 \epsilon > 0$; 
\begin{eqnarray}
\frac{\b 1_p' \widehat{\b \Gamma} \widehat{\b \mu}}{p} &=& \frac{\b 1_p' \widehat{\b \Gamma} \widehat{\b \mu}}{p} - \frac{\b 1_p' \b \Gamma \b \mu }{p} + \frac{\b 1_p' \b \Gamma \b \mu }{p} \nonumber \\
& \ge &  \frac{\b 1_p' \b \Gamma \b \mu }{p} - | \frac{\b 1_p' \widehat{\b \Gamma} \widehat{\b \mu}}{p} - \frac{\b 1_p' \b \Gamma \b \mu }{p} | \nonumber \\
& \ge &  \frac{\b 1_p' \b \Gamma \b \mu }{p}  - \epsilon \nonumber \\
& \ge &  C - \epsilon> 2 \epsilon - \epsilon = \epsilon > 0, \label{ptme27}
\end{eqnarray}
where we use $E_1$ in the second inequality and the condition for the third inequality. This clearly shows that at event $E_1$, when the condition
$\b 1_p' \b \Gamma \b \mu /p \ge  C > 2 \epsilon > 0$ holds, we have $\b 1_p' \widehat{\b \Gamma} \widehat{\b \mu} / p >  \epsilon > 0$. So
\begin{equation}
p^{-1}[ \widehat{MSR}^2 1_{ \{\b 1_p' \widehat{\b \Gamma} \widehat{\b \mu} > 0\}} - MSR^2 1_{ \{ \b 1_p' \b \Gamma \b \mu > 0 \}}]
=p^{-1}[ \widehat{MSR}^2 - MSR^2],\label{npt6-1}
\end{equation}
as used in the maximum Sharpe Ratios of Theorem \ref{nwsr}.



We consider $\b 1_p' \b \Gamma \b \mu / p \le - C < - 2 \epsilon < 0$.
Assume that we use event $E_1$:
\begin{eqnarray}
\frac{\b 1_p' \b \Gamma \b \mu}{p} & = & \frac{\b 1_p' \b \Gamma \b \mu}{p} - \frac{\b 1_p' \widehat{\b \Gamma} \widehat{\b \mu}}{p} + \frac{1_p' \widehat{\b \Gamma} \widehat{\b \mu}}{p} \nonumber \\
& \ge &
 \frac{\b 1_p' \widehat{\b \Gamma} \widehat{\b \mu}}{p} - | \frac{\b 1_p' \b \Gamma \b \mu}{p} - \frac{\b 1_p' \widehat{\b \Gamma} \widehat{\b \mu}}{p} | \nonumber \\
 & \ge & \frac{\b 1_p' \widehat{\b \Gamma} \widehat{\b \mu}}{p} - \epsilon.\label{ptme211}
 \end{eqnarray}
Then, in (\ref{ptme211}), using the condition $ \b 1_p' \b \Gamma \b \mu / p \le -C < - 2 \epsilon < 0$ (note that this also implies $\b 1_p' \b \Gamma \b \mu /p < 0$)
\[ 0 > -2 \epsilon > -C  \ge \b 1_p' \b \Gamma \b \mu / p \ge 1_p' \widehat{\b \Gamma} \widehat{\b \mu}/p - \epsilon,\]
which implies that, with $C > 2 \epsilon$, adding $\epsilon$ to all sides above yields
\[ -\epsilon > -(C - \epsilon)
 \ge \b 1_p' \widehat{\b \Gamma} \widehat{\b \mu}/p,\]
which clearly shows that when $1_p' \b \Gamma \b \mu / p < 0$, we will have $1_p' \widehat{\b \Gamma} \widehat{\b \mu} / p < 0$, since $-\epsilon <0$. So,
\begin{equation}
p^{-1} [ \widehat{MSR}_c^2 1_{\{\b 1_p' \widehat{\b \Gamma} \widehat{\b \mu} < 0\}} - MSR_c^2 1_{\{ \b 1_p' \b \Gamma \b \mu < 0\} }
= p^{-1} [ \widehat{MSR}_c^2 - MSR_c^2],\label{npt6-2}
\end{equation}
as in the maximum Sharpe Ratios in Theorem \ref{tme1}. Clearly under event $E_1$ with $\b 1_p' \b \Gamma \b \mu /p  \ge C > 2 \epsilon > 0$, (\ref{ptme21}) is rewritten as
\begin{equation}
\left| \frac{(\widehat{MSR}^2 - MSR^2)/p}{MSR^2/p}
\right| = O_p (K^2 \bar{s} l_n),\label{npt6-3}
\end{equation}
where we use Theorem \ref{nwsr}. Under event $E_1$, with $1_p' \Gamma \mu /p  \le -C < -2 \epsilon <0$, (\ref{ptme21}) is rewritten as
\begin{equation}
\left| \frac{(\widehat{MSR_c}^2 - MSR_c^2)/p}{MSR_c^2/p}
\right| = O_p (K^2 \bar{s} l_n),\label{npt6-4}
\end{equation}
where we use Theorem \ref{tme1}.

Note that we can rewrite the event $E_1:=\{ | \widehat{F} - F | \le \epsilon\}$, with $\epsilon = O( K \bar{s} l_n )$. Note that event $E_1$ occurs with probability approaching one
by Lemma \ref{tl3} in Supplement B, so we have proven the desired result. 

\begin{flushright}
{\bf Q.E.D.}
\end{flushright}




\noindent
{\bf Proof of Theorem \ref{msros}}.
(A.2) of \cite{ao2019} shows that the squared ratio of the estimated maximum out-of-sample Sharpe Ratio to the theoretical ratio can be written as

\begin{equation}
 [\frac{\widehat{SR}_{mos}}{SR^*}]^2 = \frac{\frac{(\b\mu' \widehat{\b \Gamma} \widehat{\b \mu})^2}{\widehat{\b \mu}' \widehat{\b \Gamma}'  \b\Sigma_y \widehat{\b \Gamma} \widehat{\b \mu}}}{
\b\mu' \b\Gamma \b\mu}
= \frac{\left[ \frac{\b\mu' \widehat{\b \Gamma} \widehat{\b \mu}}{\b\mu' \b\Gamma\b\mu}
\right]^2}{\left[\frac{\widehat{\b \mu}' \widehat{\b \Gamma}' \b\Sigma_y \widehat{\b \Gamma} \widehat{\b \mu}}{\b\mu'\b\Gamma \b\mu}\right]}.\label{a23a}
\end{equation}

The proof will consider the numerator and the denominator of the squared maximum out-of-sample Sharpe Ratio. We start with the numerator using the definition, $\b\Gamma:= \b\Sigma_y^{-1}$
\begin{equation}
\frac{\b\mu' \widehat{\b \Gamma} \widehat{\b \mu}}{\b\mu' \b\Gamma \b\mu} = \frac{ \b\mu' \widehat{\b \Gamma} \widehat{\b \mu} - \b\mu' \b\Gamma \b\mu}{\b\mu' \b\Gamma \b\mu} +1.\label{osn0}
\end{equation}
Consider the fraction on the right-hand side. Start with the numerator in (\ref{osn0}).
\begin{eqnarray}
|\b\mu' \widehat{\b \Gamma} \widehat{\b \mu} - \b\mu'\b\Gamma \b\mu|/p & = &
| \b\mu' \widehat{\b \Gamma} \widehat{\b \mu} - \b\mu'\b\Gamma \widehat{\b \mu} + \b\mu' \b\Gamma \widehat{\b \mu} - \b\mu' \b\Gamma \b\mu|/p \nonumber \\
& \le & | \b\mu' ( \widehat{\b \Gamma} - \b\Gamma ) \widehat{\b \mu} |/p + |\b\mu'\b\Gamma (\widehat{\b \mu} - \b\mu) |/p\nonumber \\
& \le & |\b\mu' (\widehat{\b \Gamma} - \b\Gamma) (\widehat{\b \mu} - \b\mu)|/p + |\b\mu' (\widehat{\b \Gamma} -\b\Gamma)\b\mu |/p +
|\b\mu'\b\Gamma (\widehat{\b \mu} - \b\mu) |/p \nonumber \\
& = &  O_p \left(K \bar{s} l_n \max\left( K \sqrt{\ln(n)/n}, \sqrt{\ln(p)/n}\right)\right)  + O_p ( K^2 \bar{s} l_n ) \nonumber \\
&+& O_p\left( K^{5/2} \bar{s} r_n max\left(K \sqrt{\ln(n)/n}, \sqrt{\ln(p)/n}\right)\right)
 \nonumber \\
& = &  O_p ( K^2 \bar{s} l_n),\label{osn1}
\end{eqnarray}
where we use (\ref{la3.3}), (\ref{la3.4}), and (\ref{la3.5}) for the rates and the dominant rate in the last equality is by Assumption \ref{as8} and $l_n$ definition (\ref{dn}). By Assumption \ref{as8}(ii)
\begin{equation}
\frac{\b\mu' \b\Gamma \b\mu}{p} \ge c  >0.\label{d4}
\end{equation}

\noindent Then, by (\ref{osn1})(\ref{d4}) in (\ref{osn0})
\begin{equation}
\frac{\b\mu' \widehat{\b \Gamma} \widehat{\b \mu}/p}{\b\mu' \b\Gamma \b\mu/p} \le
\frac{|\b\mu' \widehat{\b \Gamma} \widehat{\b \mu} - \b\mu' \b\Gamma \b\mu|/p}{\b\mu' \b\Gamma \b\mu/p} + 1 = O_p ( K^2 \bar{s} l_n) + 1.\label{osn3}
\end{equation}
We now attempt to show that the denominator in (\ref{a23a})
\begin{equation}
\frac{\widehat{\b \mu}' \widehat{\b \Gamma} \b\Sigma_y \widehat{\b \Gamma} \widehat{\b \mu}}{\b\mu' \b\Sigma_y^{-1} \mu} \stackrel{p}{\to}1.\label{os00}
\end{equation}
In that respect, bearing in mind that $\b\Gamma = \b\Sigma_y^{-1}$ is symmetric
\begin{equation}
\frac{\widehat{\b \mu}' \widehat{\b \Gamma}' \b\Sigma_y \widehat{\b \Gamma} \widehat{\b \mu}}{\b\mu' \b\Sigma_y^{-1} \b\mu}  = \frac{\widehat{\b \mu}' \widehat{\b \Gamma}' \b\Sigma_y \widehat{\b \Gamma} \widehat{\b \mu} - \b\mu' \b\Gamma \b\Sigma_y \b\Gamma \b\mu}{ \mu' \b\Gamma \b\Sigma_y \b\Gamma  \b\mu} + 1 \ge 1 - \left| \frac{\widehat{\b \mu}' \widehat{\b \Gamma}' \b\Sigma_y \widehat{\b \Gamma}\widehat{\b \mu}- \b\mu' \b\Gamma \b\Sigma_y \b\Gamma \b\mu}{ \b\mu' \b\Gamma \b\Sigma_y \b\Gamma  \b\mu}\right|.\label{os0}
\end{equation}
We can write
\begin{equation}
\widehat{\b \Gamma} \widehat{\b \mu} - \b\Gamma \b\mu  = (\widehat{\b \Gamma} - \b\Gamma ) \widehat{\b \mu} + \b\Gamma (\widehat{\b \mu} - \b\mu).\label{os1}
\end{equation}
Using (\ref{os1})
\begin{eqnarray}
| \widehat{\b \mu}' \widehat{\b \Gamma}' \b\Sigma_y \widehat{\b \Gamma} \widehat{\b \mu} - \b\mu' \b\Gamma \b\Sigma_y \b\Gamma \b\mu | & = &
| [ (\widehat{\b \mu} \widehat{\b \Gamma} - \b\mu \b\Gamma)+ \b\mu \b\Gamma]'
\b\Sigma_y
[ (\widehat{\b \mu} \widehat{\b \Gamma} - \b\mu \b\Gamma)+ \b\mu \b\Gamma] - \b\mu'
\b\Gamma \b\Sigma_y \b\Gamma \b\mu| \nonumber \\
&\le &
| [ (\widehat{\b \Gamma} - \b\Gamma ) \widehat{\b \mu}]' \b\Sigma_y [ (\widehat{\b \Gamma} - \b\Gamma ) \widehat{\b \mu}] |  \label{os2a}\\
& + & 2 |[ (\widehat{\b \Gamma} - \b\Gamma ) \widehat{\b \mu}]' \b\Sigma_y  \b\Gamma (\widehat{\b \mu} - \b\mu)|  \label{os2b} \\
& + & 2 | [ (\widehat{\b \Gamma} - \b\Gamma ) \widehat{\b \mu}]' \b\Sigma_y \b\Gamma \b\mu | \label{os2c} \\
& + & | [ \b\Gamma (\widehat{\b \mu} - \b\mu )]' \b\Sigma_y [ \b\Gamma (\widehat{\b\mu} - \b\mu )]|\label{os2d} \\
& + & 2 |[\b\Gamma ( \widehat{\b \mu}- \b\mu)]' \b\Sigma_y \b\Gamma \b\mu| \label{os2e}
\end{eqnarray}
First, we consider (\ref{os2a}).
\begin{eqnarray}
|\widehat{\b \mu}' (\widehat{\b \Gamma} - \b\Gamma)' \b\Sigma_y (\widehat{\b \Gamma} - \b\Gamma) \widehat{\b \mu}| &\le & \textnormal{Eigmax}(\b\Sigma_y) \| (\widehat{\b \Gamma} - \b\Gamma) \widehat{\b \mu} \|_2^2 \nonumber \\
& = &  \textnormal{Eigmax}(\b\Sigma_y) [ \sum_{j=1}^p \{ (\widehat{\b \Gamma}_j - \b\Gamma_j)' \widehat{\b \mu} \}^2] \nonumber \\
& \le &  \textnormal{Eigmax}(\b\Sigma_y) p \max_{1 \le j \le p} [ (\widehat{\b \Gamma}_j - \b\Gamma_j)' \widehat{\b \mu}]^2 \nonumber \\
& \le &  \textnormal{Eigmax}(\b\Sigma_y) p  (\max_{1 \le j \le p} \| \widehat{\b \Gamma}_j - \b\Gamma_j \|_1)^2 \| \widehat{\b \mu} \|_{\infty}^2 \nonumber \\
& = & p O (p) O_p ( \bar{s}^2 l_n^2) O_p (K^2) = p O_p ( p \bar{s} l_n) O_p (K^2 \bar{s} l_n),\label{os4}
\end{eqnarray}
where we use H{\"o}lder's inequality for the third inequality and Theorem \ref{rthm2} and (\ref{maxe}), (\ref{mui}) for the rate. Now, consider (\ref{os2b}), and by definition
$\b\Gamma:= \b\Sigma_y^{-1}$.

\begin{eqnarray}
|[ (\widehat{\b \Gamma} - \b\Gamma ) \widehat{\b \mu}]' \b\Sigma_y  \Gamma (\widehat{\b \mu} - \mu)|
& = & |  (\widehat{\b \mu}- \mu)' (\widehat{\b \Gamma} - \b\Gamma) \widehat{\b \mu}| \nonumber \\
& \le & | (\widehat{\b \mu} - \b\mu )' (\widehat{\b \Gamma} - \b\Gamma)(\widehat{\b \mu} - \b\mu)| + | (\widehat{\b \mu} - \b\mu) (\widehat{\b \Gamma} - \b\Gamma) \b\mu| \nonumber \\
& = &  p [ O_p\left(max\left(K \sqrt{\ln(n)/n}, \sqrt{\ln(p)/n}\right)\right)]^2 O_p (\bar{s} l_n)] \nonumber \\
& +& [p  O( K) O_p (\bar{s} l_n) O_p \left(max \left(K \sqrt{\ln(n)/n}, \sqrt{ln(p)/n}\right)\right)] \nonumber \\
& = & p O_p \left( K \bar{s} l_n \max\left(K \sqrt{\ln(n)/n}, \sqrt{\ln(p)/n}\right)\right),\label{os5}
\end{eqnarray}
by (\ref{la3.1})(\ref{la3.4}) for the second equality, and the dominant rate in third equality can be seen from Assumption \ref{as8}.
Next, consider (\ref{os2c}), and recall that $\b\Gamma:= \b\Sigma_y^{-1}$

\begin{eqnarray}
| [ (\widehat{\b \Gamma} - \b\Gamma ) \widehat{\b \mu}]' \b\Sigma_y \b\Gamma \b\mu | & = & | \b\mu' (\widehat{\b \Gamma} - \b\Gamma) \widehat{\b \mu}| \nonumber \\
& \le & | \b\mu' (\widehat{\b \Gamma} - \b\Gamma) (\widehat{\b \mu} - \b\mu) | + | \mu' (\widehat{\b \Gamma} - \b\Gamma) \b\mu | \nonumber \\
& = & p [ O(K) O_p (\bar{s} l_n) O_p\left(\max\left(K\sqrt{\ln(n)/n}, \sqrt{\ln(p)/n}\right)\right) + O_p(K^2 \bar{s} l_n)] \nonumber \\
& = & p O_p ( K^2 \bar{s} l_n ), \label{os6}
\end{eqnarray}
where we use (\ref{la3.4})(\ref{la3.5}) for the second equality, and the dominant rate in the third equality can be seen from Assumption \ref{as8}.
Consider now (\ref{os2d}) by the symmetry of $\b\Gamma = \b\Sigma_y^{-1}$
\begin{equation}
| [\b\Gamma (\widehat{\b \mu} - \b\mu )]' \b\Sigma_y \b\Gamma (\widehat{\b\mu} - \b\mu )|   =  | (\widehat{\b \mu} - \b\mu)' \Gamma (\widehat{\b \mu} - \b\mu)| 
 =  p [O_p\left(\max\left(K \sqrt{\ln(n)/n}, \sqrt{\ln(p)/n}\right)\right)]^2 O (\bar{s} r_n K^{3/2})\label{os7}
\end{equation}
by (\ref{la3.2}).
Next, analyze (\ref{os2e}) by the symmetricity of $\b\Gamma = \b\Sigma_y^{-1}$

\begin{eqnarray}
 | [\b\Gamma ( \widehat{\b \mu}- \b\mu)]' \b\Sigma_y \b\Gamma \b\mu|  =  | (\widehat{\b \mu} - \b\mu)' \b\Gamma \b\mu| 
  =  p O_p\left(\bar{s} r_n K^{5/2}\max\left(K \sqrt{\ln(n)/n}, \sqrt{\ln(p)/n}\right)\right),\label{os8}
 \end{eqnarray}
by (\ref{la3.3}). Combine the rates and terms (\ref{os4})-(\ref{os8}) in (\ref{os2a})-(\ref{os2e}) to obtain
\begin{equation}
| \widehat{\b \mu}' \widehat{\b \Gamma}' \b\Sigma_y \widehat{\b \Gamma} \widehat{\b \mu} - \b\mu' \b\Gamma \b\Sigma_y \b\Gamma \b\mu |/p  =  O_p ( K^2 \bar{s} l_n),\label{os9}
\end{equation}
by the dominant rate in (\ref{os6}), as seen in Assumption \ref{as9}: $p \bar{s} l_n \to 0$ in (\ref{os4}), and $l_n$ definition in Assumption \ref{as7}.

See that by (\ref{d4})
\begin{equation}
\frac{\b\mu'\b\Gamma\b\Sigma_y\b\Gamma\b\mu}{p} = \frac{\b\mu' \b\Gamma\b\mu}{p} \ge c >0.\label{os10}
\end{equation}

Combine (\ref{os9})(\ref{os10}), in the second right side term in (\ref{os0}) via Assumption \ref{as8}
\begin{equation}
\frac{| \widehat{\b \mu}' \widehat{\b \Gamma}' \b\Sigma_y \widehat{\b \Gamma} \widehat{\b \mu} - \b\mu'\b\Gamma \b\Sigma_y \b\Gamma\b\mu|/p}{\b\mu'\b\Gamma\b\Sigma_y\b\Gamma\b\mu/p}
\le    O_p ( K^2 \bar{s} l_n)= o_p (1).\label{a41a}
\end{equation}
Therefore, we show (\ref{os00}) via (\ref{os0}). Then, combine (\ref{osn3})(\ref{os00}) in (\ref{a23a}) to obtain the desired result.

\begin{flushright}
{\bf Q.E.D.}
\end{flushright}

\setcounter{equation}{0}\setcounter{lemma}{0}\setcounter{assum}{0}\renewcommand{\theequation}{B.%
\arabic{equation}}\renewcommand{\thelemma}{B.\arabic{lemma}}%
\renewcommand{\theassum}{B.\arabic{assum}}%
\renewcommand{\baselinestretch}{1}\baselineskip=15pt

\section*{Supplement B}

Here, we provide results that are used in proofs of Section 4. We provide a matrix norm inequality.
Let $\b x$ be a generic vector, which is $p  \times 1$. $\b M$ is a square matrix of dimension $p$, where $\b M_j'$ is the $j$th row of dimension $1 \times p$, and $\b M_j$ is the transpose of this row vector.

\begin{lemma}\label{tl1}


\[ \| \b M \b x \|_1 \le  p \max_{1 \le j \le p} \| \b M_j \|_1  \| \b x \|_{\infty}.\]

\end{lemma}

\noindent
{\bf Proof of Lemma \ref{tl1}}.


\begin{equation}
\| \b M \b x \|_1 \le  p \max_j \|\b M_j \|_1 \|\b x \|_{\infty},\label{la1.3}
  \end{equation}
where we use H{\"o}lder's inequality to obtain each inequality.

\begin{flushright}{\bf Q.E.D.}\end{flushright}

Recall the definition of
$A:= \b 1_p' \b \Gamma 1_p/p$ and $\widehat{A}:= \b 1_p' \widehat{\b \Gamma} 1_p/p$, and $\bar{s} l_n$ is the rate of convergence in Theorem \ref{rthm2} in main text, and defined in Assumption \ref{as7} with the property $\bar{s} l_n \to 0$.

\begin{lemma}\label{tl2}
Under Assumptions \ref{as1}-\ref{as4}, \ref{as6}-\ref{as7}
\[  | \widehat{A} - A | = O_p ( \bar{s} l_n)= o_p (1). \]
\end{lemma}

\noindent
{\bf Proof of Lemma \ref{tl2}}.

\begin{eqnarray}
|\b 1_p' (\widehat{\b \Gamma} - \b \Gamma) \b 1_p|/p & \le & \|(\widehat{\b \Gamma} - \b \Gamma) \b 1_p \|_1  \|\b 1_p \|_{\infty} /p\nonumber \\
& \le & \max_{1 \le j \le p} \|\widehat{\b \Gamma}_j - \b \Gamma_j\|_1 
 =   O_p ( \bar{s}  l_n) = o_p (1), \label{la1.4}
\end{eqnarray}
where H{\"o}lder's inequality is used in the first inequality, Lemma \ref{tl1} is used for the second inequality, and the last equality is obtained by using Theorem \ref{rthm2} and imposing Assumption \ref{as7}.

\begin{flushright}
{\bf Q.E.D.}
\end{flushright}

Before the next Lemma, we define $\widehat{F} := \b 1_p' \widehat{\b \Gamma} \widehat{\b \mu}/p$, and $F := \b 1_p' \b \Gamma \b \mu/p$.

\begin{lemma}\label{tl3}
Under Assumptions \ref{as1}-\ref{as4}, \ref{as6}, \ref{as7}(i), \ref{as8}(i) 
\[  | \widehat{F} - F |  = O_p ( K \bar{s} l_n)=o_p (1).\]
\end{lemma}

\noindent
{\bf Proof of Lemma \ref{tl3}}. We can decompose $\widehat{F}$ by simple addition and subtraction into
\begin{eqnarray}
\widehat{F} - F & = & [\b 1_p' (\widehat{\b \Gamma} - \b \Gamma) ( \widehat{\b \mu} - \b \mu)]/p \label{la2.a} \\
& + & [\b 1_p' (\widehat{\b \Gamma} - \b \Gamma)  \b \mu]/p \label{la2.b}\\
& + & [\b 1_p' \b \Gamma( \widehat{\b \mu} - \b \mu)]/p \label{la2.c}
\end{eqnarray}

Now, we analyze each of the terms above. 

\begin{eqnarray}
|\b 1_p' (\widehat{\b \Gamma} - \b \Gamma) ( \widehat{\b \mu} - \b \mu) |/p & \le & \| (\widehat{\b \Gamma} - \b \Gamma) 1_p \|_1 \|\widehat{\b \mu} - \b \mu \|_{\infty}/p \nonumber \\
& \le & [ \max_{1 \le j \le p} \| \widehat{\b \Gamma}_j - \b \Gamma_j \|_1] \|\widehat{\b \mu} - \b \mu \|_{\infty} \nonumber \\
& = & O_p\left(\bar{s} l_n\right) O_p\left(\max\left(K \sqrt{\ln(n)/n}, \sqrt{ln(p)/n}\right)\right),\label{la.21}
\end{eqnarray}
where we use H{\"o}lder's inequality in the first inequality and Lemma \ref{tl1} in the second inequality above, and the rate is from Theorem \ref{rthm2}. To get to the other terms, we need two extra results. Use the definition of $\b \Gamma$ in (\ref{sw3})
\begin{eqnarray}
\|\b \Gamma \|_{l_{\infty}} &\le & \| \b\Omega \|_{l_{\infty}}  + \| \b\Omega L \b\Omega \|_{l_{\infty}} \nonumber \\
& \le &  \| \b\Omega \|_{l_{\infty}}  + \| \b\Omega \|_{l_{\infty}}^2 \| \b L \|_{l_{\infty}} \nonumber \\
& = & O (\sqrt{\bar{s}}) + O (\bar{s}) O (r_n K^{3/2}) = O (\bar{s} r_n K^{3/2}),\label{E1}
\end{eqnarray}
where for the rates we use (\ref{pt2-8})(\ref{pt2-9}) and since $K$ is nondecreasing in $n$. Note that $\b \mu = \E[\b y_t] = \b B \E[f_t]$. So with $b_{j,k}$ representing $j,k$th element of $p \times k$: $\b B$ matrix, and $\E[f_{t,k}]$ representing $k$th element of $K \times 1$ vector $\E(\b f_t)$ we have that
\begin{eqnarray}
\| \b \mu \|_{\infty}& :=& \max_{1 \le j \le p} |\b b_j' \E[\b f_t] |= \max_{1 \le j \le p} | \sum_{k=1}^K b_{j,k} \E[f_{t,k}]| \nonumber \\
& \le & \max_{1 \le j \le p} \max_{1 \le k \le K} | b_{j,k}| [ K \max_{1 \le k \le K} | \E[f_{t,k}]|] 
 =  O(K),\label{mui}
\end{eqnarray}
where the rate is by Assumption \ref{as4}, \ref{as6}. Therefore, we consider (\ref{la2.b}) above.
\begin{equation}
|\b 1_p' (\widehat{\b \Gamma} - \b \Gamma) \b \mu |/p
=   O_p ( \bar{s} l_n) O(K),\label{la.22}
\end{equation}
where we use 
the same analysis that leads to (\ref{la.21}),
and the rate is from Theorem \ref{rthm2}, (\ref{mui}). Now consider (\ref{la2.c}).
\begin{eqnarray}
|\b 1_p' \b \Gamma( \widehat{\b \mu} - \b \mu) | /p& \le & \| \b \Gamma \b 1_p \|_1 \|\widehat{\b \mu} - \b \mu \|_{\infty}/p \nonumber \\
& \le &  [ \max_{1 \le j \le p} \| \b \Gamma_j \|_1] \|\widehat{\b \mu} - \b \mu \|_{\infty} \nonumber \\
& = & O\left(\bar{s} r_n K^{3/2}\right)O_p\left(\max\left(K \sqrt{\ln(n)/n}, \sqrt{\ln(p)/n}\right)\right),\label{la.23}
\end{eqnarray}
where we use H{\"o}lder's inequality in the first inequality and Lemma \ref{tl1} in the second inequality above, and the rate is from Theorem \ref{rthm2}, (\ref{E1}).
Combine (\ref{la.21})(\ref{la.22})(\ref{la.23}) in (\ref{la2.a})-(\ref{la2.c}), and note that the largest rate is coming from
(\ref{la.22}) by $\bar{s} l_n$ definition in Assumption \ref{as7}. 
\begin{flushright}
{\bf Q.E.D.}
\end{flushright}
Note that $D := \b\mu'\b\Gamma\b\mu/p$, and its estimator is $\widehat{D}:= \widehat{\b\mu}' \widehat{\b\Gamma}
\widehat{\b\mu}/p$.

\begin{lemma}\label{tl4}
Under Assumptions \ref{as1}-\ref{as4}, \ref{as6}, \ref{as7}(i), \ref{as8}(i)  
\[| \widehat{D} - D | = O_p ( K^2 \bar{s} l_n)= o_p (1).\]
\end{lemma}

\noindent
{\bf Proof of Lemma \ref{tl4}}. By simple addition and subtraction,
\begin{eqnarray}
\widehat{D} - D & = &[ (\widehat{\b \mu} - \b\mu)' (\widehat{\b \Gamma} - \b\Gamma) (\widehat{\b \mu} - \b\mu)]/p \label{la3.a} \\
& + & [(\widehat{\b \mu} - \b\mu)' \b\Gamma
(\widehat{\b \mu} - \b\mu)]/p\label{la3.b} \\
& + & [2 (\widehat{\b \mu} - \b\mu)' \b\Gamma \b\mu ]/p\label{la3.c} \\
& + & [2\b\mu' (\widehat{\b \Gamma} - \b\Gamma) (\widehat{\b \mu} - \b\mu) ]/p\label{la3.d} \\
& + &[ \b\mu' (\widehat{\b \Gamma} - \b\Gamma) \mu]/p\label{la3.e}.
\end{eqnarray}
Consider the first right side term above
\begin{eqnarray}
 |(\widehat{\b \mu} - \b\mu)' (\widehat{\b \Gamma} - \b\Gamma) (\widehat{\b \mu} - \b\mu) |/p  & \le &
\| (\widehat{\b \Gamma} - \b\Gamma) (\widehat{\b \mu} - \b\mu ) \|_1 \|\widehat{\b \mu} - \b\mu \|_{\infty}/p \nonumber \\
& \le & [\|\widehat{\b \mu} - \b\mu \|_{\infty}]^2 [\max_j \| \widehat{\b \Gamma}_j - \b\Gamma_j \|_1 ] \nonumber \\
& = &  [O_p \left(\max\left( K \sqrt{\ln(n)/n}, \sqrt{\ln(p)/n}\right)\right)]^2 O_p ( \bar{s} l_n)
\label{la3.1}
\end{eqnarray}
where H{\"o}lder's inequality is used for the first inequality above, and the inequality Lemma \ref{tl1} for the second inequality above, and for the rates we use Theorem \ref{rthm2}. We continue with  (\ref{la3.b}).
\begin{eqnarray}
 |(\widehat{\b \mu} - \b\mu)' (\b\Gamma) (\widehat{\b \mu} - \b\mu) |/p  & \le &
\| (\b\Gamma) (\widehat{\b \mu} - \b\mu ) \|_1 \|\widehat{\b \mu} - \b\mu \|_{\infty}/p \nonumber \\
& \le & [\|\widehat{\b \mu} - \b\mu \|_{\infty}]^2 [\max_j \|  \b\Gamma_j \|_1 ] \nonumber \\
& = & [O_p\left(\max\left(K\sqrt{\ln(n)/n},\sqrt{\ln(p)/n}\right)\right)]^2O(\bar{s} r_n K^{3/2}),\label{la3.2}
\end{eqnarray}
where H{\"o}lder's inequality is used for the first inequality above, and the inequality Lemma \ref{tl1} for the second inequality above, and for the rates, we use Theorem \ref{rthm2} and (\ref{E1}). Then, we consider  (\ref{la3.c})
\begin{eqnarray}
 |(\widehat{\b \mu} - \b\mu)' (\b\Gamma) ( \b\mu) |/p  & \le &
\| (\b\Gamma) (\widehat{\b \mu} - \b\mu ) \|_1 \|\b\mu \|_{\infty}/p \nonumber \\
& \le &   [\|\widehat{\b \mu} - \b\mu \|_{\infty}] [\max_j \|  \b\Gamma_j \|_1 ]  O(K)\nonumber \\
& = &  O_p\left(\max\left(K\sqrt{\ln(n)/n}, \sqrt{\ln(p)/n}\right)\right) O(\bar{s} r_n K^{3/2} ) O(K),\label{la3.3}
\end{eqnarray}
where H{\"o}lder's inequality is used for the first inequality above, and the inequality Lemma \ref{tl1} for the second inequality above, and for the rates, we use Theorem \ref{rthm2} and (\ref{mui}).
Then, we consider  (\ref{la3.d}).
\begin{eqnarray}
 |(\b\mu)' (\widehat{\b \Gamma} - \b\Gamma) ( \widehat{\b \mu} - \b\mu) |/p  & \le &
\| (\widehat{\b \Gamma}- \b\Gamma) ( \b\mu ) \|_1 \|\widehat{\b \mu}- \b\mu \|_{\infty}/p \nonumber \\
& \le &  \|  \b\mu  \|_{\infty}  \max_j \| \widehat{\b \Gamma}_j -   \b\Gamma_j \|_1 \| \widehat{\b \mu} - \b\mu \|_{\infty} \nonumber \\
& \le & \ [\max_j \| \widehat{\b \Gamma}_j-  \b\Gamma_j \|_1 ] \| (\widehat{\b \mu} - \b\mu)\|_{\infty} O(K) \nonumber \\
& = &   O_p ( K \bar{s} l_n ) O_p\left(\max\left(K \sqrt{\ln (n)/n}, \sqrt{\ln(p)/n}\right)\right), \label{la3.4}
\end{eqnarray}
where H{\"o}lder's inequality is used for the first inequality above, and the inequality Lemma \ref{tl1} for the second inequality above, for the third inequality above, we use (\ref{mui}), and for the rates, we use Theorem \ref{rthm2}.
Then, we consider  (\ref{la3.e}):
\begin{eqnarray}
 |(\b\mu)' (\widehat{\b \Gamma} - \b\Gamma) (\b\mu) |/p  & \le &
\| (\widehat{\b \Gamma}- \b\Gamma) (\b\mu ) \|_1 \|\b\mu \|_{\infty}/p \nonumber \\
& \le &  [\| \b\mu  \|_{\infty}]^2  \max_j \| \widehat{\b \Gamma}_j -   \b\Gamma_j \|_1  \nonumber \\
& \le &   [\max_j \| \widehat{\b \Gamma}_j-  \b\Gamma_j \|_1 ]  O(K^2) \nonumber \\
&=&  O_p ( K^2 \bar{s} l_n), \label{la3.5}
\end{eqnarray}
where H{\"o}lder's inequality is used for the first inequality above, and the inequality Lemma \ref{tl1} for the second inequality above, for the third inequality above, we use (\ref{mui}), and for the rate, we use Theorem \ref{rthm2}. Note that in  (\ref{la3.a})-(\ref{la3.e})  the rate in (\ref{la3.5}) is the slowest due to $l_n$ definition in (\ref{dn}) to obtain
\begin{equation}
 | \widehat{D} - D | = O_p ( K^2 \bar{s} l_n)=o_p (1).\label{lpla3}
\end{equation}
\begin{flushright}
{\bf Q.E.D.}
\end{flushright}
The following lemma establishes orders for the terms in the optimal weight, A, B, D. Note that both $A, D$ are positive by Assumption 2 and uniformly bounded away from zero.
\begin{lemma}\label{tl5}
Under Assumptions \ref{as1}, \ref{as4}, \ref{as6}
\[ A = O (1 ).\]
\[ |F| = O (K^{1/2}).\]
\[D = O (K).\]
\end{lemma}

\noindent
{\bf Proof of Lemma \ref{tl5}}.
Note that $A=\b 1_p' \b \Gamma \b 1_p/p \le \textnormal{Eigmax}( \b \Gamma)$. Then by p.221 of \cite{abamag2005}, (Exercise 8.27.b in \cite{abamag2005}),
$\b\Omega \b B [(\textnormal{cov}(\b f_t))^{-1} + \b B' \b\Omega \b B]^{-1} \b B' \b\Omega$ is positive semidefinite, so we can use Exercise 12.40b of \cite{abamag2005}, since $\b\Omega$ is symmetric, and by (\ref{sw3})
\begin{eqnarray}
\textnormal{Eigmax} (\b \Gamma)  &:= &\textnormal{Eigmax} (\b\Omega - \b\Omega \b B [(\textnormal{cov} (\b f_t))^{-1} + \b B' \b\Omega \b B]^{-1} \b B' \b\Omega) \nonumber \\
&\le & \textnormal{Eigmax} (\b\Omega) = \frac{1}{\textnormal{Eigmin} (\b\Sigma_n)} \le \frac{1}{c},\label{r81}
\end{eqnarray}
since $\b\Omega:= \b\Sigma_n^{-1}$, and by Assumption \ref{as1} $\textnormal{Eigmin} (\b\Sigma_n) \ge c >0$. This last point shows that $A = O(1)$.

Now consider $D = \b \mu' \b \Gamma \b \mu /p$. By Theorem 5.6.2b of \cite{hj2013}
\begin{eqnarray}
\| \b \mu \|_2^2 & := & \| \b B \E[\b f_t] \|_2^2 \le \| \b B \|_{l_2}^2 \| \E[\b f_t] \|_2^2 \nonumber \\
& = & O (p) O (K)= O(p K),\label{r81a}
\end{eqnarray}
by (6.3) of \cite{fan2008}, $\| \b B \|_{l_2}^2 = O(p)$ under Assumption \ref{as6}, and by Assumption \ref{as4}, $\| \E[\b f_t] \|_2^2 = O(K)$, since $\b f_t: K \times 1$ vector of factors. By (\ref{r81})(\ref{r81a})
\begin{equation}
\b \mu' \b \Gamma \b \mu /p \le \textnormal{Eigmax} (\b \Gamma) \| \b \mu \|_2^2/p = O (pK)/p = O (K).\label{r82}
\end{equation}

For the term F, the proof can be obtained by using the Cauchy-Schwartz inequality first and the same analysis as for terms A and D. 

\begin{flushright}
{\bf Q.E.D.}
\end{flushright}

Next, we need the following technical lemma, which provides the limit and the rate for the denominator in the optimal portfolio.

\begin{lemma}\label{tl6}
Under Assumptions \ref{as1}-\ref{as4}, \ref{as6}, \ref{as7}(i), \ref{as8}(i) 
\[ |(\widehat{A} \widehat{D} - \widehat{F}^2) - (AD - F^2)| = O_p ( K^2  \bar{s} l_n)= o_p (1) .\]
\end{lemma}

\noindent
{\bf Proof of Lemma \ref{tl6}}. Note that by simple addition and subtraction,
\[ \widehat{A}\widehat{D} - \widehat{F}^2= [(\widehat{A}-A)+A] [(\widehat{D} - D ) + D ] - [(\widehat{F} -F)+F]^2.\]
Then, using this last expression and simplifying, $A, D$ being both positive,
\begin{eqnarray}
 | (\widehat{A} \widehat{D} - \widehat{F}^2) - (AD - F^2)| & \le &  \{ |\widehat{A} - A | |\widehat{D} - D| + |\widehat{A} - A | D  \nonumber \\
& + & A |\widehat{D}-D| + (\widehat{F} - F)^2
+  2 | F| |\widehat{F} - F| \} \nonumber \\
& = & O_p (\bar{s} l_n) O_p (K^2  \bar{s} l_n) + O_p (\bar{s} l_n) O (K)  \nonumber \\
&+& O(1) O_p (K^2 \bar{s} l_n) + O_p ( \bar{s}^2 l_n^2 K^2) + O (K^{1/2}) O_p (\bar{s} l_n K) \nonumber \\
& = & O_p ( K^2 \bar{s} l_n)= o_p (1),\label{pla51}
\end{eqnarray}
where we use (\ref{la1.4}), Lemma \ref{tl3}, (\ref{lpla3}), Lemma \ref{tl5}, and Assumption \ref{as8}.
\begin{flushright}
{\bf Q.E.D.}
\end{flushright}

\setcounter{equation}{0}\setcounter{lemma}{0}\setcounter{assum}{0}\renewcommand{\theequation}{C.%
\arabic{equation}}\renewcommand{\thelemma}{C.\arabic{lemma}}%
\renewcommand{\theassum}{C.\arabic{assum}}%
\renewcommand{\baselinestretch}{1}\baselineskip=15pt

\section*{Supplement C}

This part covers the proofs for Corollaries 1-3 in the main text.

{\bf Proof of Corollary 1}. Rewrite the ratio of the Sharpe Ratio estimate to its target in the following way by (\ref{5.1})
\begin{equation}
\frac{\widehat{SR}_{nw,p}^2}{SR^2} = \frac{(\b 1_p' \hat{\b \Gamma}' \b \mu/p)^2/(\b 1_p' \b \Gamma \b \mu/p)^2}{(\b 1_p' \hat{\b \Gamma}' \b \Sigma_y \hat{\b \Gamma} \b 1_p/p)/
(\b 1_p' \b \Gamma \b 1_p/p)}.\label{pc1}
\end{equation}

\noindent Consider the numerator in (\ref{pc1}).
\[
\frac{(\b 1_p' \hat{\b \Gamma}' \b \mu/p)}{(\b 1_p' \b \Gamma \b \mu/p)} \le 1 + \frac{|(\b 1_p' \hat{\b \Gamma}' \b \mu/p) - (\b 1_p' \b \Gamma \b \mu/p)|}{\b 1_p' \b \Gamma \b \mu/p}.\]
Then by Holder's inequality and Lemma B.1
\begin{eqnarray}
|(\b 1_p' \hat{\b \Gamma}' \b \mu/p) - (\b 1_p' \b \Gamma \b \mu/p)| & = &\frac{1}{p} | \b \mu' \hat{\b \Gamma} \b 1_p - \b \mu' \b \Gamma \b 1_p| \nonumber \\
& = & \frac{1}{p} | \b \mu' ( \hat{\b \Gamma} - \b \Gamma) \b 1_p | \le \| \mu \|_{\infty} \| ( \hat{\b \Gamma} - \b \Gamma) \b 1_p \|_1 \nonumber \\
& \le & \| \mu \|_{\infty} \max_{1 \le j \le p} \| ( \hat{\b \Gamma}_j - \b \Gamma_j)  \|_1 \nonumber \\
& = & O(K) O_p (\bar{s} l_n),\label{e1}
\end{eqnarray}
and the rates are by (\ref{mui}), Theorem 2. Since $| \b 1_p' \b \Gamma \b \mu|/p \ge C > 0$ by Assumption, using (\ref{e1}) we have 

\begin{equation}
\frac{(\b 1_p' \hat{\b \Gamma}' \b \mu/p)^2}{(\b 1_p' \b \Gamma \b \mu/p)^2} = 1 + O_p ( K \bar{s} l_n).\label{pc5}
\end{equation}
Analyze the denominator in (\ref{pc1}), 
\begin{equation}
\frac{(\b 1_p' \hat{\b \Gamma}' \b \Sigma_y \hat{\b \Gamma} \b 1_p/p)}{
(\b 1_p' \b \Gamma \b 1_p/p)} \ge 1 - \frac{ | \b 1_p' {\hat{\b \Gamma}}' \b \Sigma_y \hat{\b \Gamma} \b 1_p/p - \b 1_p' \b \Gamma \b 1_p/p|}{
\b 1_p' \b \Gamma \b 1_p/p}.\label{pc5a}
\end{equation}
Next, see that by adding and subtracting and via triangle inequality in (\ref{pc5a}) numerator

\begin{eqnarray}
| \b 1_p' {\hat{\b \Gamma}}' \b \Sigma_y \hat{\b \Gamma} \b 1_p/p &-& \b 1_p' \b \Gamma \b 1_p/p|  \le 
\frac{1}{p} | \b 1_p' ({\hat{\b \Gamma}}- \b \Gamma)' \b \Sigma_y (\hat{\b \Gamma}- \b \Gamma) \b 1_p| \nonumber \\
& + & \frac{2}{p} |  | \b 1_p' ({\hat{\b \Gamma}}- \b \Gamma)' \b \Sigma_y \b \Gamma \b 1_p|.\label{pc6}
\end{eqnarray}
Consider the first term in right side of (\ref{pc6})
\begin{eqnarray}
\frac{1}{p} | \b 1_p' ({\hat{\b \Gamma}}- \b \Gamma)' \b \Sigma_y (\hat{\b \Gamma}- \b \Gamma) \b 1_p|  & \le & 
\frac{1}{p}[Eigmax( \b \Sigma_y) \| (\hat{\b \Gamma} - \b \Gamma ) \b 1_p\|_2^2 ] \nonumber \\
& \le & Eigmax( \b \Sigma_y) \max_{1 \le j \le p} \| \hat{\b \Gamma}_j- \b \Gamma_j \|_1^2 \nonumber \\
& = & O (p) O_p ( \bar{s}^2 l_n^2) = O_p (p \bar{s} l_n) O_p ( \bar{s} l_n)= o_p (1),\label{pc7}
\end{eqnarray}
by Theorem 2, (\ref{maxe}), and Assumption 9. Then in (\ref{pc6}), take the second right side term, with $\b \Gamma:= \b \Sigma_y^{-1}$, 
\begin{eqnarray}
\frac{2}{p} |  | \b 1_p' ({\hat{\b \Gamma}}- \b \Gamma)' \b \Sigma_y \b \Gamma \b 1_p| &=&
\frac{2}{p} | \b 1_p' (\hat{\b \Gamma} - \b \Gamma)' \b 1_p | \nonumber \\
& = &  \frac{2}{p} | \b 1_p' (\hat{\b \Gamma} - \b \Gamma) \b 1_p | = O_p ( \bar{s} l_n) = o_p (1),\label{pc8}
\end{eqnarray}
by Lemma B.2, Assumption 9. Use (\ref{pc7})(\ref{pc8}) in (\ref{pc5a})(\ref{pc6}) by Assumption 8(ii)
\begin{equation}
\frac{(\b 1_p' \hat{\b \Gamma}' \b \Sigma_y \hat{\b \Gamma} \b 1_p/p)}{
(\b 1_p' \b \Gamma \b 1_p/p)} \ge 1 - o_p (1).\label{pc9}
\end{equation}
Combine (\ref{pc5})(\ref{pc9}) in (\ref{pc1}) to have the result.{\bf Q.E.D}

{\bf Proof of Corollary 2}. Define the following terms
\begin{equation}
\hat{r}_1:= \hat{D} - \rho_1 \hat{F}, \quad \hat{r}_2:=\rho_1 \hat{A} -  \hat{F} \label{pc2-0}
\end{equation}
and 
\begin{equation}
r_1:= D - \rho_1 F, \quad \hat{r}_2:=\rho_1 A -  F \label{pc2-0}
\end{equation}
The estimate of the portfolio return
\begin{equation}
\hat{\b w}_{mv}' \b\mu = \frac{\hat{r}_1 \b 1_p' \hat{\b \Gamma}' \b \mu /p + \hat{r}_2 \hat{\b \mu}' \hat{\b \Gamma}' \b \mu/p}{\hat{A} \hat{D} - \hat{F}^2}.\label{pc2-1}
\end{equation}
The target portfolio return
\begin{equation}
\b w_{mv}' \b\mu = \frac{r_1 \b 1_p' \b \Gamma' \b \mu /p + r_2 \b \mu' \b \Gamma' \b \mu/p}{AD - F^2}.\label{pc2-1a}\end{equation}
The estimate of variance of the portfolio, with $\b \Sigma_y$ constant, is
\begin{equation}
\hat{\b w}_{mv}' \b \Sigma_y \hat{\b w}_{mv} = \frac{\hat{r}_1^2 \left( \frac{\b 1_p' \hat{\b \Gamma}' }{p} \b \Sigma_y \frac{\hat{\b \Gamma} \b 1_p}{p}
\right) + 
2 \hat{r}_1 \hat{r}_2 \left( \frac{\b 1_p' \hat{\b \Gamma}' }{p} \b \Sigma_y \frac{\hat{\b \Gamma} \hat{\b \mu}}{p}
\right)
 + \hat{r}_2^2 \left( \frac{\hat{\b \mu}' \hat{\b \Gamma}' }{p} \b \Sigma_y \frac{\hat{\b \Gamma}  \hat{\b \mu}}{p}
\right)}{[\hat{A} \hat{D} - \hat{F}^2]^2}.\label{pc2-2} 
\end{equation}
Target variance is
\begin{equation}
\b w_{mv}' \b \Sigma_y \b w_{mv} = \frac{r_1^2 \left( \frac{\b 1_p' \b \Gamma' }{p} \b \Sigma_y \frac{\b \Gamma \b 1_p}{p}
\right) + 
2 r_1 r_2 \left( \frac{\b 1_p' \b \Gamma' }{p} \b \Sigma_y \frac{\b \Gamma \b \mu}{p}
\right)
 + r_2^2 \left( \frac{\b \mu' \b \Gamma' }{p} \b \Sigma_y \frac{\b \Gamma  \b \mu}{p}
\right)}{[AD - F^2]^2}.\label{pc2-2a} 
\end{equation}

Start with the estimate of square of the Sharpe Ratio:
\[ \widehat{SR}_{mv}^2 :=  \frac{(\hat{\b w}_{mv}' \b \mu)^2}{\hat{\b w}_{mv}' \b \Sigma_y \hat{\b w}_{mv}} =
\frac{(\hat{r}_1 \b 1_p' \hat{\b \Gamma}' \b \mu /p + \hat{r}_2 \hat{\b \mu}' \hat{\b \Gamma}' \b \mu/p)^2}{\hat{r}_1^2 \left( \frac{\b 1_p' \hat{\b \Gamma}' }{p} \b \Sigma_y \frac{\hat{\b \Gamma} \b 1_p}{p}
\right) + 
2 \hat{r}_1 \hat{r}_2 \left( \frac{\b 1_p' \hat{\b \Gamma}' }{p} \b \Sigma_y \frac{\hat{\b \Gamma} \hat{\b \mu}}{p}
\right)
 + \hat{r}_2^2 \left( \frac{\hat{\b \mu}' \hat{\b \Gamma}' }{p} \b \Sigma_y \frac{\hat{\b \Gamma}  \hat{\b \mu}}{p}
\right)}.\]

Then the target Sharpe Ratio is:
\[ SR_{mv}^2:=\frac{(\b w_{mv}' \b \mu)^2}{\b w_{mv}' \b \Sigma_y \b w_{mv}}=
\frac{(r_1 \b 1_p' \b \Gamma' \b \mu /p + r_2 \b \mu' \b \Gamma' \b \mu/p)^2}{
r_1^2 \left( \frac{\b 1_p' \b \Gamma' }{p} \b \Sigma_y \frac{\b \Gamma \b 1_p}{p}
\right) + 
2 r_1 r_2 \left( \frac{\b 1_p' \b \Gamma' }{p} \b \Sigma_y \frac{\b \Gamma \b \mu}{p}
\right)
 + r_2^2 \left( \frac{\b \mu' \b \Gamma' }{p} \b \Sigma_y \frac{\b \Gamma  \b \mu}{p}
\right)}.\]
Take the ratio of the estimate to the target Sharpe Ratio, and scaling variances by $p$
{\small
\begin{equation}
\frac{ \widehat{SR}_{mv}^2}{SR_{mv}^2}=\frac{(\hat{r}_1 \b 1_p' \hat{\b \Gamma}' \b \mu /p + \hat{r}_2 \hat{\b \mu}' \hat{\b \Gamma}' \b \mu/p)^2
/(r_1 \b 1_p' \b \Gamma' \b \mu /p + r_2 \b \mu' \b \Gamma' \b \mu/p)^2}{p[\hat{r}_1^2 \left( \frac{\b 1_p' \hat{\b \Gamma}' }{p} \b \Sigma_y \frac{\hat{\b \Gamma} \b 1_p}{p}
\right) + 
2 \hat{r}_1 \hat{r}_2 \left( \frac{\b 1_p' \hat{\b \Gamma}' }{p} \b \Sigma_y \frac{\hat{\b \Gamma} \hat{\b \mu}}{p}
\right)
 + \hat{r}_2^2 \left( \frac{\hat{\b \mu}' \hat{\b \Gamma}' }{p} \b \Sigma_y \frac{\hat{\b \Gamma}  \hat{\b \mu}}{p}
\right)]/p[r_1^2 \left( \frac{\b 1_p' \b \Gamma' }{p} \b \Sigma_y \frac{\b \Gamma \b 1_p}{p}
\right) + 
2 r_1 r_2 \left( \frac{\b 1_p' \b \Gamma' }{p} \b \Sigma_y \frac{\b \Gamma \b \mu}{p}
\right)
 + r_2^2 \left( \frac{\b \mu' \b \Gamma' }{p} \b \Sigma_y \frac{\b \Gamma  \b \mu}{p}
\right)]}.\label{c.14a}
\end{equation}}

{\bf Step 1}\\

Start with the terms in numerator in (\ref{c.14a})
 which will be upper bounded by 
\begin{eqnarray}
(\hat{r}_1 \b 1_p' \hat{\b \Gamma}' \b \mu /p + \hat{r}_2 \hat{\b \mu}' \hat{\b \Gamma}' \b \mu/p) &-& (r_1 \b 1_p' \b \Gamma' \b \mu /p + r_2 \b \mu' \b \Gamma' \b \mu/p) \nonumber \\
& \le & 
 |  \hat{r}_1 \b 1_p' \hat{\b \Gamma}' \b \mu /p - r_1 \b 1_p' \b \Gamma \b \mu /p| 
+ |  \hat{r}_2 \hat{\b \mu}' \hat{\b \Gamma}' \b \mu /p - r_2 \b \mu' \b \Gamma \b \mu /p| 
.\label{pc2-4}
\end{eqnarray}

First term on the right side of (\ref{pc2-4}), and $\b \Gamma$ is symmetric
\begin{eqnarray}
|  \hat{r}_1 \b 1_p' \hat{\b \Gamma}' \b \mu /p - r_1 \b 1_p' \b \Gamma \b \mu /p|  & \le & 
| \hat{r}_1 \b 1_p' \hat{\b \Gamma}' \b \mu /p - r_1 \b 1_p' \hat{\b \Gamma}' \b \mu /p| \nonumber \\
& + & |r_1 \b 1_p' \hat{\b \Gamma}' \b \mu /p - r_1 \b 1_p' \b \Gamma' \b \mu /p|.\label{pc2-5}
\end{eqnarray}

Take the first term on the right side of (\ref{pc2-5})
\begin{equation}
| \hat{r}_1 \b 1_p' \hat{\b \Gamma}' \b \mu /p - r_1 \b 1_p' \hat{\b \Gamma}' \b \mu /p| \le | \hat{r}_1 - r_1 | 
| \b 1_p' \hat{\b \Gamma}' \b \mu/p|.\label{pc2-6}
\end{equation}

Then by Lemma B.3-B.6
\begin{equation}
| \hat{r}_1 - r_1 | = \left| (\hat{D}- \rho_1 \hat{F}) - (D - \rho_1 F) 
\right| = O_p (K^2 \bar{s} l_n).\label{pc2-7}
\end{equation}

Next by (\ref{e1}), Lemma B.3, $F:= \b 1_p' \b \Gamma \b \mu/p$
\begin{eqnarray}
\frac{\b 1_p' \hat{\b \Gamma}' \mu}{p}&=& \frac{\b \mu' \hat{\b \Gamma} \b 1_p}{p}  \le 
\frac{| \b \mu' \hat{\b \Gamma} \b 1_p - \b \mu' \b \Gamma \b 1_p|}{p} + F \nonumber \\
& = & O_p ( K \bar{s} l_n) + O (K^{1/2}) = O( K^{1/2}),\label{pc2-9}
\end{eqnarray}
where the last equality is by Assumption 8, $K \bar{s} l_n \to 0$. Combine (\ref{pc2-7})(\ref{pc2-9}) in (\ref{pc2-6})
 \begin{equation}
| \hat{r}_1 \b 1_p' \hat{\b \Gamma}' \b \mu /p - r_1 \b 1_p' \hat{\b \Gamma}' \b \mu /p|  = O_p (K^{5/2} \bar{s} l_n).\label{pc2-10}
\end{equation}

Then consider the second term on right side of  (\ref{pc2-5})

\begin{eqnarray}
|r_1 \b 1_p' \hat{\b \Gamma}' \b \mu /p - r_1 \b 1_p' \b \Gamma' \b \mu /p| & \le & | r_1| \left| \frac{\b 1_p' (\hat{\b \Gamma} - \b \Gamma)' \b \mu }{p}
\right| \nonumber \\
& = & O (K) O_p ( K \bar{s} l_n) = O_p ( K^2 \bar{s} l_n),\label{pc2-11}
\end{eqnarray}
by Lemma B.5
\begin{equation}
| r_1| = O(K),\label{pc2-10a}
\end{equation}
and (\ref{e1}). So use (\ref{pc2-10})(\ref{pc2-11}) in (\ref{pc2-5})
\begin{equation}
|  \hat{r}_1 \b 1_p' \hat{\b \Gamma}' \b \mu /p - r_1 \b 1_p' \b \Gamma \b \mu /p| = O_p ( K^{5/2} \bar{s} l_n).\label{pc2-11a}
\end{equation}

Consider the second term in (\ref{pc2-4})

\begin{eqnarray}
| \hat{r}_2 ( \hat{\b \mu}' \hat{\b \Gamma}' \b \mu /p) - r_2 (\b \mu' \b \Gamma \b \mu /p) |
&\le& | \hat{r}_2 ( \hat{\b \mu}' \hat{\b \Gamma}' \b \mu /p) - r_2 (\hat{\b \mu}' \hat{\b \Gamma}' \b \mu /p)|\nonumber \\
& + & | r_2 (\hat{\b \mu}' \hat{\b \Gamma}' \b \mu /p) - r_2 (\b \mu' \b \Gamma \b \mu/p) |.\label{pc2-12}
\end{eqnarray}
Take the first term on the right side of (\ref{pc2-12})
\begin{equation}
| \hat{r}_2 ( \hat{\b \mu}' \hat{\b \Gamma}' \b \mu /p) - r_2 (\hat{\b \mu}' \hat{\b \Gamma}' \b \mu /p)|
\le | \hat{r}_2 - r_2| | \frac{\hat{\b \mu}' {\hat{\b \Gamma}}' \mu }{p}|.\label{pc2-13}
\end{equation}
Analyze (\ref{pc2-13}) in the same way as in (\ref{pc2-7}) use, (\ref{pc2-0}), Lemma B.2-B.3,

\begin{equation}
| \hat{r}_2 - r_2 | = O_p ( K \bar{s} l_n).\label{pc2-13a}
\end{equation}

Then use $D:= \mu' \Gamma \mu/p$, and Lemma B.5 $D= O(K)$ with (\ref{osn0})(\ref{osn1}), by Assumption 8
\[ \hat{\b \mu}' \hat{\b \Gamma}' \b \mu/p = \b \mu' \hat{\b \Gamma} \hat{\b \mu}/p = O_p (K).\]
Next use the last two rates in (\ref{pc2-13}) 
\begin{equation}
| \hat{r}_2 ( \hat{\b \mu}' \hat{\b \Gamma}' \b \mu /p) - r_2 (\hat{\b \mu}' \hat{\b \Gamma}' \b \mu /p)|
= O_p (K^{2} \bar{s} l_n).\label{pc2-14}
\end{equation}

Then take the second term on the right side in (\ref{pc2-12})
\begin{eqnarray}
| r_2 [(\hat{\b \mu}' \hat{\b \Gamma}' \b \mu /p) -  (\b \mu' \b \Gamma \b \mu/p)] | & \le & 
| r_2| | [(\hat{\b \mu}' \hat{\b \Gamma}' \b \mu /p) -  (\b \mu' \b \Gamma \b \mu/p)] | \nonumber \\
& = & O (K^{1/2}) O_p (K^2 \bar{s} l_n) = O_p ( K^{5/2} \bar{s} l_n),\label{pc2-15}
\end{eqnarray}
since 
\begin{equation}
| r_2 | = O( K^{1/2}),\label{pc2-15a}
\end{equation}
by Lemma B.5, and $r_2$ definition, and we use (\ref{osn1}) for the other rate in (\ref{pc2-15}). Combine (\ref{pc2-14})(\ref{pc2-15}) for (\ref{pc2-12})

\begin{equation}
| \hat{r}_2 ( \hat{\b \mu}' \hat{\b \Gamma}' \b \mu /p) - r_2 (\b \mu' \b \Gamma \b \mu /p) |
= O_p ( K^{5/2} \bar{s} l_n).\label{pc2-16}
\end{equation}
Use (\ref{pc2-11a})(\ref{pc2-16}) in (\ref{pc2-4}), by Assumption 8
\[
(\hat{r}_1 \b 1_p' \hat{\b \Gamma}' \b \mu /p + \hat{r}_2 \hat{\b \mu}' \hat{\b \Gamma}' \b \mu/p) - (r_1 \b 1_p' \b \Gamma' \b \mu /p + r_2 \b \mu' \b \Gamma' \b \mu/p) =O_p (K^{5/2} \bar{s} l_n)=o_p (1).\]
Then since $\b w_{mv}' \b \mu = \rho_1 > c >0$, with $F:= \b 1_p' \b \Gamma \b \mu/p, D:= \b \mu' \b \Gamma \b \mu/p$
\[ \b w_{mv}' \b \mu= \frac{r_1 F + r_2 D}{AD - F^2} = \rho_1.\]
With $AD - F^2  \ge C_1 > 0$
we have 
\[ r_1 F + r_2 D \ge C_1 \rho_1 > 0.\]
Next numerator in (\ref{c.14a}) can be written (without squaring)
\begin{eqnarray}
\frac{(\hat{r}_1 \b 1_p' \hat{\b \Gamma}' \b \mu /p + \hat{r}_2 \hat{\b \mu}' \hat{\b \Gamma}' \b \mu/p)}{
(r_1 \b 1_p' \b \Gamma' \b \mu /p + r_2 \b \mu' \b \Gamma' \b \mu/p)} & \le & 
1 - \frac{ \left| (\hat{r}_1 \b 1_p' \hat{\b \Gamma}' \b \mu /p + \hat{r}_2 \hat{\b \mu}' \hat{\b \Gamma}' \b \mu/p) - (r_1 \b 1_p' \b \Gamma' \b \mu /p + r_2 \b \mu' \b \Gamma' \b \mu/p) \right|}{r_1 F + r_2 D} \nonumber \\
& \le & 1 + O_p ( K^{5/2} \bar{s} l_n).\label{c.32}
\end{eqnarray}

{\bf Step 2}\\

We analyze the terms in the denominator of (\ref{c.14a})
\begin{eqnarray}
\hat{r}_1^2 \left( \frac{\b 1_p' \hat{\b \Gamma}' \b \Sigma_y \hat{\b \Gamma} \b 1_p}{p}\right)& - &
r_1^2 \left( \frac{\b 1_p' \b \Gamma \b \Sigma_y \b \Gamma \b 1_p}{p}
\right) \nonumber \\
& + & 
\hat{r}_2^2 \left( \frac{ \hat{\b \mu}' \hat{\b \Gamma}' \b \Sigma_y \hat{\b \Gamma} \hat{\b \mu}}{p}\right) - 
r_2^2 \left( \frac{\b \mu' \b \Gamma \b \Sigma_y \b \Gamma \b \mu}{p}
\right) \nonumber \\
&+& 2 \hat{r}_1 \hat{r}_2 \left( \frac{\b 1_p' \hat{\b \Gamma}' \b \Sigma_y \hat{\b \Gamma} \hat{\b \mu}}{p}\right) - 
2 r_1 r_2\left( \frac{\b 1_p' \b \Gamma \b \Sigma_y \b \Gamma \b \mu}{p}
\right).\label{pc2-v0}
\end{eqnarray}

In (\ref{pc2-v0}) consider the first term on the right side by adding and subtracting

\begin{eqnarray}
\hat{r}_1^2 \left( \frac{\b 1_p' \hat{\b \Gamma}' \b \Sigma_y \hat{\b \Gamma} \b 1_p}{p}\right) - 
r_1^2 \left( \frac{\b 1_p' \b \Gamma \b \Sigma_y \b \Gamma \b 1_p}{p}
\right)& \le& (\hat{r}_1^2 - r_1^2) \left( \frac{\b 1_p' \hat{\b \Gamma}' 
 \b \Sigma_y  \hat{\b \Gamma} 1_p}{p}
\right) \nonumber \\
& + & r_1^2 \left[ \left( \frac{\b 1_p' \hat{\b \Gamma}' \b \Sigma_y \hat{\b \Gamma} \b 1_p}{p}\right) - \left( \frac{\b 1_p' \b \Gamma \b \Sigma_y 
 \b \Gamma \b 1_p}{p}
\right)\right].\label{pc2-v2}
\end{eqnarray}

In (\ref{pc2-v2}) the first right side term will be considered by adding and subtracting
\begin{eqnarray}
(\hat{r}_1^2 - r_1^2) \left( \frac{\b 1_p' \hat{\b \Gamma}' 
 \b \Sigma_y  \hat{\b \Gamma} 1_p}{p}
\right)  & \le & |\hat{r}_1^2 - r_1^2| \left[ \left( \frac{\b 1_p' \hat{\b \Gamma}' \b \Sigma_y \hat{\b \Gamma} \b 1_p}{p}\right) - \left( \frac{\b 1_p' \b \Gamma \b \Sigma_y 
 \b \Gamma \b 1_p}{p}
\right) 
\right]
\nonumber \\
& + & 
|\hat{r}_1^2 - r_1^2| \left[ \frac{\b 1_p' \b \Gamma \b 1_p}{p}
\right].\label{pc2-v3}
\end{eqnarray}

In (\ref{pc2-v3}), by (\ref{pc2-7}), and by Lemma B.5 $| r_1 | = O (K)$, and Assumption 8
\begin{eqnarray}
|\hat{r}_1^2 - r_1^2| & \le & | \hat{r}_1 - r_1 | [ |\hat{r}_1 + |r_1|] \nonumber \\
& = & O_p ( K^2 \bar{s} l_n) O_p (K) = O_p (K^3 \bar{s} l_n).\label{pc2-v4}
\end{eqnarray}
Using (\ref{pc6})-(\ref{pc8}) with (\ref{pc2-v4}) in the first term on the right side of (\ref{pc2-v3}), 
\begin{equation}
|\hat{r}_1^2 - r_1^2| \left[ \left( \frac{\b 1_p' \hat{\b \Gamma}' \b \Sigma_y \hat{\b \Gamma} \b 1_p}{p}\right) - \left( \frac{\b 1_p' \b \Gamma \b \Sigma_y 
 \b \Gamma \b 1_p}{p}
\right) 
\right] = O_p ( K^3 \bar{s} l_n) O_p (   \bar{s} l_n).\label{pc2-v4a}
\end{equation}
Next use Lemma B.5 with (\ref{pc2-v4}) on the second right side term in (\ref{pc2-v3})
\begin{equation}
|\hat{r}_1^2 - r_1^2| \left[ \frac{\b 1_p' \b \Gamma \b 1_p}{p}
\right] = O_p (K^3 \bar{s} l_n) O(1).\label{pc2-v8}
\end{equation}

In (\ref{pc2-v2}) the first right side term, use (\ref{pc2-v4a})(\ref{pc2-v8}), and since $ \bar{s}l_n^ = o(1)$ by Assumption 9
\begin{equation}
(\hat{r}_1^2 - r_1^2) \left( \frac{\b 1_p' \hat{\b \Gamma}' 
 \b \Sigma_y  \hat{\b \Gamma} 1_p}{p}
\right) = O_p (K^3 \bar{s} l_n).\label{pc2-v9}
\end{equation}

Analyze the second term on the right side of (\ref{pc2-v2}), by (\ref{pc6})-(\ref{pc8}) with Lemma B.5 $ |r_1| = O(K)$, and Assumption 9 $p \bar{s} l_n = o(1)$
\begin{equation}
r_1^2 \left[ \left( \frac{\b 1_p' \hat{\b \Gamma}' \b \Sigma_y \hat{\b \Gamma} \b 1_p}{p}\right) - \left( \frac{\b 1_p' \b \Gamma \b \Sigma_y 
 \b \Gamma \b 1_p}{p}
\right)\right] = O_p ( K^2) O_p ( \bar{s} l_n) = O_p (K^2 \bar{s} l_n).\label{pc2-v10}
\end{equation}

Clearly by (\ref{pc2-v9})(\ref{pc2-v10}) in (\ref{pc2-v2}), by Assumption 8
\begin{equation}
\hat{r}_1^2 \left( \frac{\b 1_p' \hat{\b \Gamma}' \b \Sigma_y \hat{\b \Gamma} \b 1_p}{p}\right) - 
r_1^2 \left( \frac{\b 1_p' \b \Gamma \b \Sigma_y \b \Gamma \b 1_p}{p}
\right) = O_p ( K^3 \bar{s} l_n) = o_p (1).\label{pc2-v10a}
\end{equation}

In (\ref{pc2-v0}) consider the second term on the right side
which will be upper bounded by adding and subtracting and triangle inequality
\begin{eqnarray}
\hat{r}_2^2 \left( \frac{ \hat{\b \mu}' \hat{\b \Gamma}' \b \Sigma_y \hat{\b \Gamma} \hat{\b \mu}}{p}\right) - 
r_2^2 \left( \frac{\b \mu' \b \Gamma \b \Sigma_y \b \Gamma \b \mu}{p}
\right) & \le & |\hat{r}_2^2 - r_2^2| \left| \frac{\hat{\b \mu}' \hat{\b \Gamma}' \b \Sigma_y \hat{\b \Gamma} \hat{\b \mu}}{p}
\right| \nonumber \\
& + & r_2^2  \left| \frac{\hat{\b \mu}' \hat{\b \Gamma}' \b \Sigma_y \hat{\b \Gamma} \hat{\b \mu}}{p} - 
\frac{\b \mu' \b \Gamma \b \Sigma_y \b \Gamma \b \mu}{p} \right|.\label{pc2-v12}
\end{eqnarray}
In (\ref{pc2-v12}) the first term on the right side will be analyzed by adding and subtracting
\begin{eqnarray}
|\hat{r}_2^2 - r_2^2| \left| \frac{\hat{\b \mu}' \hat{\b \Gamma}' \b \Sigma_y \hat{\b \Gamma} \hat{\b \mu}}{p}
\right| & \le & | \hat{r}_2^2 - r_2^2|  \left| \frac{\hat{\b \mu}' \hat{\b \Gamma}' \b \Sigma_y \hat{\b \Gamma} \hat{\b \mu}}{p} - 
\frac{\b \mu' \b \Gamma \b \Sigma_y \b \Gamma \b \mu}{p} \right| \nonumber \\
& + & | \hat{r}_2^2 - r_2^2| \left| \frac{\b \mu' \b \Gamma \b \Sigma_y \b \Gamma \b \mu}{p} \right|.\label{pc2-v13}
\end{eqnarray}
In (\ref{pc2-v13}) consider by (\ref{pc2-13a})(\ref{pc2-15a}) and Assumption 8
\begin{eqnarray}
 | \hat{r}_2^2 - r_2^2| & \le & | \hat{r}_2 - r_2| [| \hat{r}_2| + |r_2|] \nonumber \\
 & = & O_p ( K \bar{s} l_n) O_p ( K^{1/2}) = O_p (K^{3/2} \bar{s} l_n).\label{pc2-v14}
 \end{eqnarray}
 Then by (\ref{pc2-v14})(\ref{os9}) on the first right side term (\ref{pc2-v13})
\begin{equation}
| \hat{r}_2^2 - r_2^2|  \left| \frac{\hat{\b \mu}' \hat{\b \Gamma}' \b \Sigma_y \hat{\b \Gamma} \hat{\b \mu}}{p} - 
\frac{\b \mu' \b \Gamma \b \Sigma_y \b \Gamma \b \mu}{p} \right|
= O_p ( K^{3/2} \bar{s} l_n) O_p (K^2 \bar{s} l_n) = O_p ( K^{7/2} \bar{s}^2 l_n^2).\label{pc2-v14a}
\end{equation}
Then using Lemma B.5 for second term on the right side of (\ref{pc2-v13}) in combination with (\ref{pc2-v14})(\ref{pc2-v14a}) in (\ref{pc2-v13})
\begin{equation}
|\hat{r}_2^2 - r_2^2| \left| \frac{\hat{\b \mu}' \hat{\b \Gamma}' \b \Sigma_y \hat{\b \Gamma} \hat{\b \mu}}{p}
\right| = O_p (K^{7/2} \bar{s}^2 l_n^2) + O_p ( K^{5/2} \bar{s} l_n) = O_p (K^{5/2} \bar{s} l_n)=o_p (1),\label{pc2-v15}
\end{equation}
since $K^{5/2} \bar{s} l_n = o(1)$ by Assumption 8. Next consider the second term on right side of (\ref{pc2-v12}), with $\b \Gamma:= \b \Sigma_y^{-1}$,
and (\ref{pc2-15a}) with (\ref{pc2-v14a})
\begin{equation}
r_2^2  \left| \frac{\hat{\b \mu}' \hat{\b \Gamma}' \b \Sigma_y \hat{\b \Gamma} \hat{\b \mu}}{p} - 
\frac{\b \mu' \b \Gamma \b \Sigma_y \b \Gamma \b \mu}{p} \right| = O_p (K) O_p ( K^{7/2} \bar{s}^2 l_n^2) 
= O_p ( K^{5/2} \bar{s} l_n) O_p ( K^2 \bar{s} l_n)
= o_p (1),\label{pc2-v16}
\end{equation}
by Assumption 8. Use (\ref{pc2-v15})(\ref{pc2-v16}) in (\ref{pc2-v12}) to have 

\begin{equation}
\hat{r}_2^2 \left( \frac{ \hat{\b \mu}' \hat{\b \Gamma}' \b \Sigma_y \hat{\b \Gamma} \hat{\b \mu}}{p}\right) - 
r_2^2 \left( \frac{\b \mu' \b \Gamma \b \Sigma_y \b \Gamma \b \mu}{p}
\right) = 	O_p ( K^{5/2} \bar{s} l_n) + O_p ( K^{5/2} \bar{s} l_n) O_p (K^2 \bar{s} l_n) = o_p (1),\label{pc2-v17}
\end{equation}
by Assumption 8.

Consider the third right side term in (\ref{pc2-v0}), by adding and subtracting and triangle inequality

\begin{eqnarray}
2 \hat{r}_1 \hat{r}_2 \left( \frac{\b 1_p' \hat{\b \Gamma}' \b \Sigma_y \hat{\b \Gamma} \hat{\b \mu}}{p}\right) - 
2 r_1 r_2\left( \frac{\b 1_p' \b \Gamma \b \Sigma_y \b \Gamma \b \mu}{p}
\right) & \le & 2 |\hat{r}_1 \hat{r}_2 - r_1 r_2| \left| \frac{\b 1_p' \hat{\b \Gamma}' \b \Sigma_y \hat{\b \Gamma} \hat{\b \mu}}{p}
\right| \nonumber \\
& + & 2 | r_1| |r_2| \left| \frac{\b 1_p' \b \Gamma' \b \Sigma_y \b \Gamma \b \mu}{p}-\frac{\b 1_p' \hat{\b \Gamma}' \b \Sigma_y \hat{\b \Gamma} \hat{\b \mu}}{p} 
\right|.\label{pc2-v22}
\end{eqnarray}

Consider the first right side term in (\ref{pc2-v22})

\begin{eqnarray}
2 |\hat{r}_1 \hat{r}_2 - r_1 r_2| \left| \frac{\b 1_p' \hat{\b \Gamma}' \b \Sigma_y \hat{\b \Gamma} \hat{\b \mu}}{p}
\right| & \le & 2 |\hat{r}_1 \hat{r}_2 - r_1 r_2| \left| \frac{\b 1_p' \hat{\b \Gamma}' \b \Sigma_y \hat{\b \Gamma} \hat{\b \mu}}{p}
- \frac{\b 1_p' \b \Gamma' \b \Sigma_y \b \Gamma \b \mu}{p}\right| \nonumber \\
& + & 2 |\hat{r}_1 \hat{r}_2 - r_1 r_2| \left| \frac{\b 1_p' \b \Gamma' \b \Sigma_y \b \Gamma \b \mu}{p}\right|.\label{pc2-v23}
\end{eqnarray}

In (\ref{pc2-v23}), by (\ref{pc2-7})(\ref{pc2-10a})(\ref{pc2-13a})(\ref{pc2-15a}) and Assumption 8
\begin{eqnarray}
2 | \hat{r}_1 \hat{r}_2 - r_1 r_2| & \le &  2 | \hat{r}_1 - r_1| |\hat{r}_2| + 2 | r_1| |\hat{r}_2 - r_2| \nonumber \\
& = & O_p ( K^2 \bar{s} l_n) O_p (K^{1/2}) + O ( K) O_p (K \bar{s} l_n) = O_p (K^{5/2} \bar{s} l_n).\label{pc2-v24}
\end{eqnarray}
Next consider by adding and subtracting via triangle inequality
\begin{eqnarray}
\left| \frac{\b 1_p' \hat{\b \Gamma}' \b \Sigma_y \hat{\b \Gamma} \hat{\b \mu}}{p} - \frac{\b 1_p' \b \Gamma \b \Sigma_y \b \Gamma \b \mu}{p}
\right| &\le & \left| \frac{\b 1_p' (\hat{\b \Gamma} - \b \Gamma)' \b \Sigma_y (\hat{\b \Gamma} - \b \Gamma) (\hat{\b \mu} - \b \mu)}{p}
\right| \nonumber \\
& + & 
\left| \frac{\b 1_p' (\hat{\b \Gamma} - \b \Gamma)' \b \Sigma_y (\hat{\b \Gamma} - \b \Gamma) \b \mu}{p}
\right| 
 +  \left| \frac{\b 1_p' (\hat{\b \Gamma} - \b \Gamma)' \b \Sigma_y \b \Gamma  \b \mu)}{p}
\right| \nonumber \\
&+ & \left| \frac{\b 1_p' (\hat{\b \Gamma} - \b \Gamma)' \b \Sigma_y \b \Gamma (\hat{\b \mu} - \b \mu)}{p}
\right| 
 +  \left| \frac{\b 1_p' \b \Gamma \b \Sigma_y (\hat{\b \Gamma} - \b \Gamma) (\hat{\b \mu} - \b \mu)}{p}
\right| \nonumber \\
& + & \left| \frac{\b 1_p'  \b \Gamma \b \Sigma_y (\hat{\b \Gamma} - \b \Gamma) \b \mu}{p}
\right| 
+  \left| \frac{\b 1_p'  \b \Gamma \b \Sigma_y \b \Gamma (\hat{\b \mu} - \b \mu)}{p}
\right|.\label{pc2-v25}
 \end{eqnarray}
 Consider the first term on the right side of (\ref{pc2-v25}) via Cauchy Schwartz inequality
{\small\begin{eqnarray}
\left| \frac{\b 1_p' (\hat{\b \Gamma} - \b \Gamma)' \b \Sigma_y (\hat{\b \Gamma} - \b \Gamma) (\hat{\b \mu} - \b \mu)}{p}
\right| & \le & 
\left[ \frac{\b 1_p' (\hat{\b \Gamma} - \b \Gamma)' \b \Sigma_y (\hat{\b \Gamma} - \b \Gamma) \b 1_p}{p}
\right]^{1/2}
\left[ \frac{(\hat{\b \mu} - \b \mu)' (\hat{\b \Gamma} - \b \Gamma)' \b \Sigma_y (\hat{\b \Gamma} - \b \Gamma) (\hat{\b \mu} - \b \mu)}{p}\right]^{1/2}
\nonumber \\
& = & [ O_p (p \bar{s}^2 l_n^2)]^{1/2} [ O_p ( p \bar{s}^2 l_n^2 max( K^2 lnn/n, lnp/n))]^{1/2},\label{pc2-v26}
\end{eqnarray}}
by  the analysis in (\ref{os4}) with Theorem 2(ii). In the same way as in (\ref{pc2-v26}), using (\ref{os4}) with  (\ref{mui})
{\small\begin{eqnarray}
\left|  \frac{\b 1_p' (\hat{\b \Gamma} - \b \Gamma)' \b \Sigma_y (\hat{\b \Gamma} - \b \Gamma) \b \mu}{p}
\right| &\le& \left[\frac{\b 1_p' (\hat{\b \Gamma} - \b \Gamma)' \b \Sigma_y (\hat{\b \Gamma} - \b \Gamma) \b 1_p}{p}\right]^{1/2}
\left[\frac{\b \mu' (\hat{\b \Gamma} - \b \Gamma)' \b \Sigma_y (\hat{\b \Gamma} - \b \Gamma) \b \mu}{p}\right]^{1/2} \nonumber \\
& = & [O_p (p \bar{s}^2 l_n^2)]^{1/2} [ O_p (K^2 p \bar{s}^2 l_n^2)]^{1/2}= O_p ( p \bar{s} l_n) O_p (K \bar{s} l_n).\label{pc2-v27} 
\end{eqnarray}}

Next, take the third term on the right side of (\ref{pc2-v25}), with $\b \Gamma:= \b \Sigma_y^{-1}$

\begin{equation}
\left| \frac{\b 1_p' (\hat{\b \Gamma} - \b \Gamma)' \b \Sigma_y \b \Gamma  \b \mu}{p}
\right| = \left| \frac{\b \mu' (\hat{\b \Gamma} - \b \Gamma) \b 1_p}{p}
\right| = O_p (K \bar{s} l_n),\label{pc2-v28}
\end{equation}
by the same analysis in (\ref{la.22}). Fourth term on the right side of (\ref{pc2-v25}) can use the same analysis in (\ref{la.21}), $\b \Gamma:= \b \Sigma_y^{-1}$

\begin{equation}
\left| \frac{\b 1_p' (\hat{\b \Gamma} - \b \Gamma)' \b \Sigma_y \b \Gamma (\hat{\b \mu} - \b \mu)}{p}
\right| = \left| \frac{(\hat{\b \mu} - \b \mu)' (\hat{\b \Gamma}- \b \Gamma) \b 1_p}{p} 
\right| = O_p (\bar{s} l_n) O_p ( \max(K \sqrt{ln/n}, lnp/n)).\label{pc2-v29}
\end{equation}

Then fifth term on the right side of (\ref{pc2-v25}) is

\begin{equation}
\left| \frac{\b 1_p' \b \Gamma \b \Sigma_y (\hat{\b \Gamma} - \b \Gamma) (\hat{\b \mu} - \b \mu)}{p}
\right| = \left| \frac{\b 1_p'  (\hat{\b \Gamma} - \b \Gamma) (\hat{\b \mu} - \b \mu)}{p}
\right| =O_p (\bar{s} l_n) O_p ( \max(K \sqrt{ln/n}, lnp/n)),\label{pc2-v30}
\end{equation}
by (\ref{la.21}).

The sixth term on the right side of (\ref{pc2-v25}) is 
\begin{equation}
\left| \frac{\b 1_p'  \b \Gamma \b \Sigma_y (\hat{\b \Gamma} - \b \Gamma) \b \mu}{p}
\right|  = \left| \frac{\b 1_p'   (\hat{\b \Gamma} - \b \Gamma) \b \mu}{p}
\right| = O_p (K \bar{s} l_n),\label{pc2-v31}
\end{equation}
by (\ref{la.22}). Seventh term on the right side of (\ref{pc2-v25}) is
\begin{equation}
\left| \frac{\b 1_p'  \b \Gamma \b \Sigma_y \b \Gamma (\hat{\b \mu} - \b \mu)}{p}
\right| = \left|
\frac{\b 1_p' \b \Gamma (\hat{\b \mu} - \b \mu)}{p}
\right| = O_p ( K^{3/2} \bar{s} r_n) O_p (\max (K \sqrt{lnn/n}, \sqrt{lnp/n})),\label{pc2-v32}
\end{equation}
by (\ref{la.23}). Among all right side terms in (\ref{pc2-v25}), the slowest rate are (\ref{pc2-v28})(\ref{pc2-v31}), as can be seen by Assumption 8-9, and $l_n$ definition. Hence
\begin{equation}
\left| \frac{\b 1_p' \hat{\b \Gamma}' \b \Sigma_y \hat{\b \Gamma} \hat{\b \mu}}{p} - \frac{\b 1_p' \b \Gamma \b \Sigma_y \b \Gamma \b \mu}{p}
\right| = O_p ( K \bar{s} l_n).\label{pc2-v33}
\end{equation}
Now, combine (\ref{pc2-v23})(\ref{pc2-v24})(\ref{pc2-v33}) in the first right side term (\ref{pc2-v22}), with $\b \Gamma:= \b \Sigma_y^{-1}$, Lemma B.5 (i.e. $|F|=O(K^{1/2})$)

\begin{equation}
| 2 \hat{r}_1 \hat{r}_2 - 2 r_1 r_2| \left| \frac{\b 1_p' \hat{\b \Gamma}' \b \Sigma_y \hat{\b \Gamma} \hat{\b \mu}}{p} 
\right| = O_p (K^{5/2} \bar{s} l_n) O_p (K \bar{s} l_n) + O_p ( K^{5/2} \bar{s} l_n) O_p (K^{1/2}) =O_p (K^3 \bar{s} l_n),\label{pc2-v34}
\end{equation}
where the last rate is by Assumption 8. Consider (\ref{pc2-v33}) and $|r_1| = O(K)$, $|r_2 | = O(K^{1/2})$ by (\ref{pc2-10a})(\ref{pc2-15a}), substituted into second term in right side of (\ref{pc2-v22}) 
\begin{equation}
2 | r_1 r_2| \left| \frac{\b 1_p' \b \Gamma' \b \Sigma_y \b \Gamma \b \mu}{p}-\frac{\b 1_p' \hat{\b \Gamma}' \b \Sigma_y \hat{\b \Gamma} \hat{\b \mu}}{p} \right|
= O_p (K^{3/2}) O_p (K \bar{s} l_n)= O_p(K^{5/2} \bar{s} l_n).\label{pc2-v35}
\end{equation}
So by (\ref{pc2-v34})(\ref{pc2-v35}) in left side term in (\ref{pc2-v22}), by Assumption 8
\begin{equation}
2 \hat{r}_1 \hat{r}_2 \left( \frac{\b 1_p' \hat{\b \Gamma}' \b \Sigma_y \hat{\b \Gamma} \hat{\b \mu}}{p}\right) - 
2 r_1 r_2\left( \frac{\b 1_p' \b \Gamma \b \Sigma_y \b \Gamma \b \mu}{p}
\right) = O_p ( K^3 \bar{s} l_n)=o_p (1).\label{pc2-v36}
\end{equation}

Next clearly by (\ref{pc2-v10a})(\ref{pc2-v17})(\ref{pc2-v36}) in (\ref{pc2-v0})
\begin{eqnarray}
 |\hat{r}_1^2 \left( \frac{\b 1_p' \hat{\b \Gamma}' \b \Sigma_y \hat{\b \Gamma} \b 1_p}{p}\right)& - &
r_1^2 \left( \frac{\b 1_p' \b \Gamma \b \Sigma_y \b \Gamma \b 1_p}{p}
\right) 
 +  
\hat{r}_2^2 \left( \frac{ \hat{\b \mu}' \hat{\b \Gamma}' \b \Sigma_y \hat{\b \Gamma} \hat{\b \mu}}{p}\right) - 
r_2^2 \left( \frac{\b \mu' \b \Gamma \b \Sigma_y \b \Gamma \b \mu}{p}
\right) \nonumber \\
&+& 2 \hat{r}_1 \hat{r}_2 \left( \frac{\b 1_p' \hat{\b \Gamma}' \b \Sigma_y \hat{\b \Gamma} \hat{\b \mu}}{p}\right) - 
2 r_1 r_2\left( \frac{\b 1_p' \b \Gamma \b \Sigma_y \b \Gamma \b \mu}{p}
\right)|= O_p (K^3 \bar{s} l_n) = o_p (1).\label{c.63a}
\end{eqnarray}

Next the denominator in (\ref{c.14a}) can be written as
\begin{equation}
\frac{p[\hat{r}_1^2 \left( \frac{\b 1_p' \hat{\b \Gamma}' }{p} \b \Sigma_y \frac{\hat{\b \Gamma} \b 1_p}{p}
\right) + 
2 \hat{r}_1 \hat{r}_2 \left( \frac{\b 1_p' \hat{\b \Gamma}' }{p} \b \Sigma_y \frac{\hat{\b \Gamma} \hat{\b \mu}}{p}
\right)
 + \hat{r}_2^2 \left( \frac{\hat{\b \mu}' \hat{\b \Gamma}' }{p} \b \Sigma_y \frac{\hat{\b \Gamma}  \hat{\b \mu}}{p}
\right)]}{p[r_1^2 \left( \frac{\b 1_p' \b \Gamma' }{p} \b \Sigma_y \frac{\b \Gamma \b 1_p}{p}
\right) + 
2 r_1 r_2 \left( \frac{\b 1_p' \b \Gamma' }{p} \b \Sigma_y \frac{\b \Gamma \b \mu}{p}
\right)
 + r_2^2 \left( \frac{\b \mu' \b \Gamma' }{p} \b \Sigma_y \frac{\b \Gamma  \b \mu}{p}
\right)]}
\label{c.64}
\end{equation}
The ratio in (\ref{c.64}) is  greater than equal to the following term
\begin{equation}
1 - \frac{\left| \hat{r}_1^2 \left( \frac{\b 1_p' \hat{\b \Gamma}' \b \Sigma_y \hat{\b \Gamma} \b 1_p}{p}\right) - 
r_1^2 \left( \frac{\b 1_p' \b \Gamma \b \Sigma_y \b \Gamma \b 1_p}{p}
\right) 
 +  
\hat{r}_2^2 \left( \frac{ \hat{\b \mu}' \hat{\b \Gamma}' \b \Sigma_y \hat{\b \Gamma} \hat{\b \mu}}{p}\right) - 
r_2^2 \left( \frac{\b \mu' \b \Gamma \b \Sigma_y \b \Gamma \b \mu}{p}
\right) 
+ 2 \hat{r}_1 \hat{r}_2 \left( \frac{\b 1_p' \hat{\b \Gamma}' \b \Sigma_y \hat{\b \Gamma} \hat{\b \mu}}{p}\right) - 
2 r_1 r_2\left( \frac{\b 1_p' \b \Gamma \b \Sigma_y \b \Gamma \b \mu}{p}
\right)\right|}{ [r_1^2 \left( \frac{\b 1_p' \b \Gamma \b \Sigma_y \b \Gamma \b 1_p}{p}
\right) +r_2^2 \left( \frac{\b \mu' \b \Gamma \b \Sigma_y \b \Gamma \b \mu}{p}
\right)+2 r_1 r_2\left( \frac{\b 1_p' \b \Gamma \b \Sigma_y \b \Gamma \b \mu}{p}
\right)]}.\label{c.64a}
\end{equation}
By  $r_1, r_2, A, F,D$ definitions, and Assumption $AD - F^2 \ge C_1 > 0,  A \rho_1^2 - 2 \rho_1 F + D \ge C_1 > 0$
\[ [r_1^2 \left( \frac{\b 1_p' \b \Gamma \b \Sigma_y \b \Gamma \b 1_p}{p}
\right) +r_2^2 \left( \frac{\b \mu' \b \Gamma \b \Sigma_y \b \Gamma \b \mu}{p}
\right)+2 r_1 r_2\left( \frac{\b 1_p' \b \Gamma \b \Sigma_y \b \Gamma \b \mu}{p}
\right)] = (A \rho_1^2 - 2 \rho_1 F + D)(AD - F^2) \ge C_1^2 >0,\]
the term in (\ref{c.64a}) converges in probability to one by (\ref{c.63a}).

{\bf Step 3}.

Combine steps 1-2 to  have the desired result.{\bf Q.E.D.}

{\bf Proof of Corollary 3}.

See that by $\b \Gamma:= \b \Sigma_y^{-1}$ and symmetric $\Gamma$
\begin{equation}
\frac{\widehat{MSR}_{p}^2}{MSR^2} = \frac{( \hat{\b \mu}' \hat{\b \Gamma}' \b \mu/p)/(\b \mu' \b \Gamma \b \mu/p)}{
(\hat{\b \mu}' \hat{\b \Gamma}' \b \Sigma_y \hat{\b \Gamma} \hat{\b \mu}/p)/(\b \mu' \b \Gamma \b \mu/p)}.\label{pms1}
\end{equation}

By (\ref{osn3}) in the numerator above
\begin{equation}
\frac{\hat{\b \mu}' \hat{\b \Gamma}' \b \mu/p}{\b \mu' \b \Gamma \b \mu/p} \le 1 + O_p ( K^2 \bar{s} l_n).\label{pms3}
\end{equation}

By (\ref{os0})(\ref{a41a}) with Assumption 8(ii) in the denominator of (\ref{pms1})

\begin{equation}
\frac{(\hat{\b \mu}' \hat{\b \Gamma}' \b \Sigma_y \hat{\b \Gamma} \hat{\b \mu}/p)}{(\b \mu' \b \Gamma \b \mu/p)} \ge 1 - o_p (1).\label{pms4}
\end{equation}

Use (\ref{pms3})(\ref{pms4}) in (\ref{pms1}) to get the result.{\bf Q.E.D.}

\setcounter{thm}{0}
\setcounter{equation}{0}\setcounter{lemma}{0}\setcounter{assum}{0}\renewcommand{\theequation}{D.%
\arabic{equation}}\renewcommand{\thelemma}{D.\arabic{lemma}}%
\renewcommand{\theassum}{D.\arabic{assum}}%
\renewcommand{\baselinestretch}{1}\baselineskip=15pt
\renewcommand{\thethm}{D.\arabic{thm}}%

\section*{Supplement D}

In this part we consider mean-variance efficiency of large portfolio in an out-of-sample context, and also we add a simulation to show the effects of sparsity on our and other methods.

\subsection*{Mean-Variance Efficiency}

This Supplement formally shows that we can obtain mean-variance efficiency in an out-of-sample context. \cite{ao2019} show that this is possible when $p \le n$, when both $p$, and $n$ are large. That article is a significant contribution since they also demonstrate that other methods before theirs could not obtain that result, and it is a difficult issue to address. We are interested in maximized out-of-sample expected return $\b\mu'\b w_{mos}$ and its estimate $\b\mu'\widehat{\b w}_{mos}$. Additionally, we are interested in the out-of-sample variance
of the portfolio returns $\b w_{mos}'\b\Sigma_y w_{mos}$ and its estimate $\widehat{\b w}_{mos}'\b\Sigma_y \widehat{\b w}_{mos}$. Note also that by the formula for weights
$\b w_{mos}' \b\Sigma_y \b w_{mos} = \sigma^2$, given $\b\Gamma:=\b\Sigma_y^{-1}$.

Below, we show that our estimates based on nodewise regression are consistent, and furthermore, we also provide the rate of convergence results.

\begin{thm}\label{mvef}
(i). Under Assumptions \ref{as1}-\ref{as4}, \ref{as6},\ref{as7}(i),\ref{as8}
\[ \left|  \frac{\b\mu' \widehat{\b w}_{mos}}{\b\mu' \b w_{mos}} - 1
\right| = O_p ( K^2 \bar{s} l_n) = o_p(1).\]

(ii). Under Assumptions \ref{as1}-\ref{as4},\ref{as6},\ref{as7}(i),\ref{as8}, \ref{as9}
\[\left| \widehat{\b w}_{mos}' \b\Sigma_y \widehat{\b w}_{mos} - \sigma^2 \right| = O_p ( K^2 \bar{s} l_n) = o_p (1).\]
\end{thm}





\noindent
{\bf Proof of Theorem \ref{mvef}}. (i).
Start with definition of weights, and its estimators
\begin{eqnarray}
 \left( \frac{\b\mu' \widehat{\b \Gamma} \widehat{\b \mu}}{\b\mu' \b\Gamma \b\mu}
 \right) & \times &
 \left(
 \frac{\b\mu' \b\Gamma \b\mu}{\widehat{\b \mu}' \widehat{\b \Gamma}
 \widehat{\b \mu}}
 \right)^{1/2} -  1 \nonumber \\
 & \le &
 \left[
 \left|\frac{\b\mu' \widehat{\b \Gamma} \widehat{\b \mu}}{\b\mu' \b\Gamma \b\mu} -1
 \right|+1
 \right]
 \left[ \left|
 \left(\frac{\b\mu' \b\Gamma \b\mu}{\widehat{\b \mu}' \widehat{\b \Gamma}
 \widehat{\b \mu}}
 \right)^{1/2}
  -1 \right|
 +1
\right]
-1  \nonumber \\
 & = &
\left|\frac{\b\mu' \widehat{\b \Gamma} \widehat{\b \mu}}{\b\mu' \b\Gamma \b\mu} -1
 \right|
\left|
 \left(\frac{\b\mu' \b\Gamma \b\mu}{\widehat{\b \mu}' \widehat{\b \Gamma}
 \widehat{\b \mu}}
 \right)^{1/2}
  -1  \right| \left|\frac{\mu' \widehat{\b \Gamma} \widehat{\b \mu}}{\b\mu' \b\Gamma \b\mu} -1
 \right| + \left|
 \left(\frac{\b\mu' \b\Gamma \b\mu}{\widehat{\b \mu}' \widehat{\b \Gamma}
 \widehat{\b \mu}}
 \right)^{1/2}
  -1  \right|\label{pt3i2}
 \end{eqnarray}

\noindent By (\ref{osn3})
\begin{equation}
\left|\frac{\mu' \widehat{\b \Gamma} \widehat{\b \mu}}{\b\mu' \b\Gamma \b\mu} -1
 \right|= O_p ( K^2 \bar{s} l_n).\label{pt3i3}
 \end{equation}

\noindent Next, we have
\begin{eqnarray}
  \frac{\b\mu' \b\Gamma \b\mu}{\widehat{\b \mu} \widehat{\b \Gamma} \widehat{\b \mu}}& = & \frac{\b\mu' \b\Gamma \b\mu -
  \widehat{\b \mu}' \widehat{\b \Gamma} \widehat{\b \mu}}{\widehat{\b \mu}'
  \widehat{\b \Gamma} \widehat{\b \mu}} + 1 \nonumber \\
  & \le & \frac{|\b\mu'\b\Gamma\b\mu/p - \widehat{\b \mu}' \widehat{\b \Gamma} \widehat{\b \mu}/p|}{\b\mu'\b\Gamma\b\mu/p -
  | \widehat{\b \mu}' \widehat{\b \Gamma} \widehat{\b \mu}/p - \b\mu'\b\Gamma\b\mu/p|} + 1,\label{pt3i4}
 \end{eqnarray}
where we divided both the numerator and denominator by $p$, and
\[ \widehat{\b \mu}' \widehat{\b \Gamma} \widehat{\b \mu}/p \ge \b \mu'  \b \Gamma \b \mu/p - | \widehat{\b \mu}' \widehat{\b \Gamma} \widehat{\b \mu}/p - \b\mu'\b\Gamma\b\mu/p|.\]

\noindent By (\ref{d4}),(\ref{pt3i4}), Lemma B.4 in the Supplement B, and $K^2 \bar{s} l_n = o(1)$ via Assumption \ref{as8} in the denominator below in (\ref{pt3i5})
\begin{equation}
     \frac{\b\mu' \b\Gamma \b\mu}{\widehat{\b \mu}'
     \widehat{\b \Gamma} \widehat{\b \mu}} \le \frac{O_p (K^2 \bar{s} l_n)}{c  - O_p ( K^2 \bar{s} l_n)} + 1 =
     O_p ( K^2 \bar{s} l_n) + 1
 .\label{pt3i5}
 \end{equation}
Then,
\begin{equation}
     \left|
     \left( \frac{\b\mu' \b\Gamma \b\mu}{\widehat{\b \mu}' \widehat{\b \Gamma} \widehat{\b \mu}}
     \right)^{1/2} -1
     \right| = \{ [ 1+ O_p (K^2 \bar{s} l_n)]^{1/2} -1
     \}\label{pt3i6}
      \end{equation}
Now, use Assumption \ref{as8} in (\ref{pt3i3})(\ref{pt3i6}) and  (\ref{pt3i2}) to obtain the desired result.

\begin{flushright}
{\bf Q.E. D}
\end{flushright}

(ii). Now, we analyze the risk. See that
\[ \widehat{\b w}_{mos}' \b\Sigma_y \widehat{\b w}_{mos} - \sigma^2 = \sigma^2 \left(
\frac{\widehat{\b \mu}' \widehat{\b \Gamma}' \b\Sigma_y \widehat{\b \Gamma} \widehat{\b \mu}}{\widehat{\b \mu}' \widehat{\b \Gamma} \widehat{\b \mu}} - 1
\right) = \sigma^2 \left( \frac{ \frac{\widehat{\b \mu}' \widehat{\b \Gamma}' \b\Sigma_y \widehat{\b \Gamma} \widehat{\b \mu}}{\b \mu' \b \Gamma \b \mu}}{\frac{\widehat{\b \mu}' \widehat{\b \Gamma} \widehat{\b \mu}}{\b \mu' \b \Gamma \b \mu}}
-1 \right),\]
where we multiplied and divided by $\b \mu' \b \Gamma \b \mu $, which is positive by (\ref{d4}). By (\ref{a41a}), since $\b \Gamma:= \b\Sigma_y^{-1}$, Assumption \ref{as9}
\begin{equation}
\left| \frac{\widehat{\b \mu}' \widehat{\b \Gamma}' \b\Sigma_y \widehat{\b \Gamma} \widehat{\b \mu}}{\b\mu' \b\Gamma \b\mu} -1
\right| =O_p ( K^2 \bar{s} l_n).\label{os19}
\end{equation}
Additionally, by Lemma \ref{tl4} in Supplement B  and (\ref{d4})
\begin{equation}
| \frac{\widehat{\b \mu}' \widehat{\b \Gamma} \widehat{\b \mu}/p}{\b \mu' \b \Gamma \b \mu/p}|
 \ge \frac{|\widehat{\b \mu}' \widehat{\b \Gamma} \widehat{\b \mu}/p|}{\frac{\b \mu' \b \Gamma \b \mu}{p} - | \b \mu' \b \Gamma \b \mu/p - \widehat{\b \mu}' \widehat{\b \Gamma} \widehat{\b \mu}/p|}
  = \frac{1 + O_p (K^2 \bar{s} l_n)}{c - o_p (1)} = 1+ o_p (1).\label{os20}
\end{equation}
By (\ref{os19}), (\ref{os20}) and Assumption \ref{as8},
\[ | \widehat{w}_{oos} \b\Sigma_y \widehat{\b w}_{oos} - \sigma^2| = O_p ( K^2 \bar{s} l_n) = o_p (1).\]

\begin{flushright}
{\bf Q.E.D.}
\end{flushright}

\subsection*{Effects of Sparsity}

This section of the Supplement show a small simulation with a Block Diagonal covariance matrix for the idiosyncratic part of the dgp. The dgp is the same from section 5 but with $\boldsymbol{\widehat{\b\Sigma_n}} = \widetilde{\b\Sigma_n} \odot \textnormal{BLDiag}({\b b})$, where $\textnormal{BLDiag} ({\b b})$ is the $p \times p$ block diagonal matrix with $\b b$ blocks of ones. Moreover, this simulation was only performed for $n = 200$ and for the plug-in models with block sizes of 5, 15 and 50. The objective is to look at the behavior of Nodewise Regression on different sparsity levels in the covariance matrix. 

We analyze two questions whether our methods are doing well compared to others when the model is less sparse, and then see whether sparsity effects are uniform over analysis of various Sharpe Ratio cases in Section 4. 

First, from Table 6, our methods do well in high dimensional cases, our method has the smallest error in 5 out of 12 cases, POET method has high errors in all cases. In case of low dimensions, non-linear shrinkage is the best method, POET does again poorly. Also in less sparse case of blocks with 50, in high dimensions, we get the least error in 2 cases, and the other 2 cases non-linear shrinkage gets the least errors.

Regarding the analysis of our method in various Sharpe Ratio cases, in case of the constrained maximum Sharpe Ratio (MSR), our errors are smaller with increased block size. To give an example, NW-GIC has 0.379 in high dimensional case with 5 as block size, and this decreases to 0.100 with block size of 50. In case of  Markowitz portfolio we see that  increasing the block size does not affect our errors much differently. Our method is affected by non-sparsity in maximum out-of-sample Sharpe Ratio as predicted by our Theorem 8. 

\begin{table}[htb]
\caption{Simulation Results -- Block Diagonal DGP with Real Factors}
\label{tab:sum_blk}
\begin{adjustbox}{max width=\textwidth}
\begin{threeparttable}
\begin{tabular}{lccccccccc}
\hline
         & {\ul }         & {\ul }         & {\ul }         & \multicolumn{3}{c}{{\ul Block Size = 5}}  & {\ul }         & {\ul }         & {\ul }         \\
         & \multicolumn{4}{c}{{\ul Low Dim}}                                  & {\ul } & \multicolumn{4}{c}{{\ul High Dim}}                                \\
         & MSR            & OOS-MSR        & GMV-SR         & MKW-SR          &        & MSR            & OOS-MSR        & GMV-SR         & MKW-SR         \\ \cline{2-5} \cline{7-10} 
NW-GIC   & 0.353          & \textbf{0.66}  & \textbf{0.24}  & 0.121           &        & 0.379          & \textbf{0.763} & 0.207          & 0.149          \\
NW-CV    & 0.352          & 0.667          & 0.237          & 0.121           &        & 0.379          & 0.768          & \textbf{0.206} & 0.149          \\
POET     & 0.412          & 1.041          & 0.432          & 0.347           &        & 0.438          & 2.364          & 0.485          & 0.403          \\
NL-LW    & \textbf{0.33}  & 0.980          & 0.289          & \textbf{0.08}   &        & \textbf{0.340} & 1.458          & 0.222          & \textbf{0.119} \\
SF-NL-LW & 0.348          & 0.683          & 0.237          & 0.139           &        & 0.367          & 0.799          & 0.208          & 0.145          \\
         &                &                &                &                 &        &                &                &                &                \\
         &                &                &                & \multicolumn{3}{c}{{\ul Block Size = 15}} &                &                &                \\
NW-GIC   & 0.324          & 0.906          & 0.253          & 0.135           &        & 0.346          & 1.075          & 0.239          & 0.157          \\
NW-CV    & 0.329          & 0.894          & 0.252          & 0.140           &        & 0.353          & \textbf{1.065} & 0.244          & 0.164          \\
POET     & 0.380          & 1.366          & 0.444          & 0.361           &        & 0.405          & 2.774          & 0.504          & 0.408          \\
NL-LW    & \textbf{0.304} & 1.058          & 0.283          & \textbf{0.071}  &        & \textbf{0.307} & 1.701          & 0.241          & \textbf{0.103} \\
SF-NL-LW & 0.317          & \textbf{0.863} & \textbf{0.230} & 0.132           &        & 0.332          & 1.097          & \textbf{0.228} & 0.139          \\
         &                &                &                &                 &        &                &                &                &                \\
         &                &                &                & \multicolumn{3}{c}{{\ul Block Size = 50}} &                &                &                \\
NW-GIC   & \textbf{0.149} & 2.362          & 0.316          & 0.107           &        & 0.100          & 2.947          & 0.260          & \textbf{0.070} \\
NW-CV    & 0.228          & 1.770          & 0.319          & 0.159           &        & 0.163          & 2.506          & \textbf{0.253} & 0.125          \\
POET     & 0.190          & 3.444          & 0.426          & 0.313           &        & 0.151          & 5.617          & 0.393          & 0.320          \\
NL-LW    & 0.176          & 1.312          & 0.571          & 0.103           &        & \textbf{0.092} & 2.843          & 0.359          & 0.240          \\
SF-NL-LW & 0.159          & \textbf{1.225} & \textbf{0.308} & \textbf{0.080}  &        & 0.094          & \textbf{2.331} & 0.268          & 0.074          \\ \hline
\end{tabular}%
\begin{tablenotes}
\item The table shows the simulation results for the block DGP. Each simulation was done with 100 iterations. We used a single sample size of $n=200$ and the number of stocks was either $n/2$ or $1.5n$ for the low-dimensional and the high-dimensional case, respectively. Each block of rows shows the results for a different block size (5, 15, 50) in the block diagonal DGP. The values in each cell show the average absolute estimation error for estimating the square of the Sharpe Ratio.
\end{tablenotes}
\end{threeparttable}

\end{adjustbox}
\end{table}

\end{document}